\documentclass[3p, twocolumn, number]{elsarticle}
\usepackage[utf8]{inputenc}
\usepackage{textcomp}
\usepackage{multirow}
\usepackage{nidanfloat}
\usepackage{graphicx}
\usepackage{mathtools}
\usepackage{amssymb}
\usepackage{amsmath}
\usepackage{bm}
\usepackage{siunitx}
\usepackage{longtable,tabularx}
\usepackage{float}
\usepackage{booktabs}
\usepackage{etoolbox}
\usepackage{appendix}
\usepackage{subcaption}
\usepackage{xcolor}
\usepackage{picins}

\usepackage{url}
\usepackage{tikz}
\usepackage{pgfplots}
\usetikzlibrary{calc}
\usetikzlibrary{shapes, arrows}
\usetikzlibrary{positioning}
\usepackage{color}
\usepackage{circuitikz}

\newcommand{\isdef}{\stackrel{\triangle}{=}}
\newcommand{\nn}{\nonumber}

\usetikzlibrary{shapes,arrows,calc,positioning}
\tikzstyle{bigblock} = [draw, fill=blue!20, rectangle,
minimum height=6em, minimum width=8em]
\tikzstyle{medblock} = [draw, fill=blue!20, rectangle,
minimum height=4em, minimum width=4em]
\tikzstyle{mux} = [draw, fill=black, rectangle,
minimum height=5em, minimum width=0.01em]
\tikzstyle{smallblock} = [draw, fill=blue!20, rectangle, minimum height=2em, minimum width=3em]
\tikzstyle{sum} = [draw, fill=blue!20, circle, node distance=1cm]
\tikzstyle{signal} = [coordinate]
\tikzstyle{pinstyle} = [pin edge={to-,thin,black}]
\tikzstyle{block} = [draw, fill=blue!20, rectangle,
minimum height=3em, minimum width=6em]
\tikzstyle{blockS} = [draw, fill=blue!20, rectangle,
minimum height=3em, minimum width=4em]
\tikzstyle{sum} = [draw, fill=blue!20, circle, node distance=1.5cm]
\tikzstyle{gain} = [draw, fill=blue!20, regular polygon, regular polygon sides = 3, node distance=1.25cm, shape border rotate = -90]
\tikzstyle{mult} = [draw, fill=blue!20, circle, node distance=1.25cm,inner sep=0pt, minimum size = 0.3cm]
\tikzstyle{input} = [coordinate]
\tikzstyle{output} = [coordinate]
\tikzstyle{saturation block} = [draw,fill=blue!20, 
path picture={
	\pgfpointdiff{\pgfpointanchor{path picture bounding box}{north east}}%
	{\pgfpointanchor{path picture bounding box}{south west}}
	\pgfgetlastxy\x\y
	\tikzset{x=\x*.4, y=\y*.4}
	\draw (-1,0) -- (1,0) (0,-1) -- (0,1); 
	\draw (-1,-.5) -- (-.5,-.5) -- (.5,.5) -- (1,.5);
}]
\pgfplotsset{compat=1.15}

\begin{document}


\begin{frontmatter}

\title{Development, Implementation, and Experimental Outdoor Evaluation of Quadcopter Controllers for Computationally Limited Embedded Systems}




\author[1]{Juan Paredes\corref{cor1}%
\fnref{fn1}}
\ead{jparedes@umich.edu}

\author[1]{Prashin Sharma%
\fnref{fn2}}
\ead{prashinr@umich.edu}

\author[1]{Brian Ha%
\fnref{fn3}}
\ead{brian.stephen.ha@gmail.com}

\author[1]{Manuel Lanchares%
\fnref{fn4}}
\ead{lanchares@gatech.edu}

\author[1]{Ella Atkins%
\fnref{fn5}}
\ead{ematkins@umich.edu}

\author[1]{Peter Gaskell%
\fnref{fn6}}
\ead{pgaskell@umich.edu}

\author[1]{Ilya Kolmanovsky%
\fnref{fn7}}
\ead{ilya@umich.edu}

\cortext[cor1]{Corresponding author}
\fntext[fn1]{PhD Candidate, Aerospace Engineering, 1320 Beal Ave, Ann Arbor, MI 48109}
\fntext[fn2]{PhD Candidate, Robotics Institute, 1320 Beal Ave, Ann Arbor, MI 48109, AIAA Student Member}
\fntext[fn3]{MS Graduate, Aerospace Engineering, 1320 Beal Ave, Ann Arbor, MI 48109}
\fntext[fn4]{PhD Candidate, Aerospace Engineering, Georgia Institute of Technology, North Ave NW, Atlanta, GA 30332. This work was done during his two-year affiliation as an MS Student with the Department of Aerospace Engineering, University of Michigan, 1320 Beal Ave, Ann Arbor, MI 48109}
\fntext[fn5]{Professor, Aerospace Engineering, 1320 Beal Ave, Ann Arbor, MI 48109, AIAA Fellow}
\fntext[fn6]{Adjunct Research Investigator, Robotics Institute, 2455 Hayward St, Ann Arbor, MI 48109}
\fntext[fn7]{Professor, Aerospace Engineering, 1320 Beal Ave, Ann Arbor, MI 48109, AIAA Associate Fellow}
\address[1]{University of Michigan, Ann Arbor, Michigan, 48109, USA}

\begin{abstract}
Quadcopters are increasingly used for applications ranging from hobby to industrial products and services.
This paper serves as a tutorial on the design, simulation, implementation, and experimental outdoor testing of digital quadcopter flight controllers, including Explicit Model Predictive Control, Linear Quadratic Regulator, and Proportional Integral Derivative.
A quadcopter was flown in an outdoor testing facility and made to track an inclined, circular path at different tangential velocities under ambient wind conditions.
Controller performance was evaluated via multiple metrics, such as position tracking error, velocity tracking error, and onboard computation time.
Challenges related to the use of computationally limited embedded hardware and flight in an outdoor environment are addressed with proposed solutions.
\end{abstract}


\end{frontmatter}

\begin{table}[h!]
\caption{Nomenclature}
\label{nomenclature_table}
\scriptsize
\centering
\renewcommand{\arraystretch}{1.25} 
\begin{tabular}{p{1.6cm} p{5.275cm}}
\toprule
$\hat{i}^{\rm i}, \hat{j}^{\rm i}, \hat{k}^{\rm i}$ & Orthogonal unit vectors aligned with the inertial coordinate frame \\

$\hat{i}^{\rm b}, \hat{j}^{\rm b}, \hat{k}^{\rm b}$ & Orthogonal unit vectors aligned with the body (quadcopter) coordinate frame \\

$x^{\rm i}$, $y^{\rm i}$, $z^{\rm i}$ & Position of the quadcopter geometric center along the $\hat{i}^{\rm i}$, $\hat{j}^{\rm i}$, $\hat{k}^{\rm i}$ axes respectively[m]\\

${\dot x}^{\rm i}$, ${\dot y}^{\rm i}$, ${\dot z}^{\rm i}$ & Velocity of the quadcopter geometric center along the $\hat{i}^{\rm i}$, $\hat{j}^{\rm i}$, $\hat{k}^{\rm i}$ axes respectively [m/s]\\

$\phi, \theta, \psi$ & Euler angles, quadcopter rotations around the $\hat{i}^{\rm i}$, $\hat{j}^{\rm i}$ and $\hat{k}^{\rm i}$ axes respectively [rad]\\

$\omega_{\rm x}^{\rm b}, \omega_{\rm y}^{\rm b}, \omega_{\rm z}^{\rm b}$ & Quadcopter rotation rates about the $\hat{i}^{\rm b}$, $\hat{j}^{\rm b}$ and $\hat{k}^{\rm b}$ axes respectively [rad/s]\\

$r^{\rm i}_x$, $r^{\rm i}_y$, $r^{\rm i}_z$ & Reference signals for position of the quadcopter geometric center along the $\hat{i}^{\rm i}$, $\hat{j}^{\rm i}$, $\hat{k}^{\rm i}$ axes respectively [m]\\

$r^{\rm i}_{\dot{x}}$, $r^{\rm i}_{\dot{y}}$, $r^{\rm i}_{\dot{z}}$ & Reference signals for velocity of the quadcopter geometric center along the $\hat{i}^{\rm i}$, $\hat{j}^{\rm i}$, $\hat{k}^{\rm i}$ axes respectively[m/s]\\

$r^{\rm i}_{\psi}$, $r^{\rm i}_{\dot{\psi}}$ & Reference signals for yaw and yaw rate respectively [rad, rad/s]\\

$T$  & Total vertical thrust of quadcopter rotors in body frame [N] \\

$T_{\rm hov}$ & Total vertical thrust required by quadcopter to remain in hover condition for $\phi = \theta = 0$ rad [N] \\

$\tau_{\rm x}$, $\tau_{\rm y}$, $\tau_{\rm z}$ &    Torque of quadcopter rotors around the $\hat{i}^{\rm b}$, $\hat{j}^{\rm b}$, $\hat{k}^{\rm b}$ axes respectively [N$\cdot$m] \\

$d$      &    Distance of each rotor to the geometric center of the quadcopter [m] \\   
$\omega_{{\rm ss},i}$ &  Steady state propulsor angular velocity \textcolor{white}{zzz} $i \in \{1, 2, 3, 4\}$ [RPM] \\

$\sigma_i$ & Normalized throttle signal in [0, 1] for motor $i \in \{1, 2, 3, 4\}$\\

$\omega_{\rm b}$ &  Steady state propeller angular velocity bias [RPM] \\

$C_{\rm R}$ &  Steady state propeller angular velocity parameter [RPM] \\

$T_{\rm m}$ & Time constant of propulsor first-order dynamics \\

$\omega_i$ & Propulsor angular velocity \textcolor{white}{zzzzzzzzzzzzz} $i \in \{1, 2, 3, 4\}.$ [RPM] \\

$C_{\rm T}$    &    Rotor thrust coefficient [kg$\cdot$m] \\

$C_{\rm M}$    &    Rotor moment coefficient [kg$\cdot{\rm m}^2$] \\

$T_i$   & Thrust generated by propulsor \textcolor{white}{zzzzzzzzzzzz} $i \in \{1, 2, 3, 4\}$ [N]\\

$M_i$ & Moment magnitude generated by propulsor $i \in \{1, 2, 3, 4\}$ [N$\cdot$m]\\

$m$      &    Quadcopter mass [kg] \\

$J_{\rm xx}$, $J_{\rm yy}$, $J_{\rm zz}$  &    Quadcopter moment of inertia around the $\hat{i}^{\rm b}$, $\hat{j}^{\rm b}$, $\hat{k}^{\rm b}$ axes, respectively [kg$\cdot{\rm m}^2$] \\

$T_{\rm s}$ & Sampling time [s] \\

${\rm s}\phi, {\rm c}\phi, {\rm t}\phi,$ & Sine, cosine, and tangent of angle $\phi$, respectively.\\\bottomrule

\end{tabular}
\end{table}

\section{Introduction}
As autonomous multicopters or drones become more prevalent, emphasis on their performance and safety grows. Small quadcopters are now considered frequently for commercial use. Their performance depends not only on their design but also on their flight control system. Numerous approaches to quadcopter control have been reported in the literature. Practical implementation topics such as in situ sensor calibration, vehicle parameter modeling, controller software, and real-time execution on computationally limited platforms are of interest.

This paper investigates classical and modern techniques for quadrotor flight control.  The following control algorithms are implemented and evaluated: Proportional-Derivative (PD), Proportional-Integral-Derivative (PID), Linear-Quadratic Regulator (LQR), LQR with Integrators (LQR-I), and Explicit Model Predictive Control (E-MPC).
A thorough tutorial of quadrotor flight controller design, simulation, implementation, and evaluation is presented with the following contributions: 
\begin{itemize}
    \item Details of the design, software implementation, and tuning of the quadcopter controllers with emphasis on implementation in a computationally limited embedded platform.
    \item Experimental results from outdoor flight tests including a performance comparison of each controller's ability to follow an inclined, circular path under varying ambient conditions.
    %
    %
    \item Design and implementation of ancillary autopilot components such as state estimators to support full state feedback and actuator mapping.
\end{itemize}


Quadrotor or quadcopter modelling and control has been frequently studied in the literature.
In \citet{bouabdallah2007}, quadcopter aerodynamics were modelled using blade element and momentum theory; a back-stepping, integral controller was used for attitude and position control. Maximum reference deviations of $3$ cm and $20$ cm were observed for attitude and position, respectively.
Similarly, in \citet{hoffmann2007}, various aerodynamic effects experienced by quadcopters at high speeds with wind disturbances were presented and validated through thrust test stand experiments and flight tests. In contrast to \citet{bouabdallah2007}, a simple PID controller was used for position and attitude control.
%
\citet{Powers2015} showed a simplified lumped parameter model for quadcopter motor dynamics and described the design and implementation of continuous-time LQR and nonlinear controllers. A nano quadcopter with a linear controller was commanded in tests to fly in a circular trajectory of radius $0.5$ m with a maximum acceleration of $0.008$ ${\rm m/s}^2.$ Errors of $10$ cm and $1$ cm were observed along the horizontal and vertical planes, respectively.

Quadcopter PID control is studied in \citet{dong2013} and \citet{bolandi2013}. In both references, quadcopter dynamics were linearized and treated as decoupled around each axis to facilitate the implementation of single input single output (SISO) control techniques. In \citet{dong2013}, the PID controller was designed using root locus analysis and Ziegler-Nichols tuning, while
\citet{bolandi2013} used the Direct Synthesis method. Outdoor trajectory tracking flight tests were conducted in \citet{dong2013} with a maximum position error of $50$ cm, but no discussion of velocity tracking error was presented. 

To achieve improved performance, multivariable control techniques such as LQR have also been considered. \citet{reyes2013} designed a gain scheduled LQR controller with gains calculated for two different situations: (1) Quadcopter far from the reference, and (2) Quadcopter was already closely tracking the reference state. Another implementation of LQR can be found in \citet{heng2015} which addressed actuator saturation. 

Due to its ability to handle constraints, Model Predictive Control (MPC) has been pursued for multicopters, especially in cases when substantial computational power is available.
\citet{Kamel2017} implemented linear, nonlinear, and robust MPC schemes for a hexacopter with a NUCi7 computer onboard.
\citet{liu2015} developed an explicit MPC scheme for trajectory tracking and verified it in simulation. Their approach exploited differential flatness and Bezier curves and was solved offline using a quadratic programming technique.
An efficient MPC scheme with a reduced computational footprint was proposed in \citet{abdolhosseini2013} for controlling quadcopters.
An unconstrained MPC law was implemented on a computer embedded on a quadcopter in \citet{bangura2014}.
MPC was applied to a simplified, linear dynamics model derived via feedback linearization. This yielded desired thrust, rotation rates, and angular accelerations, which were subsequently regulated by a high-gain attitude controller obtained by exploiting quadcopter attitude dynamics.

The robustness of quadcopter controllers to disturbances such as wind gusts has also been studied. In \citet{waslander2009}, a wind compensator was added to the outer PID loop to reject wind effects. This decreased position error from 40 cm to 10 cm. A constrained, finite-time optimal control scheme was developed and experimentally validated in \citet{alexis2010}. The quadcopter was able to effectively achieve the desired set point in the presence of wind gusts.

This paper synthesizes multiple quadcopter control laws and evaluates them in simulation and in flight testing.  Though organized as a tutorial, this paper also offers novelty in its careful experimental evaluation and comparison of a diverse control law suite. The organization of this paper is as follows. Quadcopter design and system modelling are presented in Section \ref{section:system_Description} followed by guidance, navigation, and control strategies in Section \ref{section:guidance}. The state estimation algorithm is presented in Section \ref{section:state_estimation}. Formulations for all tested controllers are provided in Section \ref{section:linear_controllers} with simulation results of each controller presented in Section \ref{section:simulation}. The experimental setup and results from outdoor flight tests are provided in Sections \ref{section:setup} and \ref{section:results}, respectively. Results are discussed in Section \ref{section:discussion} with a brief conclusion provided in Section \ref{section:conclusion}.
Nomenclature used in this paper is summarized in Table \ref{nomenclature_table}.


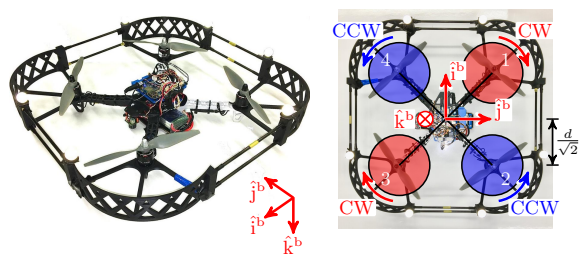
\begin{figure}[b!]
    \centering
     \resizebox{\columnwidth}{!}{%
        \begin{tikzpicture}[>={stealth'}, line width = 0.25mm]
            \def\centerarc[#1](#2)(#3:#4:#5)
		{ \draw[#1] ($(#2)+({#5*cos(#3)},{#5*sin(#3)})$) arc (#3:#4:#5); }
		\node [bigblock,fill = none ,draw=none](bg_quad){\centering \includegraphics[width=10em]{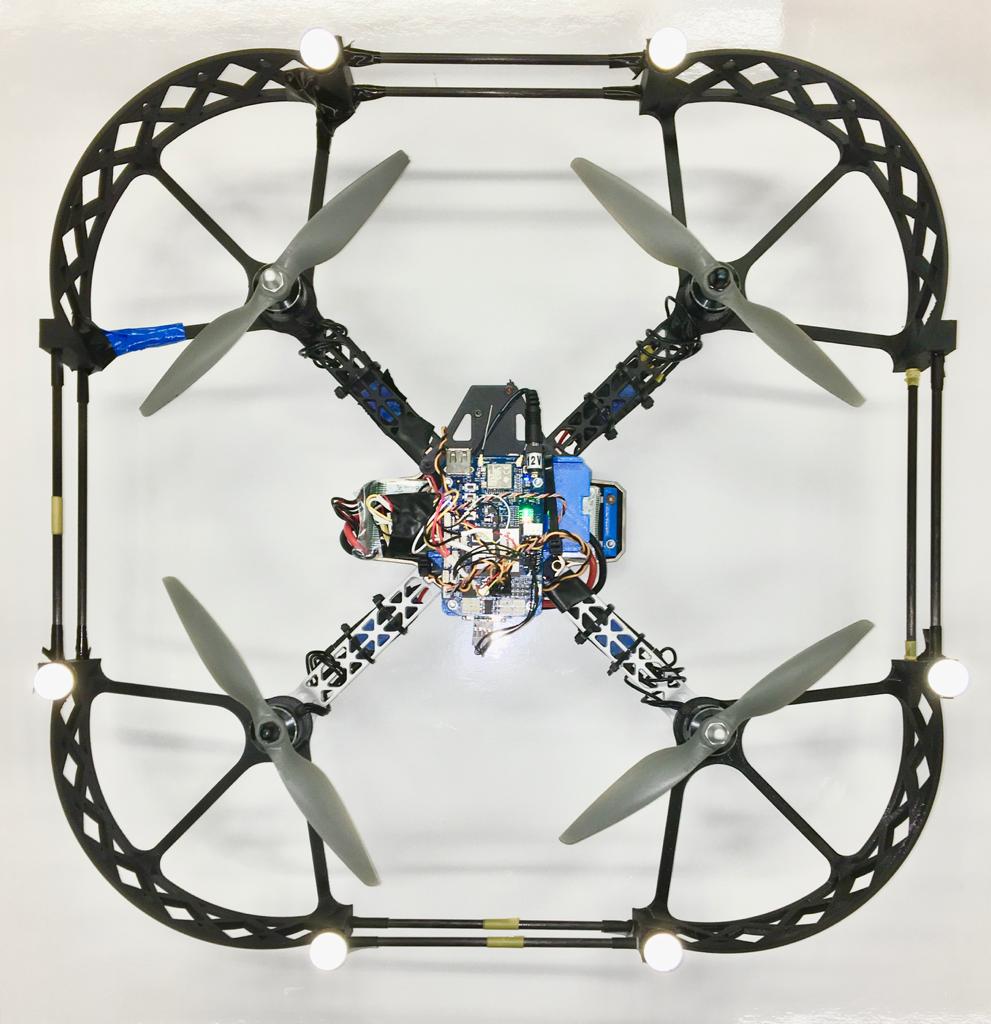}};
		\node [coordinate](quad_center) at (bg_quad.center){};
		\node [draw, fill=blue, fill opacity=0.5, circle, minimum size=1cm, below right = 0.4cm and 0.4cm of quad_center](circ_dr){};
		\node [draw, fill=red, fill opacity=0.5, circle, minimum size=1cm, below left = 0.4cm and 0.4cm of quad_center](circ_dl){};
		\node [draw, fill=red, fill opacity=0.5, circle,minimum size=1cm, above right = 0.4cm and 0.4cm of quad_center](circ_ur){};
		\node [draw, fill=blue, fill opacity=0.5, circle, minimum size=1cm, above left = 0.4cm and 0.4cm of quad_center](circ_ul){};
		\draw[-,thick] (circ_ul.center) -- (circ_dr.center);
		\draw[-,thick] (circ_ur.center) -- (circ_dl.center);
		
		\draw[->,white,ultra thick](quad_center) -- ([yshift=0.75cm]quad_center);
		\draw[->,red,very thick](quad_center) -- ([yshift=0.75cm]quad_center);
	    \draw[->,white,ultra thick]([xshift=-0.01cm]quad_center) -- ([xshift=0.75cm]quad_center);
		\draw[->,red,very thick]([xshift=-0.01cm]quad_center) -- ([xshift=0.75cm]quad_center);
		
		\node [draw, red, fill=white, circle, inner sep=0pt, minimum size=0.25cm, left = 0.2cm of quad_center](coord_z){};
		\draw[-,red,thick] (coord_z.45)--(coord_z.225);
		\draw[-,red,thick] (coord_z.135)--(coord_z.315);
		\centerarc[->,white,ultra thick](circ_ur.center)(80:10:0.6cm);
		\centerarc[->,red,very thick](circ_ur.center)(80:10:0.6cm);
		\centerarc[->,white,ultra thick](circ_dl.center)(260:190:0.6cm);
		\centerarc[->,red,very thick](circ_dl.center)(260:190:0.6cm);
		\centerarc[->,white,ultra thick](circ_ul.center)(100:170:0.6cm);
		\centerarc[->,blue,very thick](circ_ul.center)(100:170:0.6cm);
		\centerarc[->,white,ultra thick](circ_dr.center)(280:350:0.6cm);
		\centerarc[->,blue,very thick](circ_dr.center)(280:350:0.6cm);
		
		\node [draw = none,text=red,fill=white,rounded corners=0.5pt,inner sep=0.25pt] at ([xshift = 0.2cm,yshift = 0.8cm]quad_center) {\small $\hat{\rm i}^{\rm b}$};
		\node [draw = none,text=red,fill=white,rounded corners=0.5pt,inner sep=0.25pt] at ([xshift = 0.95cm,yshift = 0.15cm]quad_center) {\small $\hat{\rm j}^{\rm b}$};
		\node [draw = none,text=red,fill=white,rounded corners=0.5pt,inner sep=0.25pt] at ([xshift = -0.675cm]quad_center) {\small $\hat{\rm k}^{\rm b}$};
		\draw[<->,thick]([xshift=1.75cm]quad_center)--([xshift=1.75cm,yshift=-0.76cm]quad_center);
		\draw[-] ([xshift=1.65cm]quad_center)--([xshift=1.85cm]quad_center);
		\draw[-] ([xshift=1.65cm,yshift=-0.76cm]quad_center)--([xshift=1.85cm,yshift=-0.76cm]quad_center);
		\node [draw = none] at ([xshift=2cm,yshift=-0.38cm]quad_center) {\small $\frac{d}{\sqrt{2}}$};
		\node [draw = none, text = white] at ([xshift=-0.22cm,yshift=0.22cm]circ_ul) {\small 4};
		\node [draw = none, text = white] at ([xshift=0.22cm,yshift=0.22cm]circ_ur) {\small 1};
		\node [draw = none, text = white] at ([xshift=-0.22cm,yshift=-0.22cm]circ_dl) {\small 3};
		\node [draw = none, text = white] at ([xshift=0.22cm,yshift=-0.22cm]circ_dr) {\small 2};

		\node [draw = none, text = blue,fill=white,rounded corners=1pt,inner sep=1pt] at ([xshift=-0.7cm,yshift=0.7cm]circ_ul) {\small CCW};
		\node [draw = none, text = red,fill=white,rounded corners=1pt,inner sep=1pt] at ([xshift=0.7cm,yshift=0.7cm]circ_ur) {\small CW};
		\node [draw = none, text = red,fill=white,rounded corners=1pt,inner sep=1pt] at ([xshift=-0.7cm,yshift=-0.7cm]circ_dl) {\small CW};
		\node [draw = none, text = blue,fill=white,rounded corners=1pt,inner sep=1pt] at ([xshift=0.7cm,yshift=-0.7cm]circ_dr) {\small CCW};
		\node [bigblock, fill = none, draw = none, left = 6 em of quad_center](quad_left){\centering \includegraphics[width=14em]{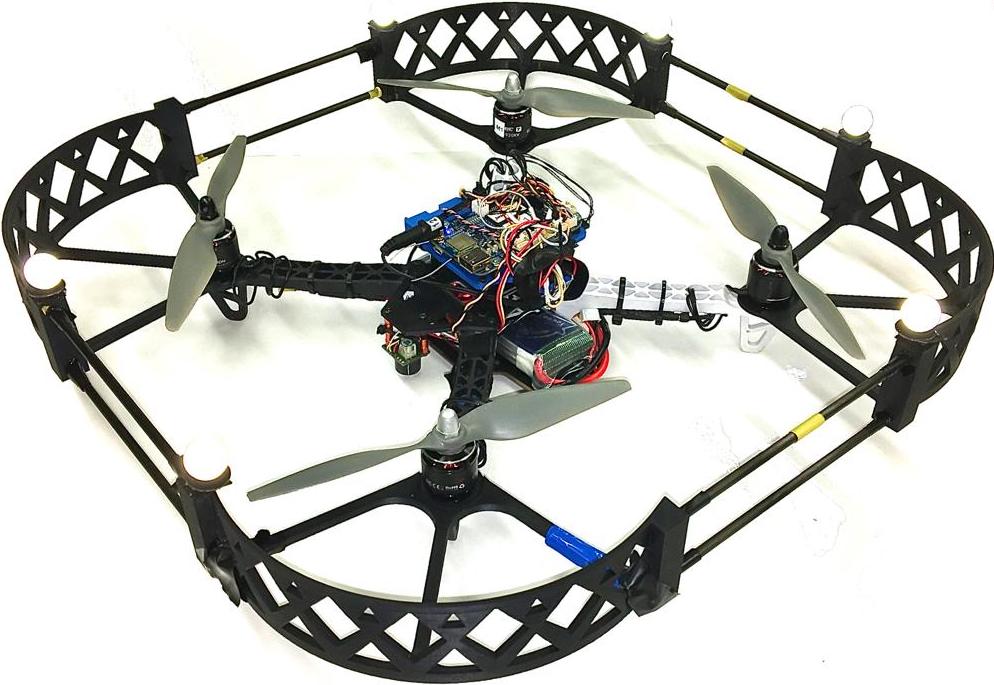}};
		
		\node [coordinate](quadL_center) at ([xshift=-2.5cm, yshift = -1.3cm]quad_center){};
		\draw[->,red,very thick]([yshift=0.01cm]quadL_center) -- ([yshift=-0.6cm]quadL_center);
		\draw[->,red,very thick](quadL_center) -- ([xshift=-0.48cm,yshift=0.336cm]quadL_center);
		\draw[->,red,very thick](quadL_center) -- ([xshift=-0.48cm,yshift=-0.336cm]quadL_center);
		\node [draw = none,text=red] at ([xshift=-0.575cm,yshift=-0.45cm]quadL_center) {\small $\hat{\rm i}^{\rm b}$};
		\node [draw = none,text=red] at ([xshift=-0.575cm,yshift=0.1cm]quadL_center) {\small $\hat{\rm j}^{\rm b}$};
		\node [draw = none,text=red] at ([yshift=-0.78cm]quadL_center) {\small $\hat{\rm k}^{\rm b}$};
        \end{tikzpicture}
    }
    \caption{Experimental quadcopter based on a DJI Flamewheel F450 frame with custom propeller guards and avionics mounts. Motion capture markers are placed on propeller guards for increased camera visibility.}
    \label{Fig:quad_hardware}
\end{figure}

\section{System Description}
\label{section:system_Description}
\subsection{Hardware}
This section describes the tested hardware platform. The assembled quadcopter is shown in Fig. \ref{Fig:quad_hardware}.
The onboard Beaglebone Blue Linux computer features a 1GHz ARM \textregistered \ Cortex-A8 processor with eight servomotor outputs, an MPU9250 Inertial Measurement Unit (IMU) with three-axis angular rate gyro, accelerometer, and magnetic field sensors, a WiFi 2.4 GHz transceiver module, and UART interfaces for serial communications. The pre-installed Robot Control Library was used to facilitate controller C code implementation.
%
%
The controller code was required to complete all calculations within a 0.005 second time interval between system interrupts ($T_{\rm s} = 0.005$ s).
The Beaglebone Blue interfaced with all peripherals including sensors, Pulse Width Modulation (PWM) signals to control motor speeds, and communication links.
An IMU measured angular rates and orientation of the body frame with respect to a flat-Earth inertial frame gravity vector. The Wi-Fi transceiver module allowed two-way communication between the embedded board and a ground station computer. An xBee modem connected to a UART interface to reliably update position and heading measurements from a Qualisys Motion Capture (MOCAP) camera system. Finally, a Radio Control (RC) receiver obtained commands from a RC Controller to initiate/terminate tests and for manual backup control as described further in Section \ref{section:setup}.

A DJI Flamewheel F450 ARF kit was used for the quadcopter frame, with AIR 2213/920KV brushless DC motors and APC (8 x 4.5) propellers for the propulsion units. The system was powered by a 3S 3000 mAh 35C LiPo battery. A Power Distribution Board (PDB) distributed 5V and 12V to the embedded board and the motors, respectively. Each motor was connected to an AIR 20A Electronic Speed Controller (ESC) controlled by Beaglebone PWM signals.
3D printed propeller guards, originally designed in \citet{mattquad2019}, were attached at the base of each motor and connected by carbon fiber tubes to provide protection for propellers and users. MOCAP was used to determine the 3D location and heading of the quadcopter with targets placed on top of the propeller guards. The quadcopter and body-fixed coordinate frame are shown in Fig. \ref{Fig:quad_hardware}.
\subsection{Quadcopter Dynamics and System Modelling} \label{QuadDynSysMod}

This section describes the quadcopter model and is based on content from Chapters 5 and 6 of \citet{quanquan2017}.

\subsubsection{Quadcopter Configuration}

The body-fixed frame (Fig. \ref{Fig:quad_hardware}) is composed of orthogonal unit vectors $\hat{i}^{\rm b}, \hat{j}^{\rm b}, \hat{k}^{\rm b}$. The $\hat{i}^{\rm b}$ axis points forward on the quadcopter, the $\hat{k}^{\rm b}$ axis points down, and the $\hat{j}^{\rm b}$ axis points right to satisfy a right-hand convention. 
Let $\hat{i}^{\rm i},$ $\hat{j}^{\rm i},$ $\hat{k}^{\rm i}$ denote orthogonal unit vectors defining an inertial coordinate frame, such that $\hat{k}^{\rm i}$ points toward the ground.
Let $\psi, \theta, \phi$ denote yaw, pitch, and roll Euler angles, respectively. Following a $\psi \to \theta \to \phi$ rotation ordering convention, the relation between inertial axes $\hat{i}^{\rm i},$ $\hat{j}^{\rm i},$ $\hat{k}^{\rm i}$ and the body-fixed axes $\hat{i}^{\rm b},$ $\hat{j}^{\rm b},$ $\hat{k}^{\rm b}$ is given by
\begin{equation}
    \begin{bmatrix} \hat{i}^{\rm i} \\ \hat{j}^{\rm i} \\ \hat{k}^{\rm i} \end{bmatrix} = 
    {\rm R}_{\rm b/i} \begin{bmatrix} \hat{i}^{\rm b} \\ \hat{j}^{\rm b} \\ \hat{k}^{\rm b} \end{bmatrix}.
\end{equation}
where ${\rm R}_{\rm b/i}$ is the rotation matrix from body-fixed frame to inertial frame, such that

\vspace{-0.75em}
\footnotesize
\begin{equation}
{\rm R}_{\rm b/i}\triangleq \begin{bmatrix}  {\rm c}{\theta} \ {\rm c}{\psi} &
    {\rm c}{\psi} \ {\rm s}{\theta} \ {\rm s}{\phi} - {\rm s}{\psi} \ {\rm c}{\phi} & 
    {\rm c}{\psi} \ {\rm s}{\theta} \ {\rm c}{\phi}+{\rm s}{\psi} \ {\rm s}{\phi}
    \\
    {\rm c}{\theta} \ {\rm s}{\psi} & 
    {\rm s}{\psi} \ {\rm s}{\theta} \ {\rm s}{\phi} + {\rm c}{\psi} \ {\rm c}{\phi} &
    {\rm s}{\psi} \ {\rm s}{\theta} \ {\rm c}{\phi} - {\rm c}{\psi} \ {\rm s}{\phi}
    \\
    -{\rm s}{\theta} &
    {\rm s}{\phi} \ {\rm c}{\theta} &  
    {\rm c}{\phi} \ {\rm c}{\theta}
    \end{bmatrix}.
\end{equation}
\normalsize

Fig. \ref{Fig:quad_hardware} displays the X-configuration chosen for the quadcopter and the direction of rotation of each propeller. Propellers 2 and 4 (in blue) rotate counterclockwise (CCW) and propellers 1 and 3 (in red) rotate clockwise (CW). 
%

%
%

\subsubsection{Propulsor Dynamics}

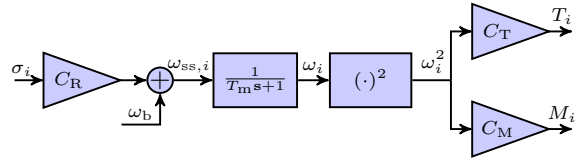
\begin{figure}[t!]
    \centering
    \resizebox{\columnwidth}{!}{%
    \begin{tikzpicture}[>={stealth'}, line width = 0.25mm]
		\def\centerarc[#1](#2)(#3:#4:#5)
		{ \draw[#1] ($(#2)+({#5*cos(#3)},{#5*sin(#3)})$) arc (#3:#4:#5); }
		\node [gain, xscale = 2, yscale=1.25](Cr){};
		\node [draw = none] at (Cr.center) {\scriptsize$C_{\rm R}$};
		\node [sum, right = 1cm of Cr.center](sumWss){};
		\node [draw = none] at (sumWss.center) {$+$};
		\node [smallblock, right = 0.5cm of sumWss.east](motDyn){\scriptsize $\frac{1}{T_{\rm m} {\bf{s}} + 1}$};
		
		
		\node [smallblock, right = 0.4cm of motDyn.east](matMix){\scriptsize $(\cdot)^2$};
		\node [gain, xscale = 2, yscale=1.25, above right = 0.05 cm and 0.75cm of matMix](CT){};
		\node [draw = none] at (CT.center) {\scriptsize$C_{\rm T}$};
		\node [gain, xscale = 2, yscale=1.25, below right = 0.05 cm and 0.75cm of matMix](CM){};
		\node [draw = none] at (CM.center) {\scriptsize$C_{\rm M}$};
		
		\draw [->] ([xshift=-0.35cm]Cr.west) -- node [above, near start, yshift = -0.07cm]{\scriptsize$\sigma_i$} ([xshift=0.02cm]Cr.west);
		\draw [->] ([xshift=-0.5cm, yshift = -0.4cm]sumWss.south) -| node [above, near start, yshift = -0.07cm]{\scriptsize$\omega_{\rm b}$}(sumWss.south);
		\draw [->] ([xshift=-0.06cm]Cr.east)--(sumWss.west);
		\draw [->] (sumWss.east)--node [above, xshift = -0.05cm, yshift = -0.035cm]{\scriptsize$\omega_{{\rm ss},i}$}(motDyn.west);
		\draw [->] (motDyn.east)--node [above, yshift = -0.035cm]{\scriptsize$\omega_i$}(matMix.west);
		\draw [-] (matMix.east)--node [above, near end, xshift = -0.1cm, yshift = -0.035cm]{\scriptsize$\omega_i^{2}$}([xshift=0.5cm]matMix.east);
		\draw [->] ([xshift=0.5cm]matMix.east) |- ([xshift=0.02cm]CT.west);
		\draw [->] ([xshift=0.5cm]matMix.east) |- ([xshift=0.02cm]CM.west);
		\draw [->] ([xshift=-0.06cm]CT.east)--node [above, near end, xshift = -0.05cm, yshift = -0.035cm]{\scriptsize$T_i$}([xshift=0.25cm]CT.east);
		\draw [->] ([xshift=-0.06cm]CM.east)--node [above, near end, xshift = -0.05cm, yshift = -0.035cm]{\scriptsize$M_i$}([xshift=0.25cm]CM.east);
		
	\end{tikzpicture}
    }
    \caption{Propulsor dynamics block diagram. The normalized control signal $\sigma_i$ (PWM duty cycle) is used to determine  output thrust magnitude $T_i$ and torque magnitude $\tau_i.$}
    \label{propulsor_dynamics_block}
\end{figure}

The dynamics model for each propulsion unit (motor with propeller) is depicted by the block diagram in Fig. \ref{propulsor_dynamics_block}. Let indices $i \in \{1, 2, 3, 4\}$ denote each propulsor as shown in Fig. \ref{Fig:quad_hardware}. Each motor was driven by a normalized throttle signal $\sigma_i \in [0, 1],$ such that $\sigma_i = 0$ gives no rotation and $\sigma_i = 1$ yielded the maximum rotation speed for propulsor $i.$ 
$\sigma_i$ represents the duty cycle of the PWM signal sent by the controller to propulsor $i.$
The steady state angular velocity of propulsor $i$ in RPM, $\omega_{{\rm ss},i},$ is given by
\begin{equation}
\omega_{{\rm ss},i} = C_{\rm R} \sigma_i + \omega_{\rm b}, \label{sigma2WssEq}
\end{equation}
where $C_{\rm R}$ and $\omega_{\rm b}$ define the steady state characteristics of each propulsor. The transient response of  propulsor angular velocity $\omega_i$ is modeled by a first-order, low pass filter with time constant $T_{\rm m},$ such that the transfer function of a model with input $\omega_{{\rm ss},i}$ and output $\omega_i$ is given by
\begin{equation}
    \omega_i = \frac{1}{T_{\rm m} \textbf{s} + 1}\omega_{{\rm ss},i},
\end{equation}
where $\textbf{s}$ is the Laplace transform complex variable.
Finally, the thrust and torque magnitudes generated by the $i^{th}$ propulsor, $T_i$ and $M_i$ respectively, are given by
\begin{align}
    T_i &= C_{\rm T} \omega_i^2, \label{Tieq}\\
    M_i &= C_{\rm M} \omega_i^2, \label{Mieq}
\end{align}
where $C_{\rm T}$ and $C_{\rm M}$ are parameters that characterize each propeller's aerodynamic properties.

\subsubsection{Quadcopter Inputs and Mixing Matrix} \label{qMixMatSec}

Assuming the quadcopter body is rigid, the propulsors generate a total thrust $T$ in direction of $-\hat{k}^{\rm b}$ with three torque components $\tau_{\rm x}$, $\tau_{\rm y}$ and $\tau_{\rm z}$ about $\hat{i}^{\rm b},$ $\hat{j}^{\rm b},$ and $\hat{k}^{\rm b}$ axes, respectively.
Each propulsor generates a thrust in the direction of $-\hat{k}^{\rm b}$ and a torque aligned with $\hat{k}^{\rm b},$ in the direction opposite to its rotation. Hence, it follows from \eqref{Tieq} and \eqref{Mieq} that

\vspace{-0.75em}
\footnotesize
\begin{align} 
    \begin{bmatrix} T \\ \tau_{\rm x} \\ \tau_{\rm y} \\ \tau_{\rm z} \end{bmatrix} &= 
    \begin{bmatrix} T_1 + T_2 + T_3 + T_4 \\ \frac{\sqrt{2}}{2} d (- T_1 - T_2 + T_3 + T_4) \notag \\ \frac{\sqrt{2}}{2} d (T_1 - T_2 - T_3 + T_4) \\ -M_1 + M_2 - M_3 + M_4 \end{bmatrix} \\ &=
    \begin{bmatrix} C_{\rm T} & C_{\rm T} & C_{\rm T} & C_{\rm T} \\ -C_{\tau} & -C_{\tau} & C_{\tau} & C_{\tau} \\ C_{\tau} & -C_{\tau} & -C_{\tau} & C_{\tau} \\ -C_{\rm M} & C_{\rm M} & -C_{\rm M} & C_{\rm M} \end{bmatrix} \begin{bmatrix} \omega_1^2 \\ \omega_2^2 \\ \omega_3^2 \\ \omega_4^2 \end{bmatrix}, \label{eqMixMatrix}
\end{align}
\normalsize
where $d = 0.17$ m is the distance from each rotor to the geometric center and $C_{\tau} \isdef \frac{\sqrt{2}}{2} d C_{\rm T}.$
The rightmost matrix of \eqref{eqMixMatrix} is denoted the mixing matrix and describes the linear transformation from squared motor angular velocities to quadcopter system input.

\subsubsection{Quadcopter System Dynamics} \label{subsec:quad_model}

The quadcopter dynamics model is derived under the assumptions that the quadcopter body is rigid, its mass properties remain constant during operation, and the geometric center and the center of gravity are located in the same position.  It is assumed that  no aerodynamic forces or moments act upon the quadcopter apart from from the thrust and torque generated by each propulsor.
In practice, the quadcopter body drag and blade flapping have a significant impact on the dynamics of the quadcopter at high flight velocities and in outdoor environments, as is stated in \citet{huang2009}.
Nonetheless, these assumptions allow the resulting model to achieve a balance between simplicity and accuracy for the purpose of designing a control strategy.
Let $m$ be the mass the quadcopter in kg,
$J \isdef {\rm diag}\left({\arraycolsep=1.2pt\def\arraystretch{0.5}\left[\begin{array}{ccc}J_{\rm xx} & J_{\rm yy} & J_{\rm zz} \end{array}\right]}^{\rm T}\right)$ be the inertia tensor with respect to the body fixed frame in $\text{kg}\cdot\text{m}^2.$
Let $X \in \mathbb{R}^{12}$ be the system state vector and $U \in \mathbb{R}^4$ be the system input vector, such that

\vspace{-0.75em}
\footnotesize
\begin{align}
    X & \isdef {\arraycolsep=4pt\def\arraystretch{0.5}\left[\begin{array}{cccccccccccc}  x^{\rm i} & y^{\rm i} & z^{\rm i} & \dot{x}^{\rm i} & \dot{y}^{\rm i} & \dot{z}^{\rm i} & \phi & \theta & \psi & \omega_{\rm x}^{\rm b} & \omega_{\rm y}^{\rm b} & \omega_{\rm z}^{\rm b} \end{array}\right]}^{\rm T}, \nn \\
    U &\isdef {\arraycolsep=4pt\def\arraystretch{0.5}\left[\begin{array}{cccc} T & \tau_{\rm x} & \tau_{\rm y} & \tau_{\rm z} \end{array}\right]}^{\rm T} \nn,
\end{align}
\normalsize
where $x^{\rm i}, y^{\rm i}, z^{\rm i}$ represent quadcopter position in the inertial frame, $\dot{x}^{\rm i}, \dot{y}^{\rm i}, \dot{z}^{\rm i}$ represent its velocity vector in the same frame, $\phi, \theta, \psi$ are the Euler angles representing the orientation of body-fixed frame with respect to inertial frame axes, and $\omega_{\rm x}^{\rm b}, \omega_{\rm y}^{\rm b}, \omega_{\rm z}^{\rm b}$ are the components of quadcopter angular velocity expressed in  the body-fixed frame.
The nonlinear model for the quadcopter dynamics is
\begin{equation}
\frac{d}{dt}X = f_{\rm quad}(X, U). \label{NLDynQuad}
\end{equation}
Assuming the only forces acting on the quadcopter center of gravity
are its weight $mg$ and total thrust $T$, Newton's Second Law yields: 

\vspace{-0.75em}
\footnotesize
\begin{equation} \label{eqSS1}
\frac{d}{dt}\begin{bmatrix} \dot{x}^{\rm i} \\ \dot{y}^{\rm i} \\ \dot{z}^{\rm i} \end{bmatrix} = \begin{bmatrix} 0 \\ 0 \\ g \end{bmatrix} - {\rm R}_{\rm b/i} \begin{bmatrix} 0 \\ 0 \\
\frac{T}{m} \end{bmatrix} =
\begin{bmatrix} 0 \\ 0 \\ g \end{bmatrix} - \begin{bmatrix} {\rm c}{\psi} \ {\rm s}{\theta} {\rm c}{\phi} + {\rm s}{\psi} {\rm s}{\phi} \\ {\rm s}{\psi} {\rm s}{\theta} {\rm c}{\phi} - {\rm c}{\psi} {\rm s}{\phi} \\ {\rm c}{\theta} {\rm c}{\phi} \end{bmatrix}
\frac{T}{m}.
\end{equation}
\normalsize
Furthermore, considering the torques $\tau_{\rm x}, \tau_{\rm y}, \tau_{\rm z}$ generated by the propulsors, it follows from the classical Euler equations of rotational dynamics that
\begin{equation} \label{eqSS2}
\frac{d}{dt}\begin{bmatrix}\omega_{\rm x}^{\rm b} \\ \omega_{\rm y}^{\rm b}  \\ \omega_{\rm z}^{\rm b}  \end{bmatrix} = \begin{bmatrix} \frac{\tau_{\rm x}}{J_{\rm xx}} \\ \frac{\tau_{\rm y}}{J_{\rm yy}} \\ \frac{\tau_{\rm z}}{J_{\rm zz}} \end{bmatrix} + \begin{bmatrix} \frac{J_{\rm yy}-J_{\rm zz}}{J_{\rm xx}} \ \omega_{\rm y}^{\rm b} \ \omega_{\rm z}^{\rm b} \\ \frac{J_{\rm zz}-J_{\rm xx}}{J_{\rm yy}} \ \omega_{\rm x}^{\rm b} \ \omega_{\rm z}^{\rm b} \\ \frac{J_{\rm xx}-J_{\rm yy}}{J_{\rm zz}} \ \omega_{\rm x}^{\rm b} \ \omega_{\rm y}^{\rm b} \end{bmatrix}.
\end{equation}
Finally, the relation between angular velocity vector components and time rate of changes of 3-2-1 Euler angles is given by
%
%
\begin{equation} \label{eqSS3}
\begin{bmatrix}\dot{\phi} \\ \dot{\theta} \\ \dot{\psi} \end{bmatrix} =
\begin{bmatrix} 1 & {\rm t}{\theta} {\rm s}{\phi} & {\rm t}{\theta} {\rm c}{\phi} \\ 0 & {\rm c}{\phi} & -{\rm s}{\phi} \\ 0 & {\rm s}{\phi}/{\rm c}{\theta} & {\rm c}{\phi}/{\rm c}{\theta} \end{bmatrix}
\ \begin{bmatrix} \omega_{\rm x}^{\rm b} \\ \omega_{\rm y}^{\rm b} \\ \omega_{\rm z}^{\rm b} \end{bmatrix}.
\end{equation}

Equation \eqref{eqSS1} describes the effects of gravity, attitude, and total thrust $T$ on three-dimensional quadcopter motion. 
Note that total thrust vector $T$ is along the $\hat{k}^{\rm b}$ vector so its projection onto inertial frame axes requires multiplication by a rotation matrix per Eq. \eqref{eqSS1}.
%
%
Equation \eqref{eqSS2} describes the effect of torques $\tau_x$, $\tau_y$ and $\tau_z$ on the angular velocity. Gyroscopic torques caused by motor rotations are considered negligible and are thus not included in the present model. Eq. \eqref{eqSS3} relates Euler angle rates and angular velocity components as a function of quadcopter attitude. %
%
These nonlinear equations are used for simulations and to derive linearized equations for controller design.


\subsubsection{Linearized System Dynamics} \label{lin_quad_dynamics}
%
%
To design LQR, LQR-I, and MPC controllers,
the nonlinear model presented above is linearized around an equilibrium point corresponding to hover.
As will be shown in test results, a linearized model is sufficient as a basis for controller design even when tracking a trajectory which
deviates substantially from hover conditions.
Let $X_{\rm hov} \in \mathbb{R}^{12}$ be the system state vector in hover conditions and $U_{\rm hov} \in \mathbb{R}^4$ be the system input vector in hover conditions, such that

\vspace{-1.25em}
\footnotesize
\begin{align}
    %
    X_{\rm hov} & \isdef {\arraycolsep=3pt\def\arraystretch{0.5}\left[\begin{array}{cccccc}  x_{\rm hov}^{\rm i} & y_{\rm hov}^{\rm i} & z_{\rm hov}^{\rm i} & \dot{x}_{\rm hov}^{\rm i} & \dot{y}_{\rm hov}^{\rm i} & \dot{z}_{\rm hov}^{\rm i} \end{array} \right.} \nn \\ & {\arraycolsep=3pt\def\arraystretch{0.5}\left.\begin{array}{cccccc}\phi_{\rm hov} & \theta_{\rm hov} & \psi_{\rm hov} & \omega_{\rm x, hov}^{\rm b} & \omega_{\rm y, hov}^{\rm b} & \omega_{\rm z, hov}^{\rm b} \end{array} \right]^{\rm T}} \nn \\
    &= {\arraycolsep=3pt\def\arraystretch{0.5}\left[\begin{array}{cccccccccccc}  x_{\rm hov}^{\rm i} & y_{\rm hov}^{\rm i} & z_{\rm hov}^{\rm i} & 0 & 0 & 0 & 0 & 0 & 0 & 0 & 0 & 0 \end{array}\right]}^{\rm T}, \nn \\
    U_{\rm hov} & \isdef {\arraycolsep=3pt\def\arraystretch{0.5}\left[\begin{array}{cccc} T_{\rm hov} & \tau_{\rm x, hov} & \tau_{\rm y, hov} & \tau_{\rm z, hov} \end{array}\right]}^{\rm T} = {\arraycolsep=3pt\def\arraystretch{0.5}\left[\begin{array}{cccc} mg & 0 & 0 & 0 \end{array}\right]}^{\rm T} \nn.
\end{align}
\normalsize
Let $\delta X \in \mathbb{R}^{12}$ and $\delta U \in \mathbb{R}^4$ be the state and input deviations from hover, respectively, such that $\delta X = X - X_{\rm hov}$ and $\delta U = U - U_{\rm hov}.$ Since $\phi_{\rm hov} = \theta_{\rm hov} = \psi_{\rm hov} = 0,$ the linearization is performed assuming that the body-fixed and the inertial frames are aligned. Furthermore, it follows from Eq. \eqref{eqSS3} that, under hover conditions, the Euler angle rates and the body angular velocities are the same. Hence, it follows that

\vspace{-1.25em}
\footnotesize
\begin{align}
    \delta X & \isdef {\arraycolsep=4pt\def\arraystretch{0.5}\left[\begin{array}{cccccc}  \delta x^{\rm b} & \delta y^{\rm b} & \delta z^{\rm b} & \delta \dot{x}^{\rm b} & \delta %
    \dot{y}^{\rm b} & \delta \dot{z}^{\rm b} \end{array}\right.} \nn \\  
    & {\arraycolsep=4pt\def\arraystretch{0.5}\left.\begin{array}{cccccc} \delta \phi & \delta \theta & \delta \psi & \delta \dot{\phi} & \delta \dot{\theta} & \delta \dot{\psi} \end{array}\right]}^{\rm T} \nn \\
    \delta U &\isdef {\arraycolsep=4pt\def\arraystretch{0.5}\left[\begin{array}{cccc} \delta T & \delta \tau_{\rm x} & \delta \tau_{\rm y} & \delta \tau_{\rm z} \end{array}\right]}^{\rm T} = {\arraycolsep=4pt\def\arraystretch{0.5}\left[\begin{array}{cccc} \delta T &\tau_{\rm x} & \tau_{\rm y} & \tau_{\rm z} \end{array}\right]}^{\rm T} \nn.
\end{align}
\normalsize
Given the nonlinear representation of the dynamics in \eqref{NLDynQuad}, the linearized quadcopter dynamics model takes the form

\begin{align}\label{LinDynXEq0}
    \frac{d}{dt} \delta X &= \left. \frac{\partial f_{\rm quad} (X, U)}{\partial X} \right\rvert_{X = X_{\rm hov}, U = U_{\rm hov}} \delta X \nn \\
    &+ \left. \frac{\partial f_{\rm quad} (X, U)}{\partial U} \right\rvert_{X = X_{\rm hov}, U = U_{\rm hov}} \delta U.
\end{align}
Note that linearization decomposes the dynamics of the nonlinear model into six sub-models. Each sub-model is defined by its generalized state $\delta\mathbb{X},$ its time derivative $\delta \dot{\mathbb{X}},$ a parameter $K_\delta,$ a generalized input $\delta \mathbb{U},$ and dynamic equations of the form
\begin{equation} \label{LinDynXEq}
     \frac{d}{dt} \begin{bmatrix}\delta \mathbb{X} \\ \delta \dot{\mathbb{X}}\end{bmatrix} = \begin{bmatrix} 0 & 1 \\ 0 & 0 \end{bmatrix} \begin{bmatrix}\delta \mathbb{X} \\ \delta \dot{\mathbb{X}}\end{bmatrix} + \begin{bmatrix} 0 \\ K_\delta \end{bmatrix} \delta \mathbb{U}.
\end{equation}
%
%
For $\delta \mathbb{X} = \delta x^{\rm b}, \delta \mathbb{U} = \delta \theta$ and $K_\delta = -g.$
For $\delta \mathbb{X} = \delta y^{\rm b}, \delta \mathbb{U} = \delta \phi$ and $K_\delta = g.$
For $\delta \mathbb{X} = \delta z^{\rm b}, \delta \mathbb{U} = \delta T$ and $K_\delta = -\frac{1}{m}.$
For $\delta \mathbb{X} = \delta \phi, \delta \mathbb{U} = \tau_{\rm y}$ and $K_\delta = \frac{1}{J_{\rm yy}}.$
For $\delta \mathbb{X} = \delta \theta, \delta \mathbb{U} = \tau_{\rm x}$ and $K_\delta = \frac{1}{J_{\rm xx}}.$
For $\delta \mathbb{X} = \delta \psi, \delta \mathbb{U} = \tau_{\rm z}$ and $K_\delta = \frac{1}{J_{\rm zz}}.$

\subsubsection{Discretized Linearized System Dynamics}

The linearized model in the previous section is discretized in order to design digital controllers.
Let $k$ be the sample index, and define discrete-time state vector $X_k$ and input vector $U_k$ such that
\begin{align}
    \delta X_k & \isdef {\arraycolsep=4pt\def\arraystretch{0.5}\left[\begin{array}{cccccc} \delta x_k^{\rm b} & \delta y_k^{\rm b} & \delta z_k^{\rm b} & \delta \dot{x}_k^{\rm b} & \delta \dot{y}_k^{\rm b} & \delta \dot{z}_k^{\rm b}\end{array}\right.} \nn \\
    &{\arraycolsep=4pt\def\arraystretch{0.5}\left.\begin{array}{cccccc}\delta \phi_k & \delta \theta_k & \delta \psi_k & \delta \dot{\phi}_k & \delta \dot{\theta}_k & \delta \dot{\psi}_k \end{array}\right]}^{\rm T}, \nn \\
    \delta U_k &\isdef {\arraycolsep=4pt\def\arraystretch{0.5}\left[\begin{array}{cccc} \delta T_k & \tau_{{\rm x},k} & \tau_{{\rm y},k} & \tau_{{\rm z},k} \end{array}\right]}^{\rm T}. \nn
\end{align}
Assuming a zero-order hold for the input and a sampling period of $T_{\rm s}$, 
each of the six sub-models \eqref{LinDynXEq} is converted to discrete-time, with the discrete-time models given by
%
%
%
\begin{equation}\label{DTLinDynEq}
    \begin{bmatrix} \delta \mathbb{X}_{k+1} \\ \delta \dot{\mathbb{X}}_{k+1}\end{bmatrix} = \begin{bmatrix} 1 & T_{\rm s} \\ 0 & 1 \end{bmatrix} \begin{bmatrix} \delta \mathbb{X}_k \\ \delta \dot{\mathbb{X}}_k\end{bmatrix} + B_{T_{\rm s}} \delta \mathbb{U}_k,
\end{equation}
where $B_{T_{\rm s}}$ is a 2-by-1 matrix.
For $\delta \mathbb{X}_k = \delta x^{\rm b}_k, \delta \mathbb{U}_k = \delta \theta_k$ and $B_{T_{\rm s}} = - g {\arraycolsep=2pt\def\arraystretch{0.5}\begin{bmatrix} \tfrac{T_{\rm s}^2}{2} & T_{\rm s} \end{bmatrix}^{\rm T}}.$
For $\delta \mathbb{X}_k = \delta y^{\rm b}_k, \delta \mathbb{U}_k = \delta \phi_k$ and $B_{T_{\rm s}} = g {\arraycolsep=2pt\def\arraystretch{0.5}\begin{bmatrix} \tfrac{T_{\rm s}^2}{2} & T_{\rm s} \end{bmatrix}^{\rm T}}.$
For $\delta \mathbb{X}_k = \delta z^{\rm b}_k, \delta \mathbb{U}_k = \delta T_k$ and $B_{T_{\rm s}} = - {\arraycolsep=2pt\def\arraystretch{0.5}\begin{bmatrix} \tfrac{T_{\rm s}^2}{2m} & \tfrac{T_{\rm s}}{m} \end{bmatrix}^{\rm T}}.$
For $\delta \mathbb{X}_k = \delta \phi_k, \delta \mathbb{U}_k = \tau_{{\rm y} ,k}$ and $B_{T_{\rm s}} = {\arraycolsep=2pt\def\arraystretch{0.5}\begin{bmatrix} \tfrac{T_{\rm s}^2}{2 J_{\rm yy}} & \tfrac{T_{\rm s}}{J_{\rm yy}} \end{bmatrix}^{\rm T}}.$
For $\delta \mathbb{X}_k = \delta \theta_k, \delta \mathbb{U}_k = \tau_{{\rm x},k}$ and $B_{T_{\rm s}} = {\arraycolsep=2pt\def\arraystretch{0.5}\begin{bmatrix} \tfrac{T_{\rm s}^2}{2 J_{\rm xx}} & \tfrac{T_{\rm s}}{J_{\rm xx}} \end{bmatrix}^{\rm T}}.$
For $\delta \mathbb{X}_k = \delta \psi_k, \delta \mathbb{U}_k = \tau_{{\rm z},k}$ and $B_{T_{\rm s}} = {\arraycolsep=2pt\def\arraystretch{0.5}\begin{bmatrix} \tfrac{T_{\rm s}^2}{2 J_{\rm zz}} & \tfrac{T_{\rm s}}{J_{\rm zz}}  \end{bmatrix}^{\rm T}}.$

\subsubsection{System Model Parameter Estimation}

A summary of the estimated model parameters is displayed in Table \ref{parameters_system}. Parameters $C_{\rm T}, C_{\rm M}, C_{\rm R}, \omega_{\rm b}$ and $T_{\rm m}$ were obtained via dynamometer tests performed on all propulsor systems, as described in Section 6.3.4 of \citet{quanquan2017}. The value of $d$ was determined by measuring the distance from the center of a propulsor to the geometric center of the quadcopter. The value of $m$ was measured using a scale and $J_{\rm xx}, J_{\rm yy}$ and $J_{\rm zz}$ were determined by performing a bifilar pendulum test, as described in Section 6.3.3 of \citet{quanquan2017}. Finally, the value of the sampling period $T_{\rm s}$ was chosen as the fastest one available on the chosen embedded system. More details of parameter estimation techniques are given in Appendix \ref{propapp}.


\begin{table}[h!]
\caption{System Model Parameters}
\label{parameters_system}
\centering
\begin{tabular}{@{}ccc@{}}
\toprule
\textbf{Parameter}  & \textbf{Value}           & \textbf{Units}\\ \midrule
$C_{\rm T}$         & $5.724165\cdot 10^{-8}$  & kg.m\\
$C_{\rm M}$         & $8.881631\cdot 10^{-10}$ & kg.$\text{m}^2$\\
$C_{\rm R}$         & $9.9573\cdot 10^{3}$     & 1/s\\
$\omega_{\rm b}$        & $12.3517$                & 1/s\\
$T_{\rm m}$         & $0.245217$               & N/A\\
$d$                 & $0.17$                    & m \\
$m$                 & $1.062$  & kg\\
$J_{\rm xx}$        & $1.07\cdot 10^{-2}$ & kg.$\text{m}^2$\\
$J_{\rm yy}$        & $1.11\cdot 10^{-2}$     & kg.$\text{m}^2$\\
$J_{\rm zz}$        & $2.29\cdot 10^{-2}$                & kg.$\text{m}^2$\\
$T_{\rm s}$         & $0.05$               & s\\\bottomrule
\end{tabular}
\end{table}

\section{Guidance, Navigation, and Control}
\label{section:guidance}
\begin{figure*}[!b]
    \centering
    \resizebox{0.81\textwidth}{!}{%
        \begin{tikzpicture}[>={stealth'}, line width = 0.25mm]
    	\def\centerarc[#1](#2)(#3:#4:#5)
    	{ \draw[#1] ($(#2)+({#5*cos(#3)},{#5*sin(#3)})$) arc (#3:#4:#5); }

    \node [smallblock, fill=blue!20, minimum height = 0.6cm , minimum width = 0.6cm] (Abs2Rel) {\scriptsize\begin{tabular}{c} Yaw \\ Alignment \end{tabular}};
        \node [smallblock, right =0.65cm of Abs2Rel, fill=green!20, minimum height = 0.6cm , minimum width = 0.6cm] (controller) {\scriptsize\begin{tabular}{c} Digital \\ Controller \end{tabular}};
        \node [input, name=ref, above left = -0.175cm and 1.5cm of Abs2Rel]{};
    	\node [smallblock, right =0.65cm of controller, minimum width = 0.6cm, minimum height = 0.6cm] (comm_map) {\scriptsize \begin{tabular}{c} Command \\ Mapping \end{tabular}};
    	\node [smallblock, right = 0.65cm of comm_map, minimum height = 0.6cm , minimum width = 0.5cm] (DA) {\scriptsize ZOH};
    	\node [smallblock, fill=white, right = 1cm of DA, xshift = -0.35cm, minimum height = 0.6cm , minimum width = 0.7cm, label={\scriptsize Quadcopter}] (system) {\centering \includegraphics[width=5em]{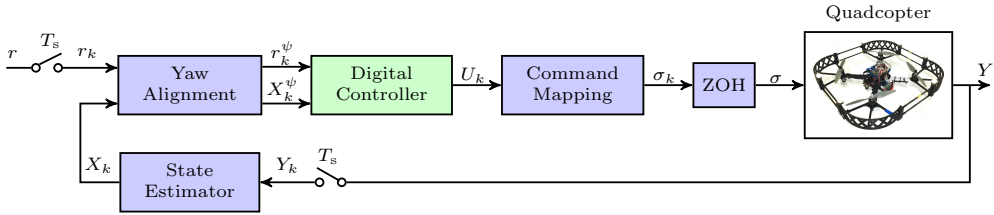}};
    	\node [output, right = 0.5cm of system](output){};

    	\node [smallblock, below =  0.5cm of Abs2Rel, minimum width = 0.6cm, minimum height = 0.6cm] (estimator) {\scriptsize \begin{tabular}{c} State \\ Estimator \end{tabular}};
    	
    	\draw [->] (controller) -- node [xshift = -0.05cm, yshift = -0.1cm,above] (du_k) {\scriptsize$U_k$} (comm_map);
    	\draw [->] (comm_map) -- node [xshift = -0.05cm, yshift = -0.1cm,above] (du) {\scriptsize$\sigma_k$} (DA);
    	
    	\node[circle,draw=black, fill=white, inner sep=0pt,minimum size=3pt] (rc1) at ([xshift=0.4cm]ref) {};
    	\node[circle,draw=black, fill=white, inner sep=0pt,minimum size=3pt] (rc2) at ([xshift=0.7cm]ref) {};
    	\draw [-] (rc1.north east) --node[below, yshift=.475cm]{\scriptsize$T_{\rm s}$} ([xshift=.3cm,yshift=.15cm]rc1.north east) {};
    	
    	\node[circle,draw=black, fill=white, inner sep=0pt,minimum size=3pt] (rc11) at ([xshift=1.1cm]estimator.east) {};
    	\node[circle,draw=black, fill=white, inner sep=0pt,minimum size=3pt] (rc21) at ([xshift=0.8cm]estimator.east) {};
    	\draw [-] ([yshift=.15cm, xshift=-.08cm]rc21.north east) --node[below,yshift=.475cm]{\scriptsize$T_{\rm s}$} ([xshift=.225cm, yshift=-0.025cm]rc21.north east) {};
    	
    	\draw [->] (DA) -- node [name=u]{} node [xshift = -0.05cm, yshift = -0.1cm,above] {\scriptsize$\sigma$} (system);
    	\draw [->] (system) -- node [name=y, near start, xshift = 0.1cm]{} node [very near end, above] {\scriptsize$Y$}(output);
    	
    	\draw [-] (ref) -- node [near start, above] {\scriptsize$r$} (rc1.west);
    	\draw [->] (rc2.east) -- node [near start, above, xshift = 0.15cm] {\scriptsize$r_k$}  node [near end, above] {} ([xshift = 1.5cm]ref);
    	\draw [-] (y.center) |- (rc11.east);
    	\draw [->] (rc21) -- node [above] {\scriptsize$Y_k$} (estimator.east);
    	\draw [->] (estimator.west) -| node [very near start, above, xshift = -0.15cm] {\scriptsize$X_k$} ([xshift=-0.5cm, yshift = -0.25cm]Abs2Rel.west) -- ([yshift = -0.25cm]Abs2Rel.west);
        \draw[->] ([yshift = 0.25cm]Abs2Rel.east) -- node [xshift = -0.05cm, yshift = -0.1cm,above] {\scriptsize$r_{k}^{\psi}$}([yshift = 0.25cm]controller.west);
        \draw[->] ([yshift = -0.25cm]Abs2Rel.east) -- node [xshift = -0.05cm, yshift = -0.1cm,above] {\scriptsize$X_{k}^{\psi}$}([yshift = -0.25cm]controller.west);
    	
    	\end{tikzpicture}
    }
    \caption{Control system block diagram. The digital controller yields the system input in terms of vertical thrust and torques, which must be mapped into motor commands.}
    \label{Fig:quad_blk_diag}
\end{figure*}

The block diagram representing quadcopter guidance, navigation, and control is shown in Fig. \ref{Fig:quad_blk_diag}. Let $T_{\rm s}$ be the sampling period,
$r \isdef {\arraycolsep=2pt\def\arraystretch{0.5}\left[\begin{array}{cccccccc} r_x^{\rm i} &  r_y^{\rm i} & r_z^{\rm i} & r_\psi^{\rm i} & r_{\dot{x}}^{\rm i} &  r_{\dot{y}}^{\rm i} & r_{\dot{z}}^{\rm i} & r_{\dot{\psi}}^{\rm i} \end{array}\right]}$
be the reference signal vector including setpoints for position ($r_x^{\rm i}, r_y^{\rm i}, r_z^{\rm i}$), velocity ($r_{\dot{x}}^{\rm i}, r_y^{\rm i}, r_z^{\rm i}$), yaw angle and yaw rate ($r_{\psi}^{\rm i}, r_{\dot{\psi}}^{\rm i}$),
%
%
and $r_k \isdef {\arraycolsep=2pt\def\arraystretch{0.5}\left[\begin{array}{cccccccc} r_{x,k}^{\rm i} &  r_{y,k}^{\rm i} & r_{z,k}^{\rm i} & r_{\psi,k}^{\rm i} & r_{\dot{x},k}^{\rm i} &  r_{\dot{y},k}^{\rm i} & r_{\dot{z},k}^{\rm i} & r_{\dot{\psi},k}^{\rm i} \end{array}\right]}$
be the sampled reference signal vector, such that $r_k = r(k T_{\rm s}).$ Let $Y$ be the vector of quadcopter system output signals and $Y_k$ be the vector of sampled sensor measurements such that $Y_k = Y(k T_{\rm s}).$ At each iteration, the State Estimator provides a sampled estimate of state vector $X_k,$ such that $X_k = X(k T_{\rm s}).$
Due to the decomposition resulting from model linearization, the digital controller requires position and velocity states and references to be aligned with the axes of the frame obtained by rotating the $\hat{i}^{\rm i}$ and $\hat{j}^{\rm i}$ axes by $\psi$ around the $\hat{k}^{\rm i}$ axis.
Hence, a Yaw Alignment procedure must be performed to obtain aligned reference and state vectors
$r_k^\psi \isdef {\arraycolsep=2pt\def\arraystretch{0.5}\left[\begin{array}{cccccccc} r_{x,k}^{\psi} &  r_{y,k}^{\psi} & r_{z,k}^{\rm i} & r_{\psi,k}^{\rm i} & r_{\dot{x},k}^{\psi} &  r_{\dot{y},k}^{\psi} & r_{\dot{z},k}^{\rm i} & r_{\dot{\psi},k}^{\rm i} \end{array}\right]}$
and
$X_k^{\psi} \isdef {\arraycolsep=3pt\def\arraystretch{0.5}\left[\begin{array}{cccccccccccc} x_k^{\psi} & y_k^{\psi} & z_k^{\rm i} & \dot{x}_k^{\psi} & \dot{y}_k^{\psi} & \dot{z}_k^{\rm i} & \phi_k & \theta_k & \psi_k & \dot{\phi}_k & \dot{\theta}_k & \dot{\psi}_k \end{array}\right]}^{\rm T},$
respectively.
The following transforms ${\arraycolsep=2pt\def\arraystretch{0.5}\begin{bmatrix} x_k^{\rm i} & y_k^{\rm i}\end{bmatrix}^{\rm T}}$ into ${\arraycolsep=2pt\def\arraystretch{0.5}\begin{bmatrix} x_k^{\psi} & y_k^{\psi}\end{bmatrix}^{\rm T}}$:
\begin{equation}\label{RefAlignEq}
    %
    \begin{bmatrix} x_k^{\psi} \\ y_k^{\psi}\end{bmatrix} = \begin{bmatrix} {\rm c}{\psi} & {\rm s}{\psi} \\  -{\rm s}{\psi} &  {\rm c}{\psi} \end{bmatrix} \begin{bmatrix} x_k^{\rm i} \\ y_k^{\rm i}\end{bmatrix}.
\end{equation}
The same transformation is applied for transforming ${\arraycolsep=2pt\def\arraystretch{0.5}\begin{bmatrix} \dot{x}_k^{\rm i} & \dot{y}_k^{\rm i}\end{bmatrix}^{\rm T}},$ ${\arraycolsep=2pt\def\arraystretch{0.5}\begin{bmatrix} r_{x,k}^{\rm i} & r_{y,k}^{\rm i}\end{bmatrix}^{\rm T}},$ and ${\arraycolsep=2pt\def\arraystretch{0.5}\begin{bmatrix} r_{\dot{x},k} & r_{\dot{y},k}\end{bmatrix}^{\rm T}}$ into ${\arraycolsep=2pt\def\arraystretch{0.5}\begin{bmatrix} x_k^{\psi} & y_k^{\psi}\end{bmatrix}^{\rm T}},$ ${\arraycolsep=2pt\def\arraystretch{0.5}\begin{bmatrix} \dot{x}_k^{\psi} & \dot{y}_k^{\psi}\end{bmatrix}^{\rm T}},$ ${\arraycolsep=2pt\def\arraystretch{0.5}\begin{bmatrix} r_{x,k}^{\psi} & r_{y,k}^{\psi}\end{bmatrix}^{\rm T}},$ and ${\arraycolsep=2pt\def\arraystretch{0.5}\begin{bmatrix} r_{\dot{x},k}^{\psi} & r_{\dot{y},k}^{\psi}\end{bmatrix}^{\rm T}},$ respectively.

The Digital Controller uses $X_k^\psi$ and $r_k^\psi$ to generate the input signal
$U_k \isdef {\arraycolsep=4pt\def\arraystretch{0.5}\left[\begin{array}{cccc} T_k & \tau_{{\rm x},k} & \tau_{{\rm y},k} & \tau_{{\rm z},k} \end{array}\right]}^{\rm T}.$
Then, a Command Mapping procedure must be performed to obtain the normalized control signals $\sigma_k \isdef {\arraycolsep=4pt\def\arraystretch{0.5}\left[\begin{array}{cccc} \sigma_{k,1} & \sigma_{k,2} & \sigma_{k,3} & \sigma_{k,4}\end{array}\right]}^{\rm T}$ that will be sent as commands to the motors.
Inverting the matrix displayed in Eq. \eqref{eqMixMatrix} and defining  $C_{T_{\rm inv}} = \frac{1}{4 C_{\rm T}}$, $C_{M_{\rm inv}} = \frac{1}{4 C_{\rm M}}$ and $C_{\tau_{\rm inv}} = \frac{\sqrt{2}}{4 d C_{\rm T}}$, the squared angular velocities of the motors required to obtain the forces and torques contained in $U$ are obtained as follows

\vspace{-0.75em}
\footnotesize
\begin{equation} \label{eqInvMixMatrix}
     \begin{bmatrix} \omega_{k,1}^2 \\ \omega_{k,2}^2 \\ \omega_{k,3}^2 \\ \omega_{k,4}^2 \end{bmatrix} =
   \begin{bmatrix} C_{{\rm T}_{\rm inv}} & -C_{\tau_{\rm inv}} & C_{\tau_{\rm inv}} & -C_{{\rm M}_{\rm inv}} \\ C_{{\rm T}_{\rm inv}} & -C_{\tau_{\rm inv}} & -C_{\tau_{\rm inv}} & C_{{\rm M}_{\rm inv}} \\ C_{{\rm T}_{\rm inv}} & C_{\tau_{\rm inv}} & -C_{\tau_{\rm inv}} & -C_{{\rm M}_{\rm inv}} \\ C_{{\rm T}_{\rm inv}} & C_{\tau_{\rm inv}} & C_{\tau_{\rm inv}} & C_{{\rm M}_{\rm inv}} \end{bmatrix}
   \begin{bmatrix} T_k \\ \tau_{{\rm x},k} \\ \tau_{{\rm y},k} \\ \tau_{{\rm z},k} \end{bmatrix}.
\end{equation}
\normalsize
Finally, it follows from \eqref{sigma2WssEq} that
\begin{equation}
    \sigma_{k,i} = \frac{\omega_{k,i} - \omega_{\rm b}}{C_{\rm R}}. \label{omega2sigma}
\end{equation}
A zero-order hold is performed on this signal, such that
\begin{equation} \label{sigmaCT}
    \sigma(t) = \sigma_k, \hspace{1cm} t\in[k T_s, (k+1)T_s)].
\end{equation}
%


\subsection{Digital Controller Structure} \label{subsec:DCS}

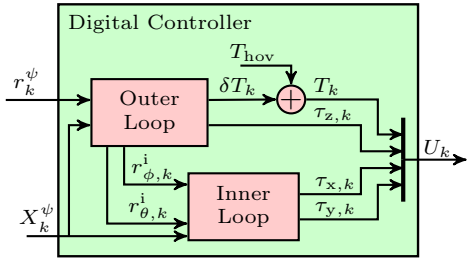
\begin{figure}[h!]
    \centering
    \resizebox{0.8\columnwidth}{!}{%
        \begin{tikzpicture}[>={stealth'}, line width = 0.25mm]

        \node [draw, thick, shape = rectangle, fill=green!20, minimum width = 4.25cm, minimum height = 3cm, anchor=center](controller){};
        \node[below, right] at ([yshift = -0.25cm]controller.north west){\scriptsize Digital Controller};
        \node [smallblock, fill=red!20, minimum height = 0.6cm , minimum width = 0.6cm, , above left = -0.2cm and 0.35cm of controller.center](OuterLoop) {\scriptsize\begin{tabular}{c} Outer \\ Loop \end{tabular}};
        \node [sum, fill=red!20, above right = 0.03cm and 0.85cm of OuterLoop.east](sumB){};
        \node[draw = none] at (sumB.center) {$+$};
        \node [smallblock, fill=red!20, minimum height = 0.6cm , minimum width = 0.6cm, below right = 0.3 cm and -0.25cm of OuterLoop](InnerLoop) {\scriptsize\begin{tabular}{c} Inner \\ Loop \end{tabular}};
        \node [coordinate, below right = 0.15cm and 2.3cm of OuterLoop](muxU){};


        \draw[->] ([xshift=-1cm, yshift=0.15cm]OuterLoop.west)-- node [above, xshift = -0.25cm, yshift = -0.075cm](vbk){\scriptsize$r_k^{\psi}$} ([yshift=0.15cm]OuterLoop.west);
        \draw[->] ([xshift=-0.3cm]OuterLoop.south) |- node[above, very near end, xshift = -0.22cm, yshift = -0.1cm](vthetak){\scriptsize$r_{\phi, k}^{\rm i}$}([yshift=0.265cm]InnerLoop.west);
        \draw[->] ([xshift=-0.5cm]OuterLoop.south) |- node[above, very near end, xshift = -0.22cm, yshift = -0.1cm](vthetak){\scriptsize$r_{\theta, k}^{\rm i}$}([yshift=-0.2cm]InnerLoop.west);

        \draw[->] ([xshift = -1.9cm, yshift = -0.35cm]InnerLoop.west) -- node [very near start, above, xshift = -0.11cm, yshift = -0.075cm](state){\scriptsize$X_k^{\psi}$} ([yshift = -0.35cm]InnerLoop.west);

        \draw[->] ([xshift = -1.4cm, yshift = -0.35cm]InnerLoop.west) |- ([yshift=-0.15cm]OuterLoop.west);

        \draw[line width=0.5mm] (muxU.center)--([yshift = 0.5cm]muxU.center);
        \draw[line width=0.5mm] (muxU.center)--([yshift = -0.5cm]muxU.center);
        \draw[->] ([yshift=0.15cm]OuterLoop.east) -- node [above, xshift = -0.05cm, yshift = -0.075cm](dT){\scriptsize$\delta T_k$} (sumB.west);
        \draw[->] ([xshift=-0.6cm, yshift = 0.225cm]sumB.north) -| node [above, very near start, yshift = -0.075cm](T0){\scriptsize$T_{\rm hov}$}(sumB.north);
        \draw[->] (sumB.east) -| node [above, xshift = -0.6cm, yshift = -0.075cm](TT){\scriptsize$T_k$} ([xshift = -0.3cm, yshift = 0.3cm]muxU.center) -- ([yshift = 0.3cm]muxU.center);

        \draw[->] ([yshift=-0.15cm]OuterLoop.east) -| node [above, xshift = -0.325cm, yshift = -0.1cm](tauz){\scriptsize$\tau_{{\rm z},k}$} ([xshift = -0.5cm, yshift = 0.1cm]muxU.center) -- ([yshift = 0.1cm]muxU.center);

         \draw[->] ([yshift=0.15cm]InnerLoop.east) -| node [above, xshift = -0.3cm, yshift = -0.1cm](taux){\scriptsize$\tau_{{\rm x},k}$} ([xshift = -0.5cm, yshift = -0.1cm]muxU.center) -- ([yshift = -0.1cm]muxU.center);

        \draw[->] ([yshift=-0.15cm]InnerLoop.east) -| node [above, xshift = -0.5cm, yshift = -0.1cm](tauy){\scriptsize$\tau_{{\rm y},k}$} ([xshift = -0.3cm, yshift = -0.3cm]muxU.center) -- ([yshift = -0.3cm]muxU.center);

        \draw[->] (muxU.center) -- node [above, very near end, xshift = -0.25cm, yshift = -0.1cm](uk){\scriptsize$U_k$} ([xshift = 0.75cm]muxU.center);
        
    \end{tikzpicture}
    }
    \caption{Controller Block Diagram}
    \label{Fig:controller_blk_diag}
\end{figure}

All digital controllers implemented in this paper follow the same structure displayed in Fig. \ref{Fig:controller_blk_diag}. The Inner Loop block features controllers that yield the required rolling and pitching moments $\tau_{\rm x}$ and $\tau_{\rm y}$ to regulate horizontal movement. The Inner Loop block requires the current quadcopter attitude as well as reference signals for roll and pitch to obtain its output, such that
\begin{align}
    \tau_{{\rm x}, k} &= C_{{\rm IL}, \theta} (\theta_k,\dot{\theta}_k,r_{\theta,k}^{\rm i}, r_{\dot{\theta},k}^{\rm i}), \label{controllerEq_1}\\
    \tau_{{\rm y}, k} &= C_{{\rm IL}, \phi} (\phi_k,\dot{\phi}_k,r_{\phi,k}^{\rm i}, r_{\dot{\phi},k}^{\rm i}).
\end{align}
The reference signals for roll and pitch are obtained from the Outer Loop block, which features controllers that yield the required thrust differential and yawing moments as well as the desired reference signals. Yaw and position are achieved with:
\begin{align}
    \delta T_k &= C_{{\rm OL}, z} (z_k^{\rm i},\dot{z}_k^{\rm i},r_{z,k}^{\rm i}, r_{\dot{z},k}^{\rm i}),\\
    \tau_{{\rm z}, k} &= C_{{\rm OL}, \psi} (\psi_k,\dot{\psi}_k,r_{\psi,k}^{\rm i}, r_{\dot{\psi},k}^{\rm i}),\\
    r_{\phi,k}^{\rm i} &= C_{{\rm OL}, y} (y_k^{\psi},\dot{y}_k^{\psi},r_{y,k}^{\psi}, r_{\dot{y},k}^{\psi}),\\
    r_{\theta,k}^{\rm i} &= C_{{\rm OL}, x} (x_k^{\psi},\dot{x}_k^{\psi},r_{x,k}^{\psi}, r_{\dot{x},k}^{\psi}). \label{controllerEq_6}
\end{align}
Total thrust is obtained by adding the required thrust differential to hover thrust:
\begin{equation}
    T_k = T_{\rm hov} + \delta T_k.
\end{equation}
Hence, a set of six control schemes is required for control of the quadcopter. Evaluated control algorithms will be discussed in Section \ref{section:linear_controllers}.
Note that all controllers saturate these signals such that $\delta T \in [-10.42, 12.34]$ N, $\tau_{\rm x} \in [-1.3679, 1.3679]$ N.m, $\tau_{\rm y} \in [-1.3679, 1.3679]$ N.m, $\tau_{\rm z} \in [-0.1766, 0.1766]$ N.m, $r_{\phi}^{\rm i}\in[-1, 1]$ rad and $r_{\theta}^{\rm i}\in[-1, 1]$ rad.
Note that while the saturation limits for $\delta T, \tau_{\rm x}, \tau_{\rm y}$ and $\tau_{\rm z}$ were derived from the thrust and torque limits of the rotors, the saturation limits for $r_{\phi}^{\rm i}$ and $r_{\theta}^{\rm i}$ were chosen arbitrarily to prevent the outer loop controllers from requesting potentially unstable reference signals for roll and pitch.

\subsection{State Estimation} 
\label{section:state_estimation}

State feedback controllers require estimates of the full state vector $X.$ For the quadcopter, $X \in \mathbb{R}^{12}$ is comprised of three position coordinates $(x^{\rm i},y^{\rm i},z^{\rm i})$, three linear velocity components $(\dot{x}^{\rm i},\dot{y}^{\rm i},\dot{z}^{\rm i})$, three Euler angles $(\phi, \theta, \psi)$, and three angular velocity components $(\omega_{\rm x}^{\rm b} , \omega_{\rm y}^{\rm b} , \omega_{\rm z}^{\rm b})$ per Section \ref{QuadDynSysMod}. 
Position and Euler angle measurements are provided by the MOCAP system. However, the following problems were encountered during experimentation which motivated utilization of a filter: 
\begin{enumerate}
    \item Neither the onboard sensors nor the MOCAP directly provided linear velocity estimates;
    \item The yaw estimate from the onboard IMU grew unbounded;
    \item Measurements were available asynchronously; the IMU and MOCAP sampling rates were 200 Hz and 100 Hz respectively.
\end{enumerate}

A Kalman Filter (KF) was chosen to filter incoming measurements for several reasons.
First, a KF reduces noise and fuses redundant sensor measurements to provide an estimate $\Hat{X}$ of the entire state vector with the minimum mean-square-error covariance (under linear KF assumptions).
Moreover, it can estimate IMU biases thus stopping the unbounded growth of estimates.
Finally, a KF can process measurements sequentially as they are sampled, regardless of their sampling rates.
In this case, an a priori update is performed every $T_{\rm s}$ seconds on all states.
If new measurements have been sampled, an a posteriori update is performed as well on all the states with new measurements.
These characteristics make the KF an ideal state estimation solution. 

Since the linearized dynamics are decoupled, three independent Kalman Filters were designed:
\begin{enumerate}
    \item Vertical position and velocity ($z^{\rm i}, \dot{z}^{\rm i}$)
    \item Horizontal position and velocity ($x^{\rm i}, y^{\rm i}, \dot{x}^{\rm i}, \dot{y}^{\rm i}$)
    \item Yaw and yaw rate ($\psi, \dot{\psi}$)
\end{enumerate}
Estimators for pitch, roll, and their respective rates were not required because onboard sensors provided acceptable estimates.
%
%
In all cases, linear KFs were based on the linearized quadcopter model.
The specific Gauss-Markov models for each estimator, as well as their corresponding covariance matrices, are defined in Appendices \ref{append_vert_estimator}, \ref{append_horiz_estimator}, and \ref{append_yaw_estimator}.
Specific calibration values can be found in Appendix \ref{append_calib_results}.

\begin{figure}[b!]
    \centering
    \includegraphics[width=\columnwidth]{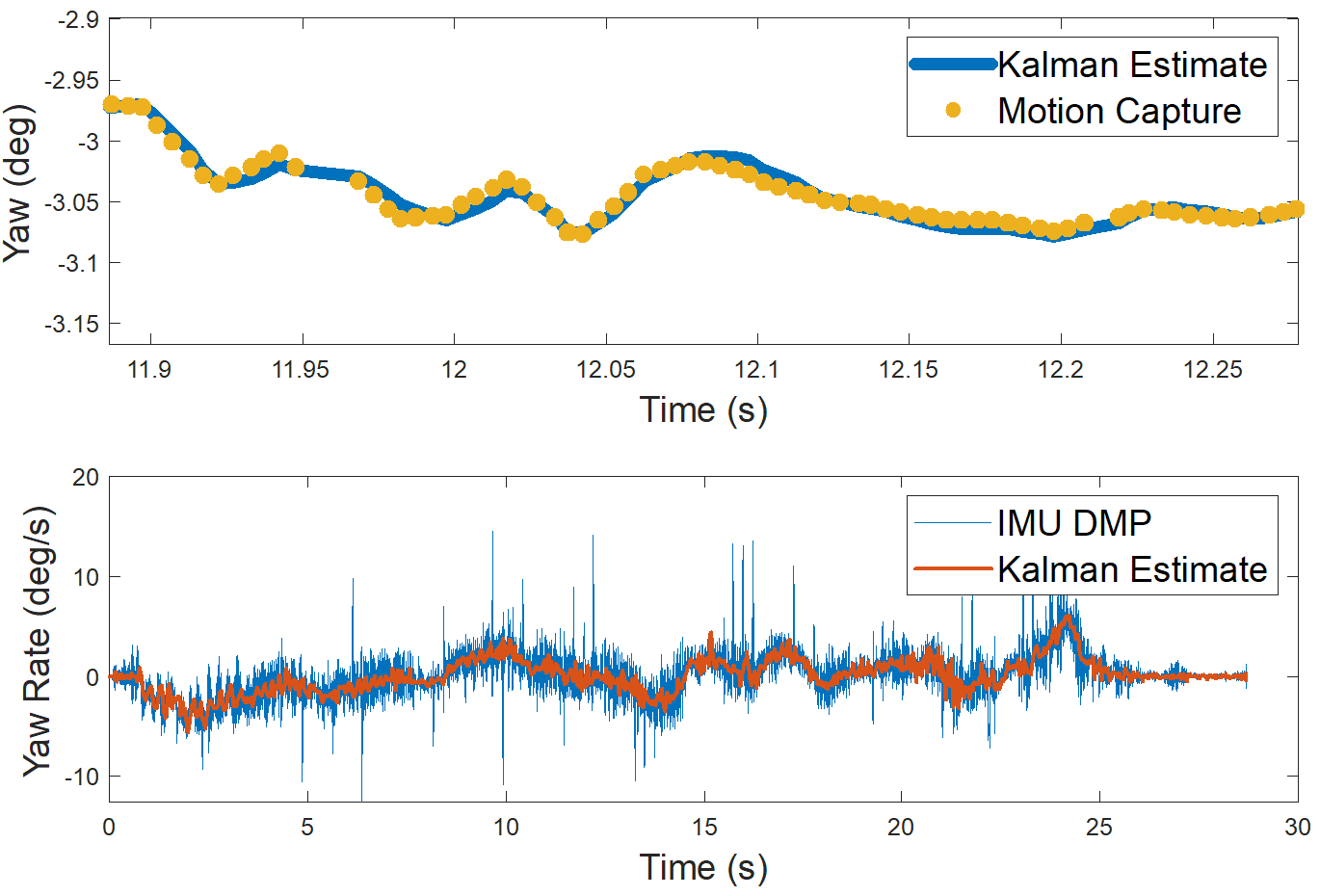}
    \caption{[Top] Yaw vs. time. MOCAP measurements are overlaid on Kalman Filter estimates. [Bottom] Yaw rate vs. time. Kalman Filter estimates are overlaid on IMU DMP measurements.}
    \label{yaw_and_yaw_rate_plot}
\end{figure}

\subsection{State Estimation Performance}
Fig. \ref{yaw_and_yaw_rate_plot} highlights the performance achieved with the proposed state estimation scheme. Note that the top and bottom plots do not share the same time scale; the top plot is zoomed in to assist with the following discussion. The top plot shows MOCAP yaw measurements overlaid on the Kalman Filter estimates. Note that, even though MOCAP data was received at a slower rate and at times had gaps in coverage, the Kalman Filter was able to maintain a smooth yaw estimate at all times. This was because the rate gyro (bottom plot) was read at 200 Hz, enabling the estimator to continue propagating both states. Note also that the Kalman yaw rate estimate has less noise than the IMU DMP measurement. These plots demonstrate how fusing measurements sequentially through a Kalman Filter yielded smoother estimates regardless of measurement asynchronicity. Similar performance was achieved with the horizontal and vertical position and velocity estimators.

\section{Digital Controllers}
\label{section:linear_controllers}

This section describes the quadcopter digital control laws along with their tuning procedures using Simulink simulations.  A brief summary of requirements and capabilities of the controllers is presented in Table \ref{Table:Controller_comparison}.  PID and LQR are classical methods well established in the literature \citep{yun2006,kwakernaak1972,sontag1990}.
MPC is also an established method, though computational overhead challenges implementation on typical real-time embedded computing components found on quadcopters. The proposed quadcopter MPC, LQR, and PID control designs are described below. 

\begin{table}[hbt!]
\caption{\label{tab:table1} Controller Comparison}
\centering
\begin{tabular}{lccccc}
\toprule
\textbf{Factors} & \textbf{PID}& \textbf{LQR} & \textbf{MPC}  \\ \midrule
Model required& No & Yes & Yes \\
Optimization Online& No &  No & Yes\\
Can be made explicit& N/A & N/A & Yes \\
Implement Constraints & No & No & Yes \\
\bottomrule
\label{Table:Controller_comparison}
\end{tabular}
\end{table}

\subsection{Explicit Model Predictive Control (E-MPC)}

Model predictive control (MPC) relies on predicting the response of the system using a model and ensures that the imposed state and control constraints are enforced. At each time step, MPC solves a constrained optimization problem, minimizing a cost function subject to the constraints. The response of the system is predicted over time period or horizon $T_h$.
%
%
The first move in the optimal control is selected as the control action thereby defining a feedback law \citep{mayne2000}. 
The main challenge in using MPC for quadcopters is its high computational cost versus limited onboard computation capabilities.
The quadcopter model shown in Section \eqref{subsec:quad_model} is nonlinear with 12 states. A direct implementation of MPC to this nonlinear model was not feasible given limited computational resources  onboard the quadrotor. 
Hence, the linearized model around hover \eqref{LinDynXEq0} is used for prediction.
Note that the linearized model decomposes into six submodels \eqref{LinDynXEq}.
For instance, the dynamics of the position state $\delta x^{\rm b}$ are described by a double integrator with pitch angle $\delta \theta$ as an input, whereas the dynamics of pitch angle $\delta \theta$ are a double integrator with input $\tau_x$. Similar considerations apply to the remaining linearized equations.
%
%
By exploiting this independence, six MPC controllers with small computational footprint can be developed based on the six double integrator models.

\begin{figure}[h!]
    \centering
    \resizebox{\columnwidth}{!}{%
    \begin{tikzpicture}[>={stealth'}, line width = 0.25mm]

        \node [smallblock,fill=blue!20, minimum height = 0.6cm , minimum width = 0.6cm] (psiMPCBlk){\begin{tabular}{c} MPC for \\ $\psi$ angle \end{tabular}};
        
        \node [smallblock,fill=blue!20, minimum height = 0.6cm , minimum width = 0.6cm, right = 0.5cm of psiMPCBlk] (zMPCBlk){\begin{tabular}{c} MPC for \\ $z$ position \end{tabular}};
        
        \node [smallblock,fill=blue!20, minimum height = 0.6cm , minimum width = 0.6cm, right = 0.5cm of zMPCBlk] (xMPCBlk){\begin{tabular}{c} MPC for \\ $x$ position \end{tabular}};
        
        \node [smallblock,fill=blue!20, minimum height = 0.6cm , minimum width = 0.6cm, right = 0.5cm of xMPCBlk] (yMPCBlk){\begin{tabular}{c} MPC for \\ $y$ position \end{tabular}};
        
        \node [smallblock,fill=blue!20, minimum height = 0.6cm , minimum width = 0.6cm, below = 1cm of xMPCBlk] (thetaMPCBlk){\begin{tabular}{c} MPC for \\ $\theta$ angle \end{tabular}};
        
        \node [smallblock,fill=blue!20, minimum height = 0.6cm , minimum width = 0.6cm, below = 1cm of yMPCBlk] (phiMPCBlk){\begin{tabular}{c} MPC for \\ $\phi$ angle \end{tabular}};
        
        \node [draw = white, above = 0.275cm of psiMPCBlk](psiVar){$\psi,\dot{\psi},r_{\psi,k}^{\rm i}, r_{\dot{\psi},k}^{\rm i}$};
        \node [draw = white, above = 0.275cm of zMPCBlk](zVar){$z^{\rm i},\dot{z}^{\rm i},r_{z}^{\rm i}, r_{\dot{z}}^{\rm i}$};
        \node [draw = white, above = 0.275cm of xMPCBlk](xVar){$x^{\psi},\dot{x}^{\psi},r_{x}^{\psi}, r_{\dot{x}}^{\psi}$};
        \node [draw = white, above = 0.275cm of yMPCBlk](yVar){$y^{\psi},\dot{y}^{\psi},r_{y}^{\psi}, r_{\dot{y}}^{\psi}$};
        
        \node [draw = white, below = 0.275cm of psiMPCBlk](tauZVar){$\tau_z$};
        \node [draw = white, below = 0.275cm of zMPCBlk](dTVar){$\delta T$};
        \node [draw = white, below = 0.275cm of thetaMPCBlk](tauyVar){$\tau_x$};
        \node [draw = white, below = 0.275cm of phiMPCBlk](tauxVar){$\tau_y$};
        
        \draw[->] ([yshift=0.075cm]psiVar.south) -- (psiMPCBlk.north);
        \draw[->] ([yshift=0.075cm]zVar.south) -- (zMPCBlk.north);
        \draw[->] ([yshift=0.075cm]xVar.south) -- (xMPCBlk.north);
        \draw[->] ([yshift=0.075cm]yVar.south) -- (yMPCBlk.north);
        
        \draw[->] ([xshift=0.4cm]xMPCBlk.south) -- node [right]{$r_\theta^{\rm i}$}([xshift=0.4cm]thetaMPCBlk.north);
        \draw[->] ([xshift=0.4cm]yMPCBlk.south) -- node [right]{$r_\phi^{\rm i}$}([xshift=0.4cm]phiMPCBlk.north);
        
        \draw[->] ([xshift=-0.25cm,yshift=0.35cm]thetaMPCBlk.north) -- ([xshift=-0.25cm]thetaMPCBlk.north);
        \draw[->] ([xshift=-0.25cm,yshift=0.35cm]phiMPCBlk.north) -- ([xshift=-0.25cm]phiMPCBlk.north);
        
        \node [draw = white, above left = 0.3cm and -0.1cm of thetaMPCBlk.north](thetaVar){$\theta,\dot{\theta}$};
        
        \node [draw = white, above left = 0.3cm and -0.1cm of phiMPCBlk.north](phiVar){$\phi,\dot{\phi}$};
        
        \draw[->] (psiMPCBlk.south) -- ([yshift=-0.05cm]tauZVar.north);
        \draw[->] (zMPCBlk.south) -- ([yshift=-0.05cm]dTVar.north);
        \draw[->] (thetaMPCBlk.south) -- ([yshift=-0.05cm]tauyVar.north);
        \draw[->] (phiMPCBlk.south) -- ([yshift=-0.05cm]tauxVar.north);
        
    \end{tikzpicture}
	}    
    \caption{MPC control architecture}
    \label{mpc_block}
\end{figure}
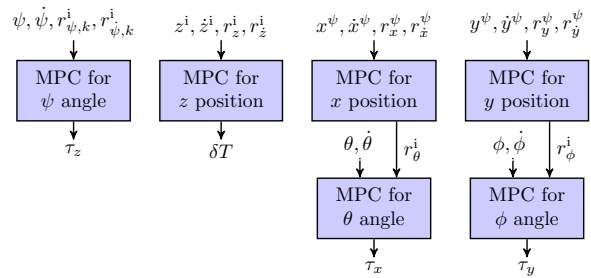

Fig. \ref{mpc_block} depicts the decoupled MPC control architecture.
This figure summarizes the inner and outer loop controller equations introduced in Eqs. \eqref{controllerEq_1} -- \eqref{controllerEq_6}.
Note that each of the MPC blocks assumes that their respective commanded inputs are applied instantly. However, the propulsion system has internal dynamics that affect generation of on $T, \tau_x, \tau_y$, and $\tau_z$. Similarly, MPC blocks for $x$ and $y$ position are designed based on the assumption that pitch and roll angles are achieved instantaneously, yet in fact they have their own dynamics determined by the MPC controllers for $\theta$ and $\phi$. These unmodelled dynamics have not produced substantive constraint violations in the conducted experiments, which justifies the use of this decomposition approach. 
 
As an example, consider MPC design for control of roll angle (other cases are similar).
It follows from \eqref{LinDynXEq} that the linearized mode for the dynamics of the roll angle has the form
\begin{equation} \renewcommand\arraystretch{1.5} 
\begin{bmatrix} \delta \dot{ \phi} \\ \delta \ddot{\phi}\end{bmatrix} = \begin{bmatrix} 0 &1 \\ 0 &0 \end{bmatrix} \begin{bmatrix} \delta \phi \\ \delta \dot{\phi}  \end{bmatrix} + \begin{bmatrix} 0 \\ \dfrac{1}{J_{yy}}  \end{bmatrix}\tau_{y},
\label{eq:roll_lin_eq}
\end{equation}
where $\delta \phi = \phi-r_\phi^{\rm i}$ and $\delta \dot{\phi} = \dot{\phi}.$ As the states of this system are the deviation with respect to roll setpoint and the roll angular velocity,
%
%
constraints may be imposed on both of these quantities.
%
%
In this case, only the roll angular velocity is constrained.
It follows from \eqref{DTLinDynEq} that discretization of \eqref{eq:roll_lin_eq} with sample time $T_s$ leads to a discrete-time model of the form
\begin{align}
x_{k+1} &= A x_k + B u_k \nn\\
&\equiv \begin{bmatrix} \delta \mathbb{\phi}_{k+1} \\ \delta \dot{\mathbb{\phi}}_{k+1}\end{bmatrix} = \begin{bmatrix} 1 & T_{\rm s} \\ 0 & 1 \end{bmatrix} \begin{bmatrix} \delta \mathbb{\phi}_k \\ \delta \dot{\mathbb{\phi}}_k\end{bmatrix} + \begin{bmatrix} \dfrac{T_{\rm s}^2}{2 J_{yy}} \\ \dfrac{T_{\rm s}}{J_{yy}} \end{bmatrix} \tau_{y,k}. \label{eq:discrete_time_mpc}
\end{align}
The roll angle MPC controller uses \eqref{eq:discrete_time_mpc} as the prediction model and determines control action based on solving the following finite-horizon discrete-time optimal control problem:
\begin{equation}
\label{eq:mpc_optimization_problem}
\begin{aligned}
\min _{u_0,u_1,\cdots,u_{N-1}} \quad & x_N^TPx_N + \sum_{k=0}^{N-1}x_k^TQx_k+u_k^TRu_k\\
\textrm{s.t.} \quad & x_{k+1}=Ax_k+Bu_k, \\
                    & x_0 \textrm{ is the current state}, \\
                    & |\dot{\phi_k}| \leq \dot{\phi}_{\rm max},
\end{aligned}
\end{equation}
where $N=T_h/T_s$, $Q,R$ are positive definite matrices and $P$ is given by the solution of the discrete time algebraic Riccati equation in the infinite horizon version of \eqref{eq:mpc_optimization_problem} without control constraints. The MPC feedback law is $u_{{\rm MPC},x}=u_0$. This MPC formulation recovers the solution of the associated LQR problem when constraints remain inactive.

As is stated in \citet{Cairano2009ModelPC}, the optimization problem \eqref{eq:mpc_optimization_problem} can be written in condensed form as
\begin{equation}
\label{qp_proble}
\begin{aligned}
\min _{U} \quad & \dfrac{1}{2}U^THU+q^TU\\
\textrm{s.t.} \quad & GU \leq W,
\end{aligned}
\end{equation}
where $U=(u_0,u_1,\cdots,u_{N-1})^T$ and $H,q,G,U$ are matrices that depend on $A,B,x_0$ and the constraints of the system. 

Solving the optimization problem as stated in Eq. \eqref{qp_proble} in real time on a Beaglebone Blue using a standard dual projection gradient algorithm implemented in C has proven to be challenging. For instance, with the prediction horizon set to $1$ s to capture all transients and a sampling period $T_s$ equal to $20$ ms, around $100$ ms was needed to solve the optimization problem \eqref{qp_proble}. Note that this time may be reduced with advanced time-distributed optimization strategies as described in \citet{liao2020}, although the problem of reducing the computation time remains difficult.

Therefore, an explicit MPC (E-MPC) approach was pursued. With E-MPC, each problem \eqref{qp_proble} is parameterized by initial condition $x_0$ which is a two dimensional vector.
%
%
The solution to problem \eqref{qp_proble} can be precomputed offline and the resulting control values can be stored in a two dimensional lookup table for online use in the Beaglebone processor.
%
%
To address the potential infeasibility of problem \eqref{qp_proble}, which can occur due to large disturbances, as was observed during flight tests, the implemented algorithm switches to an unconstrained LQR under such circumstances.

The particular structure of the quadcopter system dynamics and constraints is critical for the proposed implementation of E-MPC. When the system is linearized about a hovering condition, six two-dimensional smaller subsystems are obtained. These systems are not coupled, i.e., the control input computation is unidirectional, so there is no feedback between the six smaller subsystems (see Figure \ref{mpc_block}). Furthermore, the constraints on the system can be independently enforced on these smaller subsystems, i.e., there are no constraints that combine states/inputs of different subsystems. This convenient combination of decoupled dynamics and independent constraints is what allows the computation and storage of the control inputs as six two-dimensional look-up tables. Note that if the dimension of the smaller subsystems increases, then the size of the solution look-up table of \eqref{qp_proble} grows exponentially. Hence, E-MPC may quickly become intractable for more complex problems.
Our approach can be viewed as an instantiation of a more general idea of exploring symmetries in E-MPC design, see e.g., \cite{danielson2015}.

To tune MPC controller parameters, the sampling frequency was set to $200$ Hz, sufficiently fast for stability. To capture transients, time horizon was set to $1$ s. Constraints were set to illustrate MPC's constraint handling capability. Matrices $Q$ and $R$ were chosen by trial and error during outdoor tests. The parameters of each of six MPC controllers are summarized in Table \ref{Table:EMPCparam}.

\begin{table*}[ht!]
\footnotesize
\caption{Parameters of the six Model Predictive Controllers}
\centering
\begin{tabular}{lcccccc}
\toprule
\textbf{Parameter} & $\bm{x}$ & $\bm{y}$ & $\bm{z}$ & $\bm{\theta}$ & $\bm{\phi}$ & $\bm{\psi}$ \\ \midrule
$F_s$       & 200 Hz & 200 Hz & 200 Hz & 200 Hz & 200 Hz & 200 Hz\\
$T_h$      & 1 s & 1 s & 1 s & 1 s & 1 s & 1 s\\
$|x_{\rm max}|$  & 1000 m & 1000 m & 1000 m & 1000 rad & 1000 rad & 1000 rad\\
$|\dot{x}_{\rm max}|$  & 5 m/s & 5 m/s & 5 m/s & 45 deg/s & 45 deg/s & 90 deg/s \\
$|u_{\rm max}|$     & 45 deg  & 45 deg & 1000 N & 1000 N.m & 1000 N.m & 1000 N.m \\
{$Q$}   & diag({[}150, 200{]}) & diag({[}150, 200{]}) & diag({[}490, 117{]}) & diag({[}5, 1.5{]}) & diag({[}5,1.5{]}) & diag({[}100, 15{]}) \\
{$R$}   & 5000 & 5000 & 5.5 & 10 & 10  & 1000  \\ 
\bottomrule
\label{Table:EMPCparam}
\end{tabular}
\end{table*}

\subsection{Discrete LQR and LQR-I Control}

Discrete-time LQR exploits a linear discrete-time model $x_{k+1} = A x_k + B u_k,$
and generates the control sequence according to $u_k = -K x_k ,$ where $K = (R + B^T P B)^{-1} B^T P_k,$ to minimize performance index
\begin{equation*}
    J = \sum_{k = 0}^{\infty} x_k^T Q x_k + u_k^T R u_k.
\end{equation*}
In this expression, $P$ is the unique positive semi-definite solution to the discrete Algebraic Riccati Equation.

The design of the LQR and LQR-I controllers for each of the six subsystems is based on their models \eqref{DTLinDynEq}.
Thus, the overall LQR design is decoupled into six controllers similar to the MPC design.
%
%
Hence, in the LQR case, the controller has the form
%
%

\vspace{-0.75em}
\footnotesize
\begin{align} \label{eqLQR_basic}
\delta \mathbb{U}_k &= {\rm LQR}_{\mathbb{X}} (\mathbb{X}_k, \dot{\mathbb{X}}_k, r_{\mathbb{X},k}, r_{\dot{\mathbb{X}},k}) \nn \\
&= \begin{bmatrix} K_{\mathbb{X}} & K_{\dot{\mathbb{X}}} \end{bmatrix} \cdot \begin{bmatrix}  r_{\mathbb{X},k} - \mathbb{X}_k\\ r_{\dot{\mathbb{X}},k} - \dot{\mathbb{X}}_k \end{bmatrix} = \begin{bmatrix} K_{\mathbb{X}} & K_{\dot{\mathbb{X}}} \end{bmatrix} \cdot \begin{bmatrix} \delta \mathbb{X}_k\\ \delta \dot{\mathbb{X}}_k \end{bmatrix},
\end{align}
\normalsize
%
where $K_\mathbb{X}$ and $K_{\dot{\mathbb{X}}}$ are LQR gains, $\mathbb{X}$ is the state, $\dot{\mathbb{X}}$ is its time derivative and $r_{\mathbb{X},k}, r_{\dot{\mathbb{X}},k}$ are setpoints which are consistent with the unforced dynamics of \eqref{DTLinDynEq}. 
The state and control weighting matrices $Q_{\mathbb{X}}$ and $R_{\mathbb{X}}$ are used as tuning parameters.

\begin{figure*}[t!]
    \centering
    \resizebox{0.8\textwidth}{!}{%
    	\begin{tikzpicture}[>={stealth'}, line width = 0.25mm]
		\def\centerarc[#1](#2)(#3:#4:#5)
		{ \draw[#1] ($(#2)+({#5*cos(#3)},{#5*sin(#3)})$) arc (#3:#4:#5); }
		
		\node [sum] (sum_x) {};
		\node [draw = none] at (sum_x.center) {$+$};
		\node [sum, below = 1cm of sum_x] (sum_xdot) {};
		\node [draw = none] at (sum_xdot.center) {$+$};
		\node [smallblock, right = 0.3 of sum_x.east, minimum height = 0.1 cm, minimum width = 0.1cm](sx){\scriptsize$\frac{\bf{z}-1}{T_s \bf{z}}$};
		\node [smallblock, right = 0.3 of sum_xdot.east, minimum height = 0.1 cm, minimum width = 0.1cm](sxdot){\scriptsize$\frac{\bf{z}-1}{T_s \bf{z}}$};
		\node [gain,right = 0.3cm of sx.east, xscale = 2.5, yscale=1](Kx){};
		\node [draw = none] at ([xshift=-0.15cm]Kx.center) {\scriptsize$K_\mathbb{X}$};
		\node [gain,below = 0.475cm of Kx.center, xscale = 2.5,yscale=1](KxInt){};
		\node [draw = none] at ([xshift=-0.02cm]KxInt.center) {\scriptsize$K_{\mathbb{X},{\rm int}}$};
		\node [gain,right = 0.3cm of sxdot.east, xscale = 2.5, yscale=1](Kxdot){};
		\node [draw = none] at ([xshift=-0.15cm]Kxdot.center) {\scriptsize$K_{\dot{\mathbb{X}}}$};
		\node [sum, right = 0.35cm of KxInt.east](sumK){};
		\node [draw = none] at (sumK.center) {$+$};
		\node [sum, right = 0.25cm of sumK.east](sumTau){};
		\node [draw = none] at (sumTau.center) {$+$};
		\node [smallblock, right = 0.4 of sumTau.east, minimum height = 0.1 cm, minimum width = 0.1cm](sumInt){\scriptsize$\frac{T_s \bf{z}}{\bf{z} - 1}$};
		\node [gain, above = 0.2cm of sumInt, xscale = -1.75, yscale=1](gainTau){};
		\node [draw = none] at (gainTau.center) {\scriptsize$\tau_{\rm int}$};
		\node [sum, right = 0.5cm of gainTau.center](diffSat){};
		\node [draw = none] at (diffSat.center) {$+$};
		\node [saturation block, right = 0.75cm of sumInt, minimum width=1cm, minimum height=0.8cm] (satU) {};
		\node [draw = none] at ([xshift = 0.25cm,yshift = 0.28cm]satU.center) {\scalebox{.5}{$\delta \mathbb{U}_{\rm max}$}};
		\node [draw = none] at ([xshift = -0.25cm,yshift = -0.28cm]satU.center) {\scalebox{.5}{$\delta \mathbb{U}_{\rm min}$}};
		
		\draw [->] ([xshift=-0.35cm]sum_x.west) -- node [above, very near start, xshift = 0.04cm, yshift = -0.05cm] {\scriptsize$r_{\mathbb{X},k}$} (sum_x.west);
		\draw [->] ([xshift=-0.35cm, yshift=-0.45cm]sum_x.west) -| node [above, very near start, xshift = -0.2cm, yshift = -0.08cm] {\scriptsize$-{\mathbb{X}}_k$}(sum_x.south);
		\draw [->] ([xshift=-0.35cm]sum_xdot.west) -- node [above, very near start, xshift = 0.04cm, yshift = -0.05cm] {\scriptsize$r_{\dot{\mathbb{X}},k}$}(sum_xdot.west);
		\draw [->] ([xshift=-0.35cm, yshift=-0.45cm]sum_xdot.west) -| node [above, very near start, xshift = -0.2cm, yshift = -0.08cm] {\scriptsize$-\dot{\mathbb{X}}_k$}(sum_xdot.south);
		\draw [->] (sum_x.east) -- (sx.west);
		\draw [->] (sx.east) -- ([xshift=0.025cm]Kx.west);
		\draw [->] (sum_xdot.east) -- (sxdot.west);
		\draw [->] (sxdot.east) -- ([xshift=0.025cm]Kxdot.west);
		\draw [->] ([xshift=0.075cm]sum_x.east) |- ([xshift=0.025cm]KxInt.west);
		\draw [->] ([xshift=-0.06cm]Kx.east) --  ([xshift = 0.1cm]Kx.east) -- (sumK.120);
		\draw [->] ([xshift=-0.06cm]Kxdot.east) --  ([xshift = 0.1cm]Kxdot.east) -- (sumK.240);
		\draw [->] ([xshift=-0.06cm]KxInt.east) -- (sumK.west);
		\draw [->] (sumK.east) -- (sumTau.west);
		\draw [->] (sumTau.east) -- (sumInt.west);
		\draw [->] (sumInt.east) -- (satU.west);
		\draw [->] (sumInt.east) -| node [above,very near end, xshift=0.15cm, yshift = -0.15cm] {\tiny $-$} (diffSat.south);
		\draw [->] (diffSat.west) -- ([xshift=-0.025cm]gainTau.west);
		\draw [->] ([xshift=0.06cm]gainTau.east) -| (sumTau.north);
		\draw [->] (satU.east) -- node [above, very near end, xshift = 0.05cm,yshift = -0.05cm]{\scriptsize$\delta \mathbb{U}_k$} ([xshift=0.5cm]satU.east);
		\draw [->] ([xshift=0.175cm]satU.east) |- (diffSat.east);
	\end{tikzpicture}
	}
    \caption{Block diagram of LQR-I controller scheme with integrator anti-windup, where {\bf{z}} is the Z-transform variable.}
    \label{lqr_int_block}
\end{figure*}
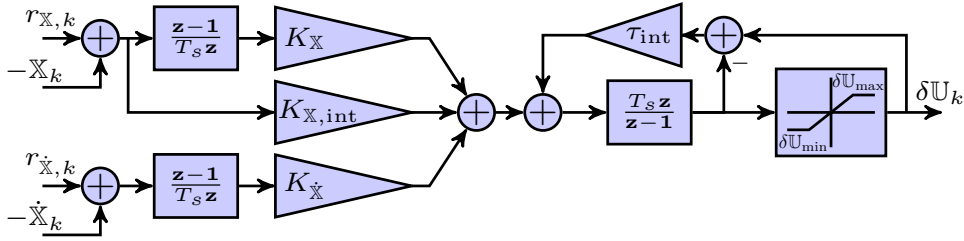

In the LQR-I case, the controller is based on the augmented model using \eqref{DTLinDynEq} with integrator state $\mathbb{X}_{\rm int},$ which has the form
\begin{align}
    \begin{bmatrix}  \delta \mathbb{X}_{k+1} \\ \delta \dot{\mathbb{X}}_{k+1} \\ \mathbb{X}_{{\rm int},k+1} \end{bmatrix} &= \begin{bmatrix} 1 & T_{\rm s} & 0 \\ 0 & 1 & 0 \\ T_{\rm s} & 0  & 1 \end{bmatrix} \begin{bmatrix}  \delta \mathbb{X}_k \\ \delta \dot{\mathbb{X}}_k \\ \mathbb{X}_{{\rm int},k} \end{bmatrix} + \begin{bmatrix} B_{T_{\rm s}} \\ 0 \end{bmatrix} \delta \mathbb{U}_k \nn \\
    &= A_{\mathbb{X},{\rm int}} \begin{bmatrix}  \delta \mathbb{X}_k \\ \delta \dot{\mathbb{X}}_k \\ \mathbb{X}_{{\rm int},k} \end{bmatrix} + B_{\mathbb{X},{\rm int}} \delta \mathbb{U}_k
\end{align}
Hence, the optimal gains $K_\mathbb{X},$ $K_{\dot{\mathbb{X}}}$ and $K_{\mathbb{X}, {\rm int}}$ are obtained by solving the discrete Algebraic Riccati Equation using $A_{\mathbb{X},{\rm int}},$ $B_{\mathbb{X},{\rm int}},$ $Q_{\mathbb{X},{\rm int}}$ and $R_{\mathbb{X},{\rm int}}.$
The LQR-I controller is augmented with anti-windup (see Fig. \ref{lqr_int_block}).
Note that the integrator is placed right before the output so that the limit established by the anti-windup is placed on the controller output rather than on the integrator state. The output $u_k$ is saturated and can only take a maximum value of $u_{\rm max}$ and a minimum value of $u_{\rm min},$ while the rate at which the integrator ``winds-down" depends on gain $\tau_{\rm int}.$
LQR-I anti-windup can be written as
\begin{equation}
    \delta \mathbb{U}_k = {\rm LQR}_{x,{\rm AW-Int}} (\mathbb{X}_k, \dot{\mathbb{X}}_k, r_{\mathbb{X},k}. r_{\dot{\mathbb{X}},k})
\end{equation}
Note that outputs obtained from traditional LQR control are also subject to output saturation.

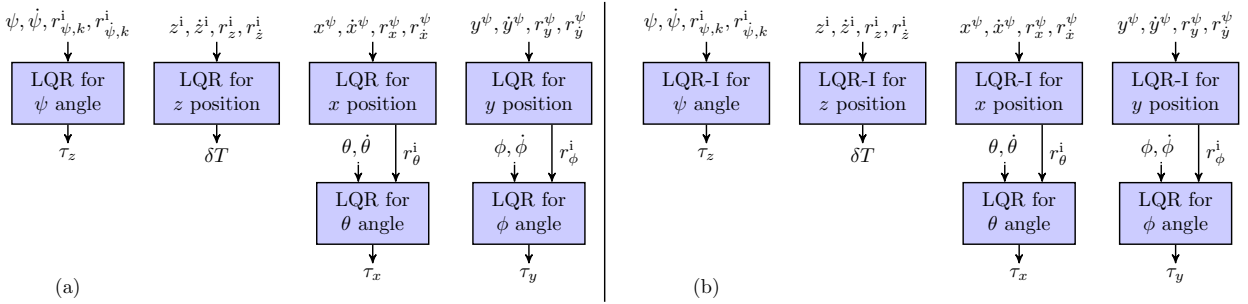
\begin{figure*}[t!]
    \centering
    \resizebox{\textwidth}{!}{%
    \begin{tikzpicture}[>={stealth'}, line width = 0.25mm]

        \node [smallblock,fill=blue!20, minimum height = 0.6cm , minimum width = 0.6cm] (psiLQRBlk){\begin{tabular}{c} LQR for \\ $\psi$ angle \end{tabular}};
        
        \node [smallblock,fill=blue!20, minimum height = 0.6cm , minimum width = 0.6cm, right = 0.5cm of psiLQRBlk] (zLQRBlk){\begin{tabular}{c} LQR for \\ $z$ position \end{tabular}};
        
        \node [smallblock,fill=blue!20, minimum height = 0.6cm , minimum width = 0.6cm, right = 0.5cm of zLQRBlk] (xLQRBlk){\begin{tabular}{c} LQR for \\ $x$ position \end{tabular}};
        
        \node [smallblock,fill=blue!20, minimum height = 0.6cm , minimum width = 0.6cm, right = 0.5cm of xLQRBlk] (yLQRBlk){\begin{tabular}{c} LQR for \\ $y$ position \end{tabular}};
        
        \node [smallblock,fill=blue!20, minimum height = 0.6cm , minimum width = 0.6cm, below = 1cm of xLQRBlk] (thetaLQRBlk){\begin{tabular}{c} LQR for \\ $\theta$ angle \end{tabular}};
        
        \node [smallblock,fill=blue!20, minimum height = 0.6cm , minimum width = 0.6cm, below = 1cm of yLQRBlk] (phiLQRBlk){\begin{tabular}{c} LQR for \\ $\phi$ angle \end{tabular}};
        
        \node [draw = white, above = 0.275cm of psiLQRBlk](psiVar){$\psi,\dot{\psi},r_{\psi,k}^{\rm i}, r_{\dot{\psi},k}^{\rm i}$};
        \node [draw = white, above = 0.275cm of zLQRBlk](zVar){$z^{\rm i},\dot{z}^{\rm i},r_{z}^{\rm i}, r_{\dot{z}}^{\rm i}$};
        \node [draw = white, above = 0.275cm of xLQRBlk](xVar){$x^{\psi},\dot{x}^{\psi},r_{x}^{\psi}, r_{\dot{x}}^{\psi}$};
        \node [draw = white, above = 0.275cm of yLQRBlk](yVar){$y^{\psi},\dot{y}^{\psi},r_{y}^{\psi}, r_{\dot{y}}^{\psi}$};
        
        \node [draw = white, below = 0.275cm of psiLQRBlk](tauZVar){$\tau_z$};
        \node [draw = white, below = 0.275cm of zLQRBlk](dTVar){$\delta T$};
        \node [draw = white, below = 0.275cm of thetaLQRBlk](tauyVar){$\tau_x$};
        \node [draw = white, below = 0.275cm of phiLQRBlk](tauxVar){$\tau_y$};
        
        \draw[->] ([yshift=0.075cm]psiVar.south) -- (psiLQRBlk.north);
        \draw[->] ([yshift=0.075cm]zVar.south) -- (zLQRBlk.north);
        \draw[->] ([yshift=0.075cm]xVar.south) -- (xLQRBlk.north);
        \draw[->] ([yshift=0.075cm]yVar.south) -- (yLQRBlk.north);
        
        \draw[->] ([xshift=0.4cm]xLQRBlk.south) -- node [right]{$r_\theta^{\rm i}$}([xshift=0.4cm]thetaLQRBlk.north);
        \draw[->] ([xshift=0.4cm]yLQRBlk.south) -- node [right]{$r_\phi^{\rm i}$}([xshift=0.4cm]phiLQRBlk.north);
        
        \draw[->] ([xshift=-0.25cm,yshift=0.35cm]thetaLQRBlk.north) -- ([xshift=-0.25cm]thetaLQRBlk.north);
        \draw[->] ([xshift=-0.25cm,yshift=0.35cm]phiLQRBlk.north) -- ([xshift=-0.25cm]phiLQRBlk.north);
        
        \node [draw = white, above left = 0.3cm and -0.1cm of thetaLQRBlk.north](thetaVar){$\theta,\dot{\theta}$};
        
        \node [draw = white, above left = 0.3cm and -0.1cm of phiLQRBlk.north](phiVar){$\phi,\dot{\phi}$};
        
        \draw[->] (psiLQRBlk.south) -- ([yshift=-0.05cm]tauZVar.north);
        \draw[->] (zLQRBlk.south) -- ([yshift=-0.05cm]dTVar.north);
        \draw[->] (thetaLQRBlk.south) -- ([yshift=-0.05cm]tauyVar.north);
        \draw[->] (phiLQRBlk.south) -- ([yshift=-0.05cm]tauxVar.north);
        
        \node [draw = white, below = 3.1cm of psiLQRBlk.center](label1){(a)};
        
        \draw[-] ([xshift=9.325cm ,yshift=1.6cm]psiLQRBlk.center) -- ([xshift=9.325cm,yshift=-3.65cm]psiLQRBlk.center);
        
        \node [smallblock,fill=blue!20, minimum height = 0.6cm , minimum width = 0.6cm, right = 9cm of psiLQRBlk] (psiLQRIBlk){\begin{tabular}{c} LQR-I for \\ $\psi$ angle \end{tabular}};
        
        \node [smallblock,fill=blue!20, minimum height = 0.6cm , minimum width = 0.6cm, right = 0.5cm of psiLQRIBlk] (zLQRIBlk){\begin{tabular}{c} LQR-I for \\ $z$ position \end{tabular}};
        
        \node [smallblock,fill=blue!20, minimum height = 0.6cm , minimum width = 0.6cm, right = 0.5cm of zLQRIBlk] (xLQRIBlk){\begin{tabular}{c} LQR-I for \\ $x$ position \end{tabular}};
        
        \node [smallblock,fill=blue!20, minimum height = 0.6cm , minimum width = 0.6cm, right = 0.5cm of xLQRIBlk] (yLQRIBlk){\begin{tabular}{c} LQR-I for \\ $y$ position \end{tabular}};
        
        \node [smallblock,fill=blue!20, minimum height = 0.6cm , minimum width = 0.6cm, below = 1cm of xLQRIBlk] (thetaLQRIBlk){\begin{tabular}{c} LQR for \\ $\theta$ angle \end{tabular}};
        
        \node [smallblock,fill=blue!20, minimum height = 0.6cm , minimum width = 0.6cm, below = 1cm of yLQRIBlk] (phiLQRIBlk){\begin{tabular}{c} LQR for \\ $\phi$ angle \end{tabular}};
        
        \node [draw = white, above = 0.275cm of psiLQRIBlk](psiIVar){$\psi,\dot{\psi},r_{\psi,k}^{\rm i}, r_{\dot{\psi},k}^{\rm i}$};
        \node [draw = white, above = 0.275cm of zLQRIBlk](zIVar){$z^{\rm i},\dot{z}^{\rm i},r_{z}^{\rm i}, r_{\dot{z}}^{\rm i}$};
        \node [draw = white, above = 0.275cm of xLQRIBlk](xIVar){$x^{\psi},\dot{x}^{\psi},r_{x}^{\psi}, r_{\dot{x}}^{\psi}$};
        \node [draw = white, above = 0.275cm of yLQRIBlk](yIVar){$y^{\psi},\dot{y}^{\psi},r_{y}^{\psi}, r_{\dot{y}}^{\psi}$};
        
        \node [draw = white, below = 0.275cm of psiLQRIBlk](tauZIVar){$\tau_z$};
        \node [draw = white, below = 0.275cm of zLQRIBlk](dTIVar){$\delta T$};
        \node [draw = white, below = 0.275cm of thetaLQRIBlk](tauyIVar){$\tau_x$};
        \node [draw = white, below = 0.275cm of phiLQRIBlk](tauxIVar){$\tau_y$};
        
        \draw[->] ([yshift=0.075cm]psiIVar.south) -- (psiLQRIBlk.north);
        \draw[->] ([yshift=0.075cm]zIVar.south) -- (zLQRIBlk.north);
        \draw[->] ([yshift=0.075cm]xIVar.south) -- (xLQRIBlk.north);
        \draw[->] ([yshift=0.075cm]yIVar.south) -- (yLQRIBlk.north);
        
        \draw[->] ([xshift=0.4cm]xLQRIBlk.south) -- node [right]{$r_\theta^{\rm i}$}([xshift=0.4cm]thetaLQRIBlk.north);
        \draw[->] ([xshift=0.4cm]yLQRIBlk.south) -- node [right]{$r_\phi^{\rm i}$}([xshift=0.4cm]phiLQRIBlk.north);
        
        \draw[->] ([xshift=-0.25cm,yshift=0.35cm]thetaLQRIBlk.north) -- ([xshift=-0.25cm]thetaLQRIBlk.north);
        \draw[->] ([xshift=-0.25cm,yshift=0.35cm]phiLQRIBlk.north) -- ([xshift=-0.25cm]phiLQRIBlk.north);
        
        \node [draw = white, above left = 0.3cm and -0.1cm of thetaLQRIBlk.north](thetaIVar){$\theta,\dot{\theta}$};
        
        \node [draw = white, above left = 0.3cm and -0.1cm of phiLQRIBlk.north](phiIVar){$\phi,\dot{\phi}$};
        
        \draw[->] (psiLQRIBlk.south) -- ([yshift=-0.05cm]tauZIVar.north);
        \draw[->] (zLQRIBlk.south) -- ([yshift=-0.05cm]dTIVar.north);
        \draw[->] (thetaLQRIBlk.south) -- ([yshift=-0.05cm]tauyIVar.north);
        \draw[->] (phiLQRIBlk.south) -- ([yshift=-0.05cm]tauxIVar.north);
        
        \node [draw = white, below = 3.1cm of psiLQRIBlk.center](label2){(b)};
        
    \end{tikzpicture}
	}    
    \caption{Control architecture for the LQR (a) and LQR-I (b) controllers.}
    \label{lqr_blocks}
\end{figure*}

A cascaded LQR/LQR-I controller scheme is employed in the quadcopter. Fig. \ref{lqr_blocks} shows the chosen architecture for these controllers.
Roll and pitch are regulated by LQR controllers since they simplify the design of the cascaded LQR-I controllers and to prevent overshoot inherent to integrators in roll and pitch.
LQR controllers are tuned from the values assigned to $Q$ and $R$ matrices. Usually, only the diagonal terms of these matrices are given nonzero values. Higher values in the diagonal of the $Q$ matrix yield a more aggressive behavior when tracking the states associated with each value, while higher values in the diagonal of the $R$ matrix yield a more conservative behavior when tracking the states associated with the respective input.
Parameters selected for the LQR and LQR-I controllers are summarized in Table \ref{Table:LQRparam}.

\begin{table*}[ht!]
\caption{Parameters of the six LQR ($Q$ = diag([$Q_p, Q_d$])) and LQR-I ($Q$ = diag([$Q_p, Q_d, Q_i$])) Controllers}
\centering
\begin{tabular}{ccccccc|cccccc}
\cmidrule{2-13}
& \multicolumn{6}{c}{\textbf{LQR}}&\multicolumn{6}{c}{\textbf{LQR-I}}\\
\midrule
\textbf{Parameter} & $\bm{x}$ & $\bm{y}$ & $\bm{z}$ & $\bm{\theta}$ & $\bm{\phi}$ & $\bm{\psi}$ & $\bm{x}$ & $\bm{y}$ & $\bm{z}$ & $\bm{\theta}$ & $\bm{\phi}$ & $\bm{\psi}$\\
\midrule
$Q_p$       & 85 & 85 & 2000 & 5.06 & 5.06 & 7& 50 & 50 & 490 & 5 & 5 & 1\\
$Q_d$      & 100 & 100 & 100 & 1.55 & 1.55 & 0.25 & 195 & 195 & 117 & 1.5 & 1.5 & 15\\
$Q_i$       & -- & -- & -- & -- & -- & -- & 25 & 25 & 12.6 & -- & -- & 10\\
$R$       &  515 & 515 & 4 & 4 & 4 & 875 & 1000 & 1000 & 5.5 & 10 & 10 & 1000\\
\bottomrule
\end{tabular}
\label{Table:LQRparam}
\end{table*}

\subsection{PID and PD Controllers} \label{subsec:PID_PD}

The PID controller in continuous-time has the following form:
\begin{equation}
    u(t) = K_p e(t) + K_i\int_{0}^{t}e(\tau)d\tau + K_d\frac{de(t)}{dt}
\end{equation}
where $e$ denotes the tracking error, $u$ is the control signal, $K_p$ , $K_i$ and $K_d$ are the controller gains.
Such a controller is implemented in discrete-time on the quadcopter using a Z-transform formulation:
\begin{equation}
    C({\rm \bf{z}}) = K_p +  K_i \frac{T_s}{{\rm \bf{z}}-1} + K_d \frac{N}{1+NT_s/({\rm \bf{z}}-1)}, \label{Eq:discrete_PID}
\end{equation}
%
%
%
\noindent where typical values of $N$ range from 8 to 20, and {\bf{z}} is the Z-transform variable. 
Note that the derivative term in Eq. \eqref{Eq:discrete_PID} is combined with a low pass filter to mitigate sensitivity of the derivative action to measurement noise \citep{aastrom1995pid}.
Furthermore, note that the PID controller given by Eq. \eqref{Eq:discrete_PID} does not include an anti-windup mechanism, unlike the integral term in the proposed LQR-I scheme. We have chosen such a controller as a benchmark for comparison because PID controllers without anti-windup are commonly offered as commercial-off-the-shelf solutions, e.g., in our case we used the PID function already provided in the Robot Control Library, which consists of a C code implementation of Eq. \eqref{Eq:discrete_PID}.

\begin{figure*}[t!]
    \centering
    \resizebox{\textwidth}{!}{%
    \begin{tikzpicture}[>={stealth'}, line width = 0.25mm]

        \node [smallblock,fill=blue!20, minimum height = 0.6cm , minimum width = 0.6cm] (psiPDBlk){\begin{tabular}{c} PD for \\ $\psi$ angle \end{tabular}};
        
        \node [smallblock,fill=blue!20, minimum height = 0.6cm , minimum width = 0.6cm, right = 0.5cm of psiPDBlk] (zPDBlk){\begin{tabular}{c} PD for \\ $z$ position \end{tabular}};
        
        \node [smallblock,fill=blue!20, minimum height = 0.6cm , minimum width = 0.6cm, right = 0.5cm of zPDBlk] (xPDBlk){\begin{tabular}{c} PD for \\ $x$ position \end{tabular}};
        
        \node [smallblock,fill=blue!20, minimum height = 0.6cm , minimum width = 0.6cm, right = 0.5cm of xPDBlk] (yPDBlk){\begin{tabular}{c} PD for \\ $y$ position \end{tabular}};
        
        \node [smallblock,fill=blue!20, minimum height = 0.6cm , minimum width = 0.6cm, below = 1cm of xPDBlk] (thetaPDBlk){\begin{tabular}{c} PD for \\ $\theta$ angle \end{tabular}};
        
        \node [smallblock,fill=blue!20, minimum height = 0.6cm , minimum width = 0.6cm, below = 1cm of yPDBlk] (phiPDBlk){\begin{tabular}{c} PD for \\ $\phi$ angle \end{tabular}};
        
        \node [draw = white, above = 0.275cm of psiPDBlk](psiVar){$\psi,r_{\psi,k}^{\rm i}$};
        \node [draw = white, above = 0.275cm of zPDBlk](zVar){$z^{\rm i},r_{z}^{\rm i}$};
        \node [draw = white, above = 0.275cm of xPDBlk](xVar){$x^{\psi},r_{x}^{\psi}$};
        \node [draw = white, above = 0.275cm of yPDBlk](yVar){$y^{\psi},r_{y}^{\psi}$};
        
        \node [draw = white, below = 0.275cm of psiPDBlk](tauZVar){$\tau_z$};
        \node [draw = white, below = 0.275cm of zPDBlk](dTVar){$\delta T$};
        \node [draw = white, below = 0.275cm of thetaPDBlk](tauyVar){$\tau_x$};
        \node [draw = white, below = 0.275cm of phiPDBlk](tauxVar){$\tau_y$};
        
        \draw[->] ([yshift=0.075cm]psiVar.south) -- (psiPDBlk.north);
        \draw[->] ([yshift=0.075cm]zVar.south) -- (zPDBlk.north);
        \draw[->] ([yshift=0.075cm]xVar.south) -- (xPDBlk.north);
        \draw[->] ([yshift=0.075cm]yVar.south) -- (yPDBlk.north);
        
        \draw[->] ([xshift=0.4cm]xPDBlk.south) -- node [right]{$r_\theta^{\rm i}$}([xshift=0.4cm]thetaPDBlk.north);
        \draw[->] ([xshift=0.4cm]yPDBlk.south) -- node [right]{$r_\phi^{\rm i}$}([xshift=0.4cm]phiPDBlk.north);
        
        \draw[->] ([xshift=-0.25cm,yshift=0.35cm]thetaPDBlk.north) -- ([xshift=-0.25cm]thetaPDBlk.north);
        \draw[->] ([xshift=-0.25cm,yshift=0.35cm]phiPDBlk.north) -- ([xshift=-0.25cm]phiPDBlk.north);
        
        \node [draw = white, above left = 0.3cm and -0.1cm of thetaPDBlk.north](thetaVar){$\theta$};
        
        \node [draw = white, above left = 0.3cm and -0.1cm of phiPDBlk.north](phiVar){$\phi$};
        
        \draw[->] (psiPDBlk.south) -- ([yshift=-0.05cm]tauZVar.north);
        \draw[->] (zPDBlk.south) -- ([yshift=-0.05cm]dTVar.north);
        \draw[->] (thetaPDBlk.south) -- ([yshift=-0.05cm]tauyVar.north);
        \draw[->] (phiPDBlk.south) -- ([yshift=-0.05cm]tauxVar.north);
        
        \node [draw = white, below = 3.1cm of psiPDBlk.center](label1){(a)};
        
        \draw[-] ([xshift=9.325cm ,yshift=1.6cm]psiPDBlk.center) -- ([xshift=9.325cm,yshift=-3.65cm]psiPDBlk.center);
        
        \node [smallblock,fill=blue!20, minimum height = 0.6cm , minimum width = 0.6cm, right = 9cm of psiPDBlk] (psiPIDBlk){\begin{tabular}{c} PID for \\ $\psi$ angle \end{tabular}};
        
        \node [smallblock,fill=blue!20, minimum height = 0.6cm , minimum width = 0.6cm, right = 0.5cm of psiPIDBlk] (zPIDBlk){\begin{tabular}{c} PID for \\ $z$ position \end{tabular}};
        
        \node [smallblock,fill=blue!20, minimum height = 0.6cm , minimum width = 0.6cm, right = 0.5cm of zPIDBlk] (xPIDBlk){\begin{tabular}{c} PID for \\ $x$ position \end{tabular}};
        
        \node [smallblock,fill=blue!20, minimum height = 0.6cm , minimum width = 0.6cm, right = 0.5cm of xPIDBlk] (yPIDBlk){\begin{tabular}{c} PID for \\ $y$ position \end{tabular}};
        
        \node [smallblock,fill=blue!20, minimum height = 0.6cm , minimum width = 0.6cm, below = 1cm of xPIDBlk] (thetaPIDBlk){\begin{tabular}{c} PD for \\ $\theta$ angle \end{tabular}};
        
        \node [smallblock,fill=blue!20, minimum height = 0.6cm , minimum width = 0.6cm, below = 1cm of yPIDBlk] (phiPIDBlk){\begin{tabular}{c} PD for \\ $\phi$ angle \end{tabular}};
        
        \node [draw = white, above = 0.275cm of psiPIDBlk](psiIVar){$\psi,r_{\psi,k}^{\rm i}$};
        \node [draw = white, above = 0.275cm of zPIDBlk](zIVar){$z^{\rm i},r_{z}^{\rm i}$};
        \node [draw = white, above = 0.275cm of xPIDBlk](xIVar){$x^{\psi},r_{x}^{\psi}$};
        \node [draw = white, above = 0.275cm of yPIDBlk](yIVar){$y^{\psi},r_{y}^{\psi}$};
        
        \node [draw = white, below = 0.275cm of psiPIDBlk](tauZIVar){$\tau_z$};
        \node [draw = white, below = 0.275cm of zPIDBlk](dTIVar){$\delta T$};
        \node [draw = white, below = 0.275cm of thetaPIDBlk](tauyIVar){$\tau_x$};
        \node [draw = white, below = 0.275cm of phiPIDBlk](tauxIVar){$\tau_y$};
        
        \draw[->] ([yshift=0.075cm]psiIVar.south) -- (psiPIDBlk.north);
        \draw[->] ([yshift=0.075cm]zIVar.south) -- (zPIDBlk.north);
        \draw[->] ([yshift=0.075cm]xIVar.south) -- (xPIDBlk.north);
        \draw[->] ([yshift=0.075cm]yIVar.south) -- (yPIDBlk.north);
        
        \draw[->] ([xshift=0.4cm]xPIDBlk.south) -- node [right]{$r_\theta^{\rm i}$}([xshift=0.4cm]thetaPIDBlk.north);
        \draw[->] ([xshift=0.4cm]yPIDBlk.south) -- node [right]{$r_\phi^{\rm i}$}([xshift=0.4cm]phiPIDBlk.north);
        
        \draw[->] ([xshift=-0.25cm,yshift=0.35cm]thetaPIDBlk.north) -- ([xshift=-0.25cm]thetaPIDBlk.north);
        \draw[->] ([xshift=-0.25cm,yshift=0.35cm]phiPIDBlk.north) -- ([xshift=-0.25cm]phiPIDBlk.north);
        
        \node [draw = white, above left = 0.3cm and -0.1cm of thetaPIDBlk.north](thetaIVar){$\theta$};
        
        \node [draw = white, above left = 0.3cm and -0.1cm of phiPIDBlk.north](phiIVar){$\phi$};
        
        \draw[->] (psiPIDBlk.south) -- ([yshift=-0.05cm]tauZIVar.north);
        \draw[->] (zPIDBlk.south) -- ([yshift=-0.05cm]dTIVar.north);
        \draw[->] (thetaPIDBlk.south) -- ([yshift=-0.05cm]tauyIVar.north);
        \draw[->] (phiPIDBlk.south) -- ([yshift=-0.05cm]tauxIVar.north);
        
        \node [draw = white, below = 3.1cm of psiPIDBlk.center](label2){(b)};
        
    \end{tikzpicture}
	}    
    \caption{Control architecture for PD (a) and PID (b) controllers.}
    \label{PID_blocks}
\end{figure*}
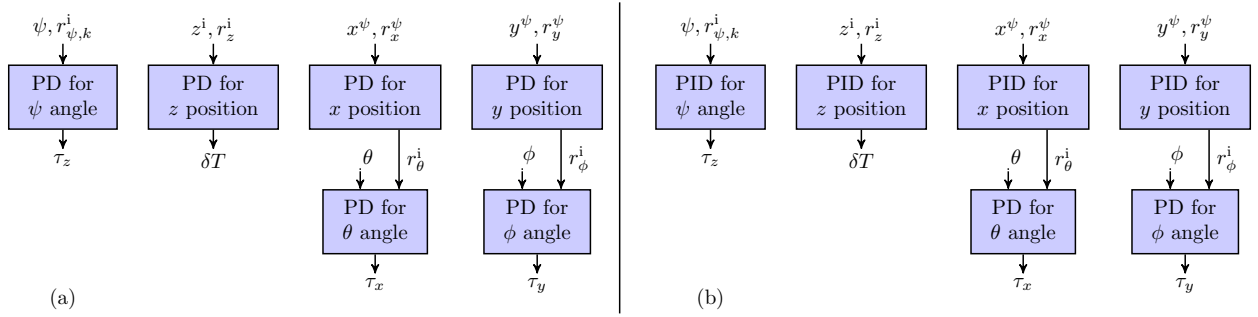

A cascaded PID/PD controller scheme is employed in the quadcopter. Fig. \ref{PID_blocks} shows the chosen architecture for these controllers.
%
%
First, the desired roll, desired pitch, and thrust are determined based on position error.
Then, required torques are calculated based on desired attitude.
Note that the controllers for roll and pitch are PD controllers since they simplify the design of the cascaded PID controllers and prevent undesired overshoot, similar to the cascaded LQR design above.
Parameters for the PD and PID controllers are summarized in Tables \ref{Table:PDparam} and \ref{Table:PIDparam}.
%
%
Note that, while the discrete PD and PID structures are the same in the simulation as in the embedded implementation, the parameters used for the simulations and for the experiments are different since the simulation parameters yielded an unstable behavior during the tests.

\begin{table*}[ht!]
\footnotesize
\caption{Parameters of the Six PD Controllers}
\centering
\begin{tabular}{ccccccc|cccccc}
\cmidrule{2-13}
& \multicolumn{6}{c}{\textbf{Simulation}}&\multicolumn{6}{c}{\textbf{Experiment}}\\
\midrule
\textbf{Parameter} & $\bm{x}$ & $\bm{y}$ & $\bm{z}$ & $\bm{\theta}$ & $\bm{\phi}$ & $\bm{\psi}$ & $\bm{x}$ & $\bm{y}$ & $\bm{z}$ & $\bm{\theta}$ & $\bm{\phi}$ & $\bm{\psi}$\\
\midrule
$K_p$       & 0.6 & 0.6 & 5 & 0.25 & 0.25 & 0.1 & 0.5 & 0.5 & 8 & 0.75 & 0.75 & 1.5\\
$K_d$       & 0.135 & 0.135 & 3.5 & 0.225 & 0.225 & 0.075 & 0.05 & 0.05 & 3.5 & 0.2 & 0.2 & 0.1\\
$N$       & 62.83 & 62.83 & 62.83 & 62.83 & 62.83 & 31.41 & 62.83 & 62.83 & 62.83 & 62.83 & 62.83 & 31.41\\
\bottomrule
\end{tabular}
\label{Table:PDparam}
\end{table*}

\begin{table*}[ht!]
\footnotesize
\caption{Parameters of the six PID Controllers}
\centering
\begin{tabular}{ccccccc|cccccc}
\cmidrule{2-13}
& \multicolumn{6}{c}{\textbf{Simulation}}&\multicolumn{6}{c}{\textbf{Experiment}}\\
\midrule
\textbf{Parameter} & $\bm{x}$ & $\bm{y}$ & $\bm{z}$ & $\bm{\theta}$ & $\bm{\phi}$ & $\bm{\psi}$ & $\bm{x}$ & $\bm{y}$ & $\bm{z}$ & $\bm{\theta}$ & $\bm{\phi}$ & $\bm{\psi}$\\
\midrule
$K_p$       & 0.6 & 0.6 & 5 & 0.25 & 0.25 & 0.1& 0.5 & 0.5 & 5 & 0.75 & 0.75 & 1.5\\
$K_i$      & 0.15 & 0.15 & 1 & -- & -- & 0.025 & 0.15 & 0.15 & 1 & -- & -- & 0.5\\
$K_d$       & 0.135 & 0.135 & 3.5 & 0.225 & 0.225 & 0.075 & 0.05 & 0.05 & 3.5 & 0.2 & 0.2 & 0.1\\
$N$       &  62.83 & 62.83 & 62.83 & 62.83 & 62.83 & 31.41 & 62.83 & 62.83 & 62.83 & 62.83 & 62.83 & 31.41\\
\bottomrule
\end{tabular}
\label{Table:PIDparam}
\end{table*}

\section{Quadcopter Simulation}
\label{section:simulation}

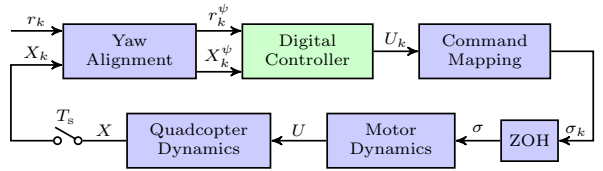
\begin{figure}[t!]
    \centering
    \resizebox{\columnwidth}{!}{%
    \begin{tikzpicture}[>={stealth'}, line width = 0.25mm]
		\def\centerarc[#1](#2)(#3:#4:#5)
		{ \draw[#1] ($(#2)+({#5*cos(#3)},{#5*sin(#3)})$) arc (#3:#4:#5); }
			\node [smallblock, fill=blue!20, minimum height = 0.6cm , minimum width = 0.6cm] (yaw_align) {\scriptsize\begin{tabular}{c} Yaw \\ Alignment \end{tabular}};
			\node [smallblock, right =0.65cm of yaw_align, fill=green!20, minimum height = 0.6cm , minimum width = 0.6cm] (controller) {\scriptsize\begin{tabular}{c} Digital \\ Controller \end{tabular}};
			\node [input, name=ref, above left = 0.3cm and 0.75cm of yaw_align.west]{};
			\node [smallblock, right =0.65cm of controller, minimum width = 0.6cm] (comm_map) {\scriptsize \begin{tabular}{c} Command \\ Mapping \end{tabular}};
			\node [smallblock, below right = 0.9cm and -0.05cm of yaw_align.center, fill=blue!20, minimum height = 0.6cm , minimum width = 0.6cm] (quadBlk) {\scriptsize\begin{tabular}{c} Quadcopter \\ Dynamics \end{tabular}};
			\node [smallblock, right = 0.8cm of quadBlk, fill=blue!20, minimum height = 0.6cm , minimum width = 0.6cm] (motorBlk) {\scriptsize\begin{tabular}{c} Motor \\ Dynamics \end{tabular}};
			\node [smallblock, right = 0.65cm of motorBlk, minimum height = 0.6cm , minimum width = 0.5cm] (DA) {\scriptsize ZOH};
			
			\node[circle,draw=black, fill=white, inner sep=0pt,minimum size=3pt] (rc11) at ([xshift=-1cm]quadBlk.west) {};
			\node[circle,draw=black, fill=white, inner sep=0pt,minimum size=3pt] (rc21) at ([xshift=-0.7cm]quadBlk.west) {};
			\draw [-] ([yshift=.15cm, xshift=-.08cm]rc11.north east) --node[below,yshift=.475cm]{\scriptsize$T_{\rm s}$} ([xshift=.225cm, yshift=-0.025cm]rc11.north east) {};
			
			\draw [->] (controller.east) -- node[above, yshift=-0.06cm]{\scriptsize$U_k$} (comm_map.west);
			\draw [->] (comm_map.east) -| ([xshift=0.5cm]DA.east) -- node[above, yshift=-0.06cm]{\scriptsize$\sigma_k$} (DA.east);
			\draw [->] (DA.west) -- node[above]{\scriptsize$\sigma$} (motorBlk.east);
			\draw [->] (motorBlk.west) -- node[above, yshift=-0.06cm]{\scriptsize$U$} (quadBlk.east);
			\draw [-] (quadBlk.west) -- node[above, yshift=-0.06cm]{\scriptsize$X$} (rc21);
			\draw [->] (rc11) -| ([xshift=-0.75cm,yshift=-0.2cm]yaw_align.west) -- node[above, yshift=-0.06cm]{\scriptsize$X_k$} ([yshift =-0.2cm]yaw_align.west);
			\draw [->] (ref.center) -- node[above, yshift=-0.06cm]{\scriptsize$r_k$} ([yshift = 0.3cm]yaw_align.west);
			\draw [->] ([yshift = 0.3cm]yaw_align.east) -- node[above, yshift=-0.06cm]{\scriptsize$r_k^{\psi}$} ([yshift = 0.3cm]controller.west);
			\draw [->] ([yshift = -0.3cm]yaw_align.east) -- node[above, yshift=-0.06cm]{\scriptsize$X_k^{\psi}$} ([yshift = -0.3cm]controller.west);
			
	\end{tikzpicture}
    }
    \caption{Simulink Model Block Diagram. The blocks are arranged such that the dynamics models run in continuous time while the controller does so at given time intervals.}
    \label{simulink_block}
\end{figure}

The block diagram of a closed loop simulation model implemented in Simulink is shown in Fig. \ref{simulink_block}.
The Motor Dynamics block simulates the dynamics of each of the four motors/propulsors used by the quadcopter, represented by Eqs. \eqref{sigma2WssEq} - \eqref{eqMixMatrix}. The inputs of this block are the PWM signals $\sigma$ and its output is the quadcopter dynamic system input vector $U.$
The Quadcopter Dynamics block simulates the dynamics represented by Eqs. \eqref{eqSS1} - \eqref{eqSS3}. The input of this block is the input vector $U$ and its output is the current state vector $X.$
The Yaw Alignment block transforms the reference setpoint vector $r_k$ and the state vector $X_k$ into $r_k^{\psi}$ and $X_k^{\psi},$ as stated in \eqref{RefAlignEq}.
The Digital Controller block contains the inner and outer loop controllers per Section \ref{subsec:DCS} and yields the discrete quadcopter system vector input vector $U_k.$ Finally, the Command Mapping block obtains $\sigma_k$ from $U_k$ per Eqs. \eqref{eqInvMixMatrix} and \eqref{omega2sigma}. The ZOH block holds $\sigma_k$ for $T_{\rm s}$ seconds, thus converting this discrete signal into the continuous signal $\sigma_k,$ as described in Eq. \eqref{sigmaCT}.

\begin{figure}[h!]
    \centering
    \includegraphics[width=\columnwidth]{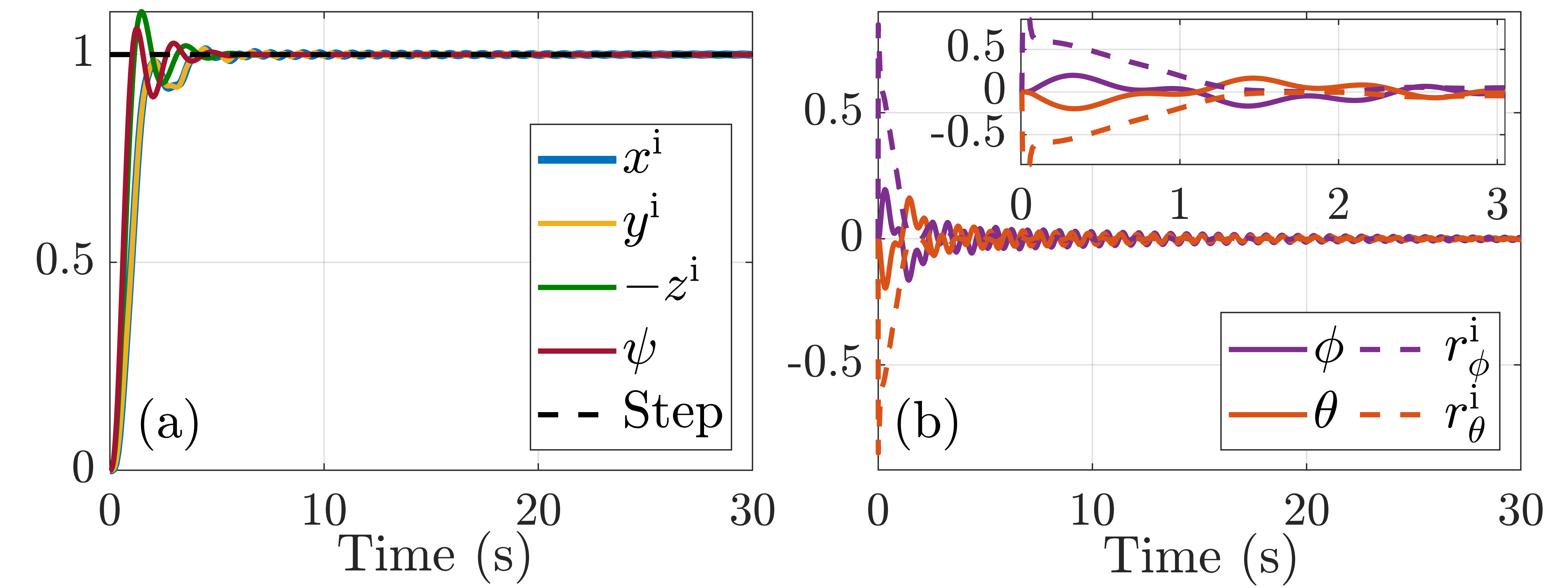}
    \caption{Simulation results for PD controller. (a) shows the response to a unit step in the position setpoint in the x-axis direction, in the y-axis direction, in the negative z-axis direction and in the positive direction of $\psi.$ (b) shows the setpoint signals and trajectories of $\phi$ and $\theta$ in response to a unit step in the position setpoint in the y-axis direction and in the x-axis direction, respectively.}
    \label{fig:Sim_Res_PD}
\end{figure}

\begin{figure}[h!]
    \centering
    \includegraphics[width=\columnwidth]{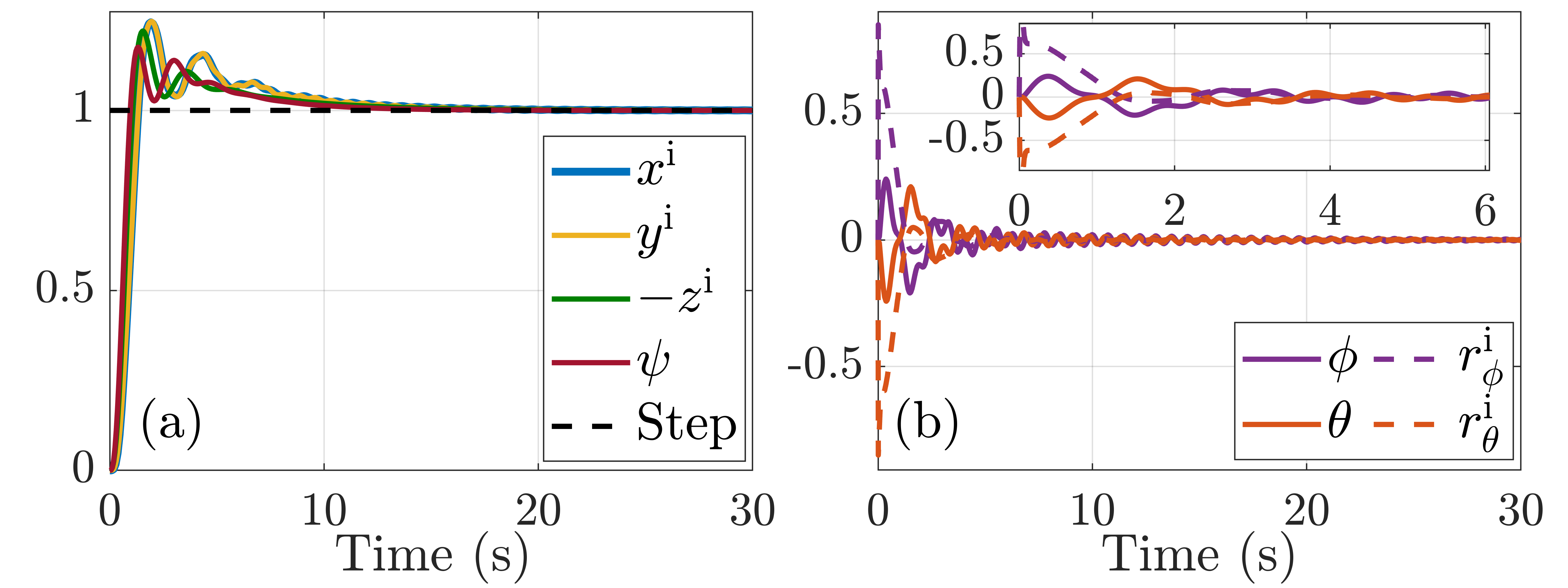}
    \caption{Simulation results for PID controller. (a) shows the response to a unit step in the position setpoint in the x-axis direction, in the y-axis direction, in the negative z-axis direction and in the positive direction of $\psi.$ (b) shows the setpoint signals and trajectories of $\phi$ and $\theta$ in response to a unit step in the position setpoint in the y-axis direction and in the x-axis direction, respectively.}
    \label{fig:Sim_Res_PID}
\end{figure}

\begin{figure}[t!]
    \centering
    \includegraphics[width=\columnwidth]{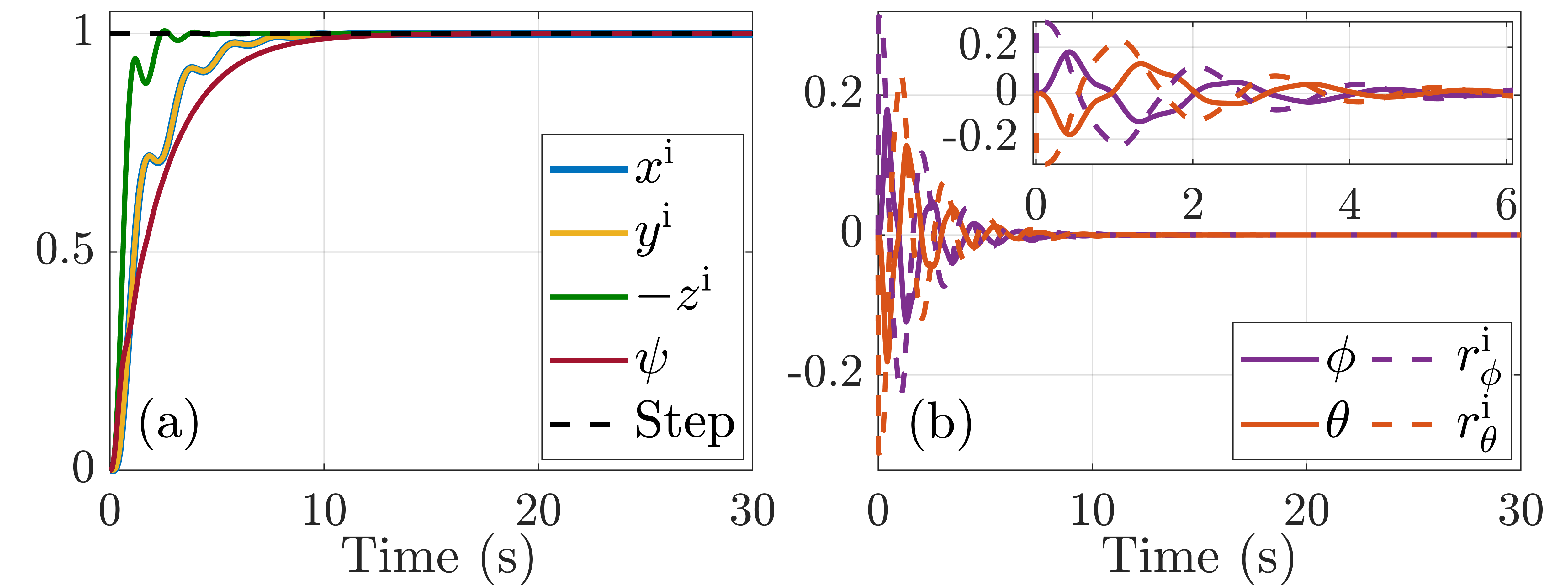}
    \caption{Simulation results for LQR controller. (a) shows the response to a unit step in the position setpoint in the x-axis direction, in the y-axis direction, in the negative z-axis direction and in the positive direction of $\psi.$ (b) shows the setpoint signals and trajectories of $\phi$ and $\theta$ in response to a unit step in the position setpoint in the y-axis direction and in the x-axis direction, respectively.}
    \label{fig:Sim_Res_LQR}
\end{figure}

\begin{figure}[t!]
    \centering
    \includegraphics[width=\columnwidth]{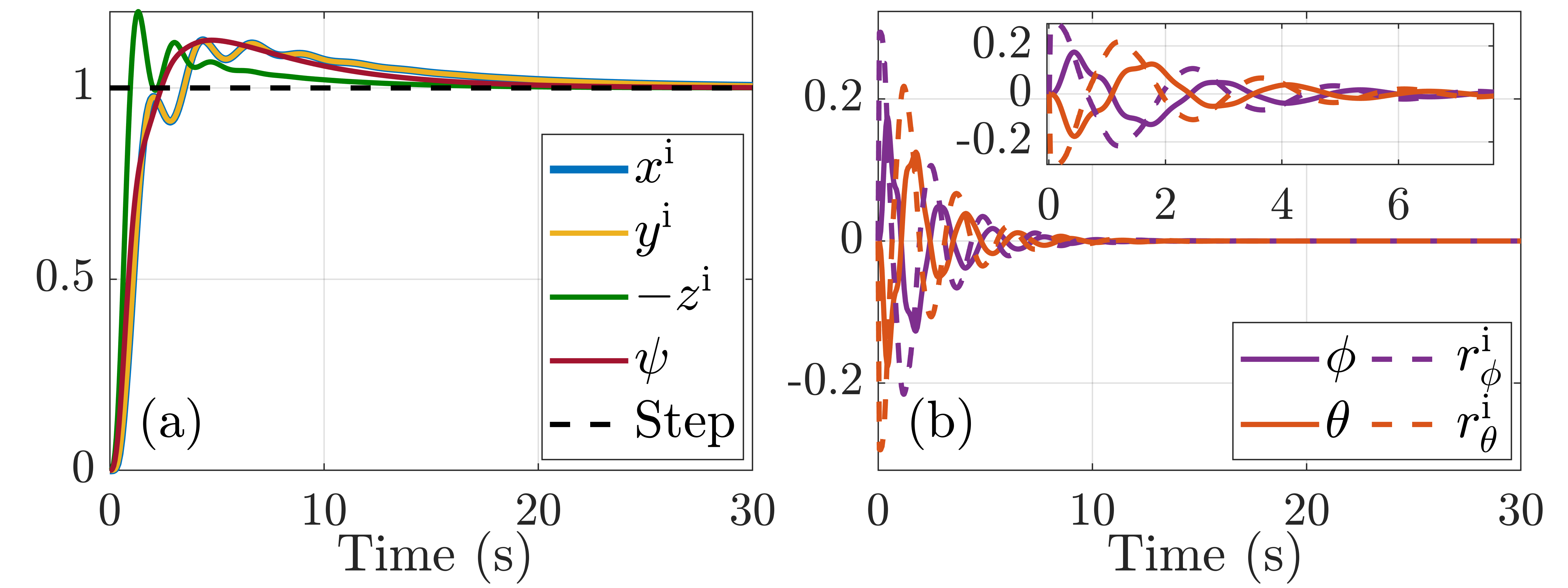}
    \caption{Simulation results for LQR-I controller. (a) shows the response to a unit step in the position setpoint in the x-axis direction, in the y-axis direction, in the negative z-axis direction and in the positive direction of $\psi.$ (b) shows the setpoint signals and trajectories of $\phi$ and $\theta$ in response to a unit step in the position setpoint in the y-axis direction and in the x-axis direction, respectively.}
    \label{fig:Sim_Res_LQR_I}
\end{figure}

\begin{figure}[t!]
    \centering
    \includegraphics[width=\columnwidth]{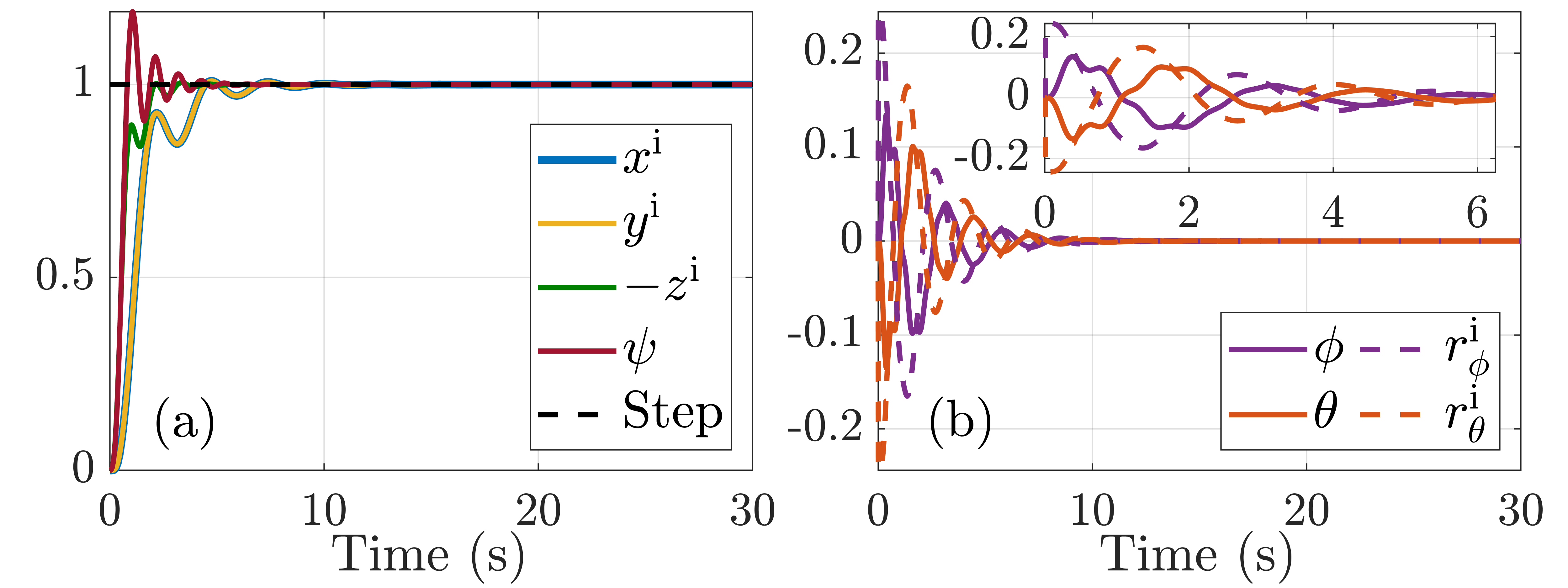}
    \caption{Simulation results for MPC controller. (a) shows the response to a unit step in the position setpoint in the x-axis direction, in the y-axis direction, in the negative z-axis direction and in the yaw angle in the positive direction of $\psi.$ (b) shows the setpoint signals and trajectories of $\phi$ and $\theta$ in response to a unit step in the position setpoint in the y-axis direction and in the x-axis direction, respectively.}
    \label{fig:Sim_Res_MPC}
\end{figure}

Figs. \ref{fig:Sim_Res_PD}, \ref{fig:Sim_Res_PID}, \ref{fig:Sim_Res_LQR}, \ref{fig:Sim_Res_LQR_I}, and \ref{fig:Sim_Res_MPC} show simulation results using step reference setpoint signals for $r_{x}^{\rm i},$ $r_{y}^{\rm i},$ $r_{z}^{\rm i}$ and $r_{\psi}^{\rm i}$ for the PD, PID, LQR, LQR-I and MPC controllers, respectively. In the next section, simulation results using an inclined, circular trajectory are compared against experimental results to evaluate performance.

\section{Experimental Setup} 
\label{section:setup}

\begin{figure*}[t!]
    \centering
    \resizebox{0.85\textwidth}{!}{%
        \begin{tikzpicture}[>={stealth'}, line width = 0.25mm]
        \node [coordinate](centerPic){};
        
        \node [bigblock, rounded corners, fill=yellow!20, right = -1.4 em of centerPic, minimum height=26em, minimum width=32.5em, yshift = -0.5em] (bGField) {};
        \node[below, right] at ([xshift = 0.1cm, yshift = -0.35cm]bGField.north west){M-Air Outdoors Netted Testing Field};
        
        \node [smallblock, rounded corners,fill=white, label = {\scriptsize \begin{tabular}{c} MOCAP \\ camera \end{tabular}}, minimum height=5.75em, minimum width=5.75em, above left = 2.5 em and 0.5 em of centerPic, xscale=-1] (cam1) {\centering \includegraphics[width=5em]{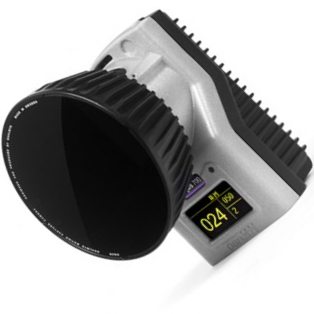}};
        \node[above, left] at ([xshift = 0.01cm, yshift = 0.2cm]cam1.south west){\scriptsize 60 Hz};
        \node [smallblock, rounded corners,fill=white, label = {\scriptsize \begin{tabular}{c} MOCAP \\ camera \end{tabular}}, minimum height=5.75em, minimum width=5.75em, right = 25 em of cam1] (cam2){\centering \includegraphics[width=5em]{Figures/Section_6/qualisys_45.png}};
        \node[above, left] at ([xshift = 0.01cm, yshift = 0.2cm]cam2.south east){\scriptsize 60 Hz};
        \node[smallblock, rounded corners,fill=white, label = {\scriptsize \begin{tabular}{c} MOCAP \\ cameras \end{tabular}}, minimum height=5.75em, minimum width=14em] at ([xshift = -12.4em]cam2.center){};]
        \node [fill=none, right = 14.1 em of cam1, xscale=-1] (cam31){\centering \includegraphics[width=5em]{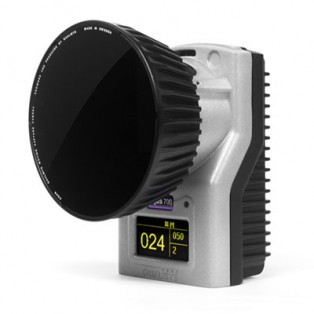}};
        \node [fill=none, left = 2.25 em of cam2] (cam32){\centering \includegraphics[width=5em]{Figures/Section_6/qualisys.png}};
         \node[above, left] at ([xshift = -0.05cm, yshift = 0.15cm]cam32.south east){\scriptsize 60 Hz};
        
        \node [smallblock, rounded corners,fill=white, label = below:{\scriptsize Quadcopter System}, minimum height=9.5em, minimum width=21em] at ([xshift = 11 em, yshift = -7.05 em]centerPic){};
         \node [fill = none, below right = 2.5 em and 8.75 em of centerPic] (quad){\centering \includegraphics[width=12em]{Figures/Section_2/quadcopter.jpg}};
         \node[above, left] at ([xshift = -0.01cm, yshift = 0.15cm]quad.south east){\scriptsize 200 Hz};

          \node [smallblock,rounded corners, fill = green!30, left = 7 em of quad.center, minimum width = 6.7em, minimum height=2em] (RC_Rec){\scriptsize WiFi Transceiver};
          \node [smallblock, rounded corners,fill = green!30, above = 1.75 em of RC_Rec.center, minimum width = 6.7em, minimum height=2em] (WiFi_Trans){\scriptsize RC Receiver};
          \node [smallblock, rounded corners,fill = green!30, below = 1.75 em of RC_Rec.center, minimum width = 6.7em, minimum height=2em] (XbeeRec){\scriptsize  Xbee Receiver};

           \node [fill = none, below right = -2.825 em and 8 em of quad.center] (WVane){\centering \includegraphics[width=7em]{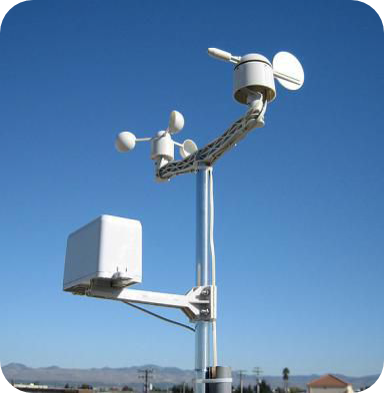}};
           \node [smallblock, rounded corners=8pt,fill=none, label = below:{\scriptsize Wind Vane}, minimum height=7.2em, minimum width=7.1em] at (WVane.center){};
           \node[fill = white,rounded corners=2pt,inner sep = 0.05cm,above, left] at ([xshift = -0.2cm, yshift = 0.4cm]WVane.south east){\scriptsize 10 Hz};

        \node [bigblock, rounded corners, fill=yellow!20, left = 8.35 em of RC_Rec.center, minimum height=18em, minimum width=11.5em, yshift = 2.5em] (bGControlPav) {};
        \node[below, right] at ([xshift = 0.1cm, yshift = -0.35cm]bGControlPav.north west){M-Air Control Pavilion};

        \node [smallblock, rounded corners,fill = green!30, left = 9.25 em of RC_Rec.center, minimum width = 9.5em, minimum height=2em] (RC_Ctrl){\scriptsize Ground Control Station};
        \node [smallblock, rounded corners,fill = green!30, above = 1.75 em of RC_Ctrl.center, minimum width = 9.5em, minimum height=2em] (WiFi_Rec){\scriptsize RC Controller};
          \node [smallblock, rounded corners, fill = green!30, below = 2.3 em of RC_Ctrl.center, minimum width = 9.5em, minimum height=2em] (XbeeTrans){\scriptsize \begin{tabular}{c} Data Trasmission and \\ Acquisition Module \\ (BBG) \end{tabular}};

        \node [smallblock, rounded corners,fill = red!30, above = 2.5 em of WiFi_Rec.center, minimum width = 9.5em, minimum height=4em] (MOCAP_sys){\scriptsize \begin{tabular}{c} MOCAP System \\ Processing and  \\ Configuration Module \end{tabular}};
        
        \draw[<->, black!15!red] (WiFi_Rec.east) -- (WiFi_Trans.west);
        \draw[<->, black!15!red] (RC_Ctrl.east) -- (RC_Rec.west);
        \draw[<-, black!15!red] (XbeeRec.west) -| ([xshift = 2em, yshift = 0.7em]XbeeTrans.east) -- ([yshift = 0.7em]XbeeTrans.east);
        \draw[<-, black!15!red] ([yshift = -0.7em]XbeeTrans.east) -| ([xshift = 2 em, yshift = -3.5em]XbeeTrans.east) -| ([xshift = -2.5 em, yshift = 0.3em]WVane.south);
        \draw[->, black!15!red] (MOCAP_sys.west)-|([xshift=-0.7em]XbeeTrans.west)--(XbeeTrans.west);
        \draw[<->,black!15!red](RC_Ctrl.south)--(XbeeTrans.north);
        \draw[<-, black!15!red] ([yshift = 1.5em]MOCAP_sys.east)-|(cam1.south);
        \draw[<-, black!15!red] ([yshift = -1.5em]MOCAP_sys.east)-|(cam2.south);
        \draw[<-, black!15!red] (MOCAP_sys.east)-|([xshift = 12.625em]cam1.south);
        \draw[<-, black!15!red] ([yshift = 0.75em]MOCAP_sys.east)-|([xshift = 11.875em]cam1.south);
        \draw[<-, black!15!red] ([yshift = -0.75em]MOCAP_sys.east)-|([xshift = 13.375em]cam1.south);
        \node[circle, fill=black, inner sep = 1pt, minimum size = 0.5 em] at ([xshift = 12.625em, yshift = 3em]cam1.south){};
        \node[circle, fill=black, inner sep = 1pt, minimum size = 0.5 em] at ([xshift = 11.875em, yshift = 3em]cam1.south){};
        \node[circle, fill=black, inner sep = 1pt, minimum size = 0.5 em] at ([xshift = 13.375em, yshift = 3em]cam1.south){};
        
          \def\x{1.0}
          \def\y{4.0}
          \def\R{\x+0.005}
          \def\yc{\y+0.04}
          \def\e{0.4}

        \begin{scope}[xshift = 4.1 em, yshift = 4.5 em]
        \begin{scope}[rotate=-135, xscale=0.6, yscale=0.6]
        \shade[left color=white,right color=blue,opacity=0.4]
          (-\x,\yc) -- (-\x,\yc) arc (-180:-360:{\R} and \e) -- (\x,\yc) -- (0,0) -- cycle;
        \end{scope}
        \end{scope}

        \begin{scope}[xshift = 25.9 em, yshift = 4.5 em]
        \begin{scope}[rotate=135, xscale=0.6, yscale=0.6]
        \shade[right color=white,left color=blue,opacity=0.4]
          (-\x,\yc) -- (-\x,\yc) arc (-180:-360:{\R} and \e) -- (\x,\yc) -- (0,0) -- cycle;
        \end{scope}
        \end{scope}
        \end{tikzpicture}
    }
    \caption{Quadcopter Flight Test Architecture.}
    \label{Fig:experimental_setup_diagram}
\end{figure*}

The quadcopter outdoor flight test architecture is shown in Fig. \ref{Fig:experimental_setup_diagram}.
The attitude and position information of the quadcopter is obtained from the MOCAP system.
This MOCAP data is serially transmitted to a Beaglebone Green (BBG), which then wirelessly relays it to the quadcopter using an XBee radio modem.
The MOCAP data is then fused by the Kalman Filter (described in Section \ref{section:state_estimation}) to produce a state estimate.

%
To mitigate the impact of short-term MOCAP data dropouts (due to varying lighting conditions, changes in ambient temperature, etc.), the Kalman Filter provided state estimates based solely on onboard sensor data.
Also, a digital wind vane was used to measure wind during outdoor flight tests.

\begin{figure}[t!]
	\centering 
        \includegraphics[width=\columnwidth]{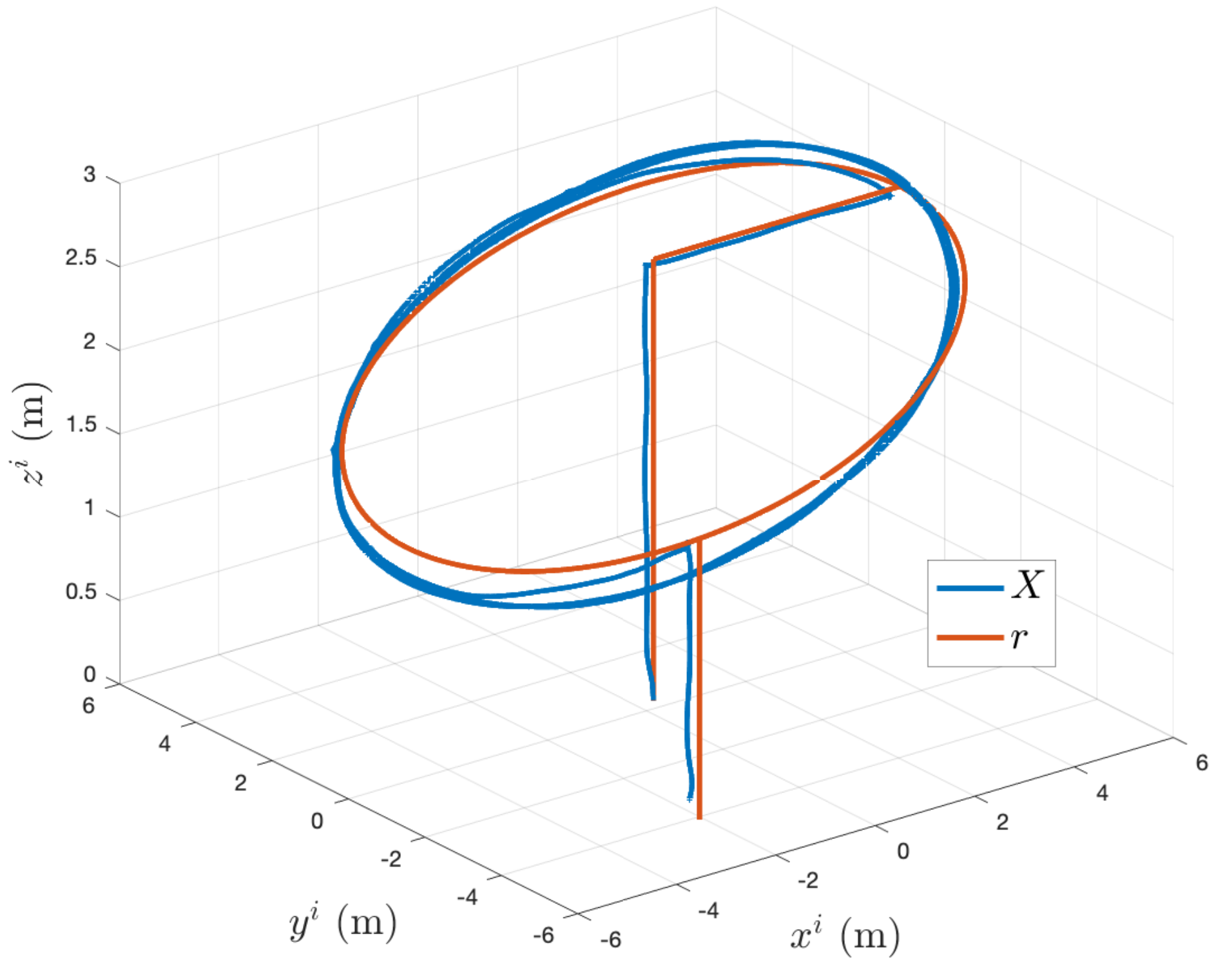}
        \caption{3D plot illustrating the inclined, circular reference trajectory and tracking results of the quadcopter LQR controller. For ease of viewing, the coordinates are North-West-Up in this plot.}
	\label{Fig:3d_position}
\end{figure}

In the experiments reported in this paper, the quadcopter tracked an inclined, circular trajectory while continuously changing its heading angle, as shown in Fig. \ref{Fig:3d_position}.
For the circular section of the trajectory, the quadcopter followed a trapezoidal velocity profile.
For the straight sections, it followed a cubic velocity profile.
All controllers were flown at tangential velocities of $1,$ $2,$ $3,$ and $4$ m/s (three times each) along a circle of radius $4.5$ m, inclined at an angle of $-7.5$ deg along the y-axis.  

\section{Results}
\label{section:results}
In this section, step responses, position and velocity tracking errors of an inclined circle, and control effort plots are presented. Each controller was manually tuned to achieve the best performance. 

\begin{figure}[t!]
    \centering
    \includegraphics[width=\columnwidth]{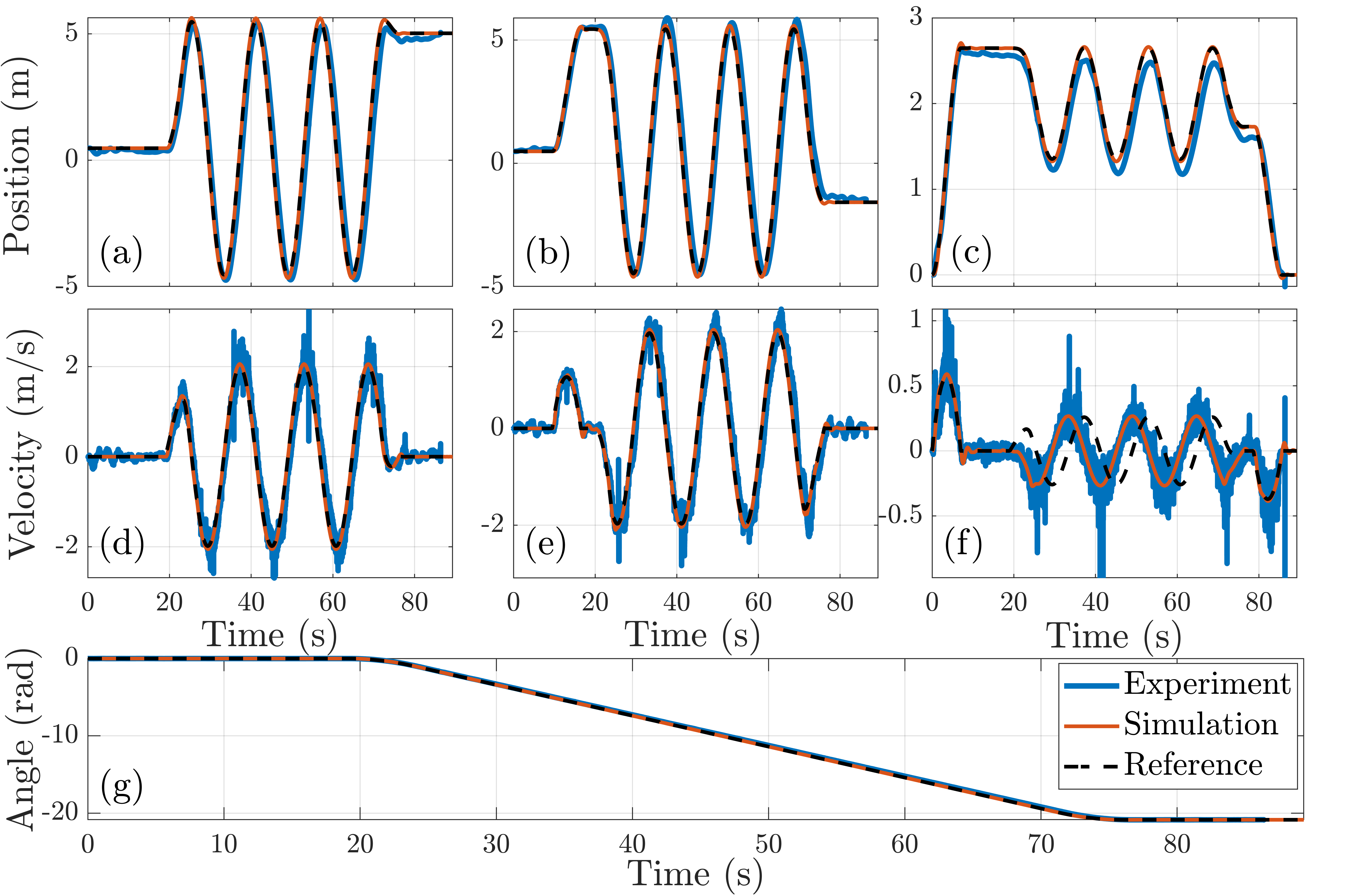}
    \caption{Experimental results (exp subscript) versus simulation results (sim subscript) for PD controller for a tangential velocity of 2 m/s. (a) x-axis position. (b) y-axis position. (c) $-$z-axis position. (d) x-axis velocity. (e) y-axis velocity. (f) $-$z-axis velocity. (g) Yaw.}
    \label{fig:Exp_Res_PD}
\end{figure}
\begin{figure}[t!]
    \centering
    \includegraphics[width=\columnwidth]{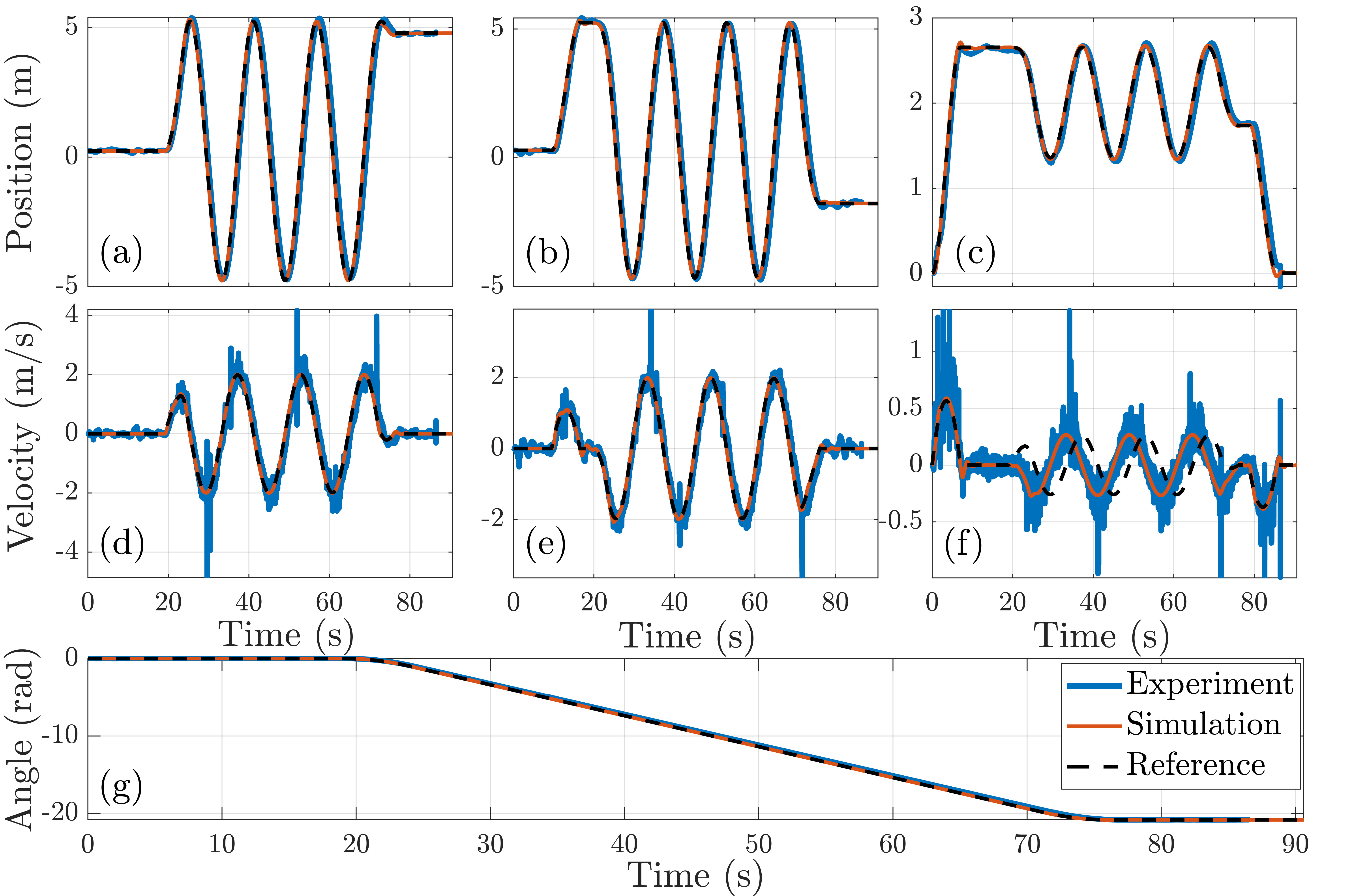}
    \caption{Experimental results (exp subscript) versus simulation results (sim subscript) for PID controller for a tangential velocity of 2 m/s. (a) x-axis position. (b) y-axis position. (c) $-$z-axis position. (d) x-axis velocity. (e) y-axis velocity. (f) $-$z-axis velocity. (g) Yaw.}
    \label{fig:Exp_Res_PID}
\end{figure}
\begin{figure}[t!]
    \centering
    \includegraphics[width=\columnwidth]{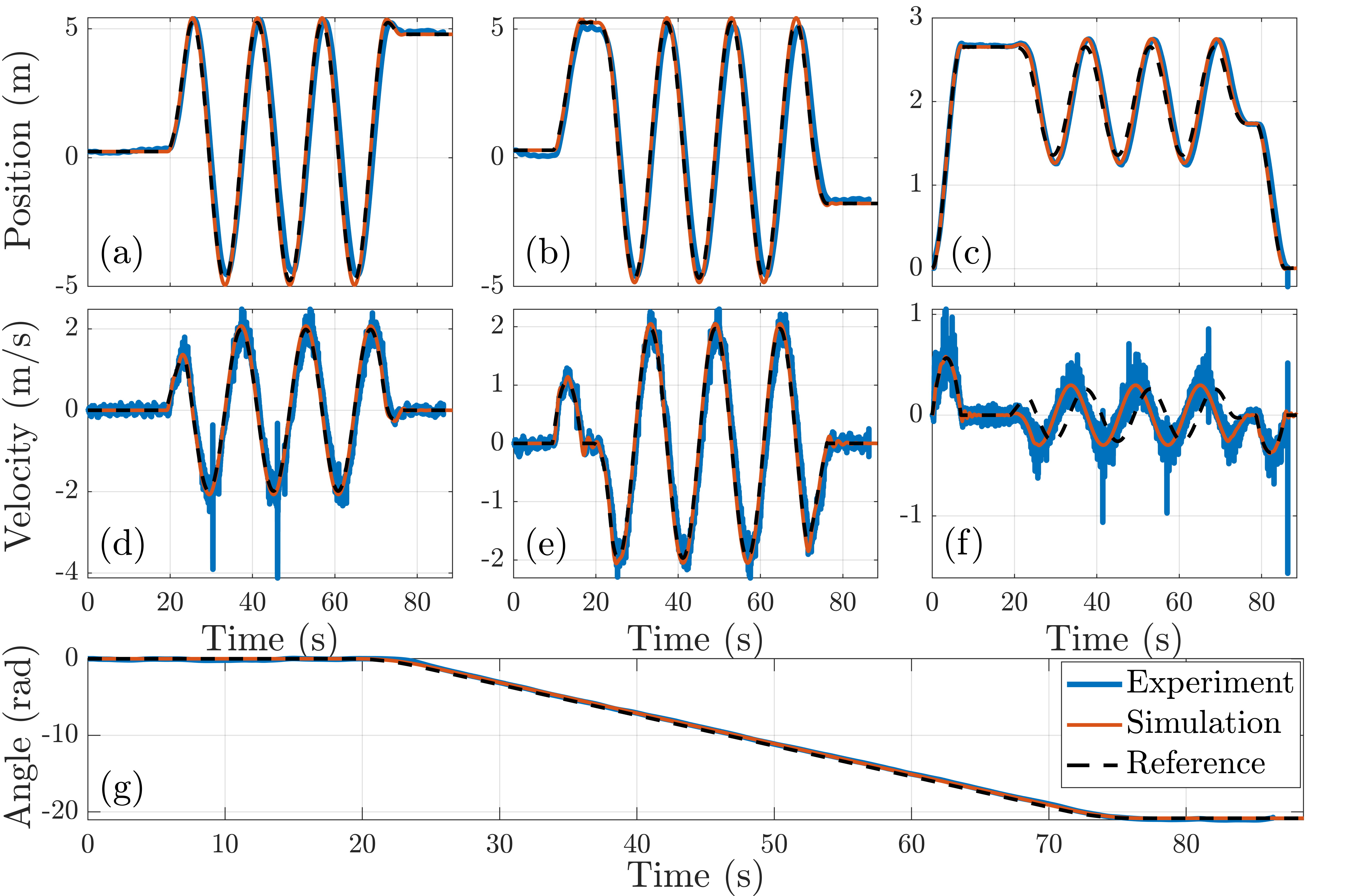}
    \caption{Experimental results (exp subscript) versus simulation results (sim subscript) for LQR controller for a tangential velocity of 2 m/s. (a) x-axis position. (b) y-axis position. (c) $-$z-axis position. (d) x-axis velocity. (e) y-axis velocity. (f) $-$z-axis velocity. (g) Yaw.}
    \label{fig:Exp_Res_LQR}
\end{figure}
\begin{figure}[t!]
    \centering
    \includegraphics[width=\columnwidth]{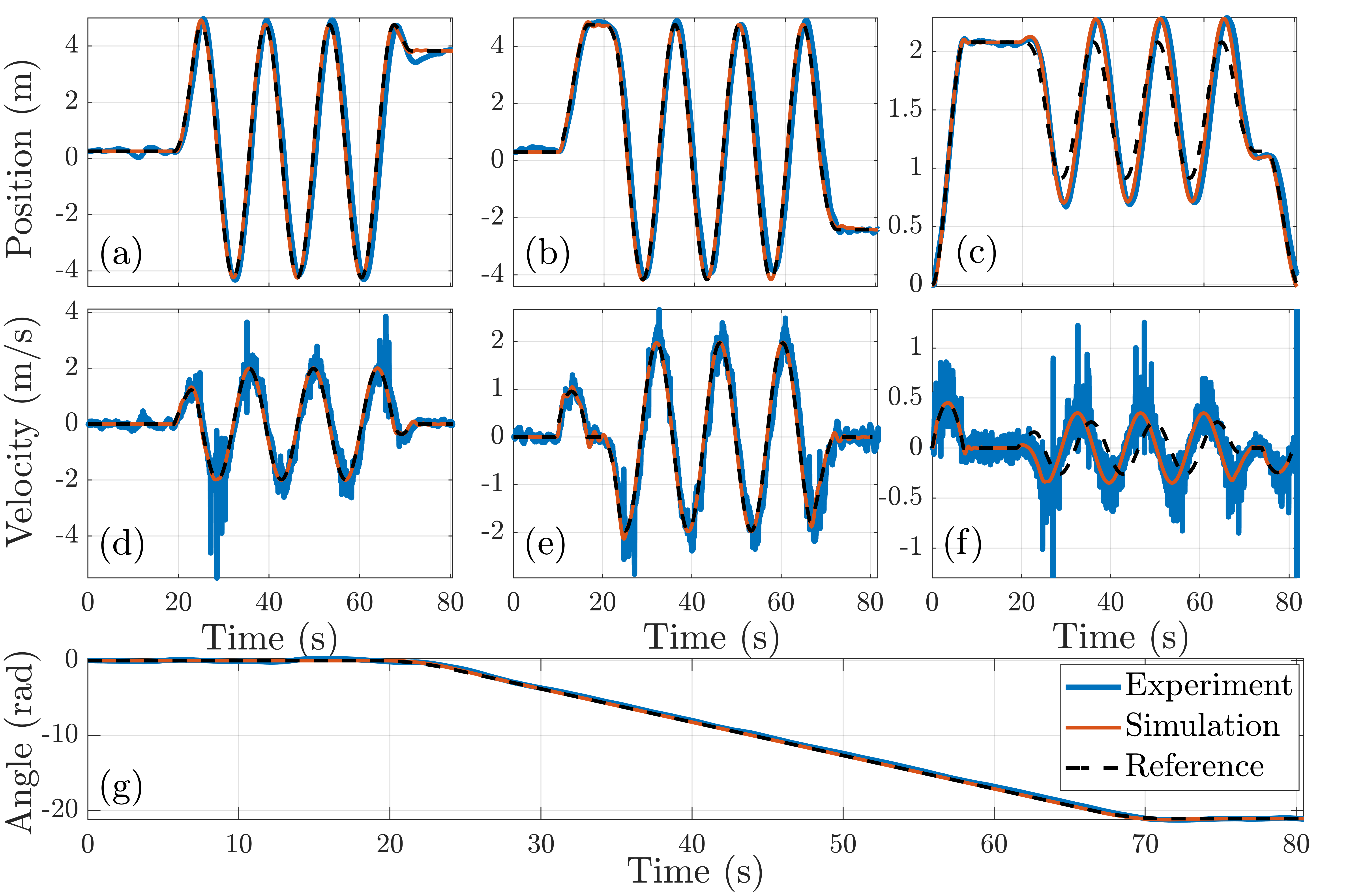}
    \caption{Experimental results (exp subscript) versus simulation results (sim subscript) for LQR-I controller for a tangential velocity of 2 m/s. (a) x-axis position. (b) y-axis position. (c) $-$z-axis position. (d) x-axis velocity. (e) y-axis velocity. (f) $-$z-axis velocity. (g) Yaw.}
    \label{fig:Exp_Res_LQR_I}
\end{figure}
\begin{figure}[t!]
    \centering
    \includegraphics[width=\columnwidth]{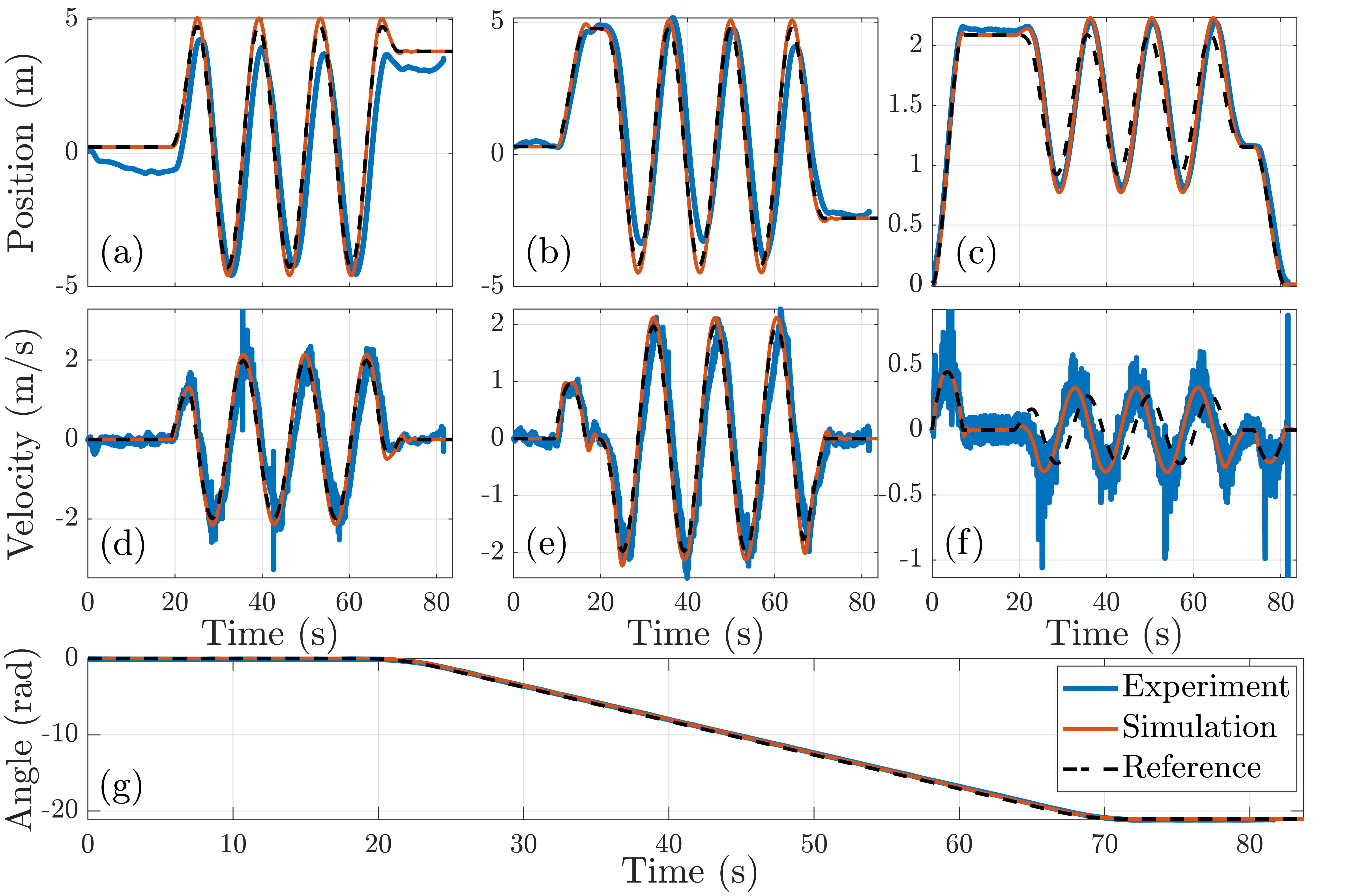}
    \caption{Experimental results (exp subscript) versus simulation results (sim subscript) for MPC controller for a tangential velocity of 2 m/s. (a) x-axis position. (b) y-axis position. (c) $-$z-axis position. (d) x-axis velocity. (e) y-axis velocity. (f) $-$z-axis velocity. (g) Yaw.}
    \label{fig:Exp_Res_MPC}
\end{figure}

Figs. \ref{fig:Exp_Res_PD} -- \ref{fig:Exp_Res_MPC} compare simulation and experimental results for a tangential velocity of 2 m/s.
Note that the noise in the velocity estimates from the experimental data was due to the Kalman Filter propagation of noisy sensor measurements.

%
Experimental results from inclined, circular trajectory tracking in M-Air over all the tangential velocities are summarized in Table \ref{Table:results}.  It is apparent that integral action reduces of maximum radial and altitude error.
%
%
Negligible wind was present.

\begin{table}[hbt!]
\caption{Position tracking performance of quadcopter controllers over all experiments}
\centering
\resizebox{\columnwidth}{!}{%
\begin{tabular}{lccc}
\toprule
\multirow{2}{*}{\textbf{Controller}} & \multirow{2}{*}{\begin{tabular}{c} \textbf{Avg. Tracking} \\ \textbf{Error (m)}\end{tabular}} & \multicolumn{2}{c}{\textbf{Max Tracking Error}} \\ \cmidrule{3-4}
 &  & \textbf{Radial(m)}  & \textbf{Altitude(m)}  \\\midrule
 PD& 0.5373 & 1.408 & 0.4507  \\
 PID& 0.1722 & 0.53462 & 0.2348 \\
 LQR& 0.4450 & 1.1852 & 0.3854  \\
 LQR-I& 0.3470 & 1.027 & 0.628 \\
 E-MPC& 0.9292 & 1.7404 & 0.708 \\
\bottomrule
\label{Table:results}
\end{tabular}
}
\end{table}

\begin{figure}[h!]
\begin{subfigure}{\columnwidth}
\includegraphics[width=\linewidth,height = 7cm]{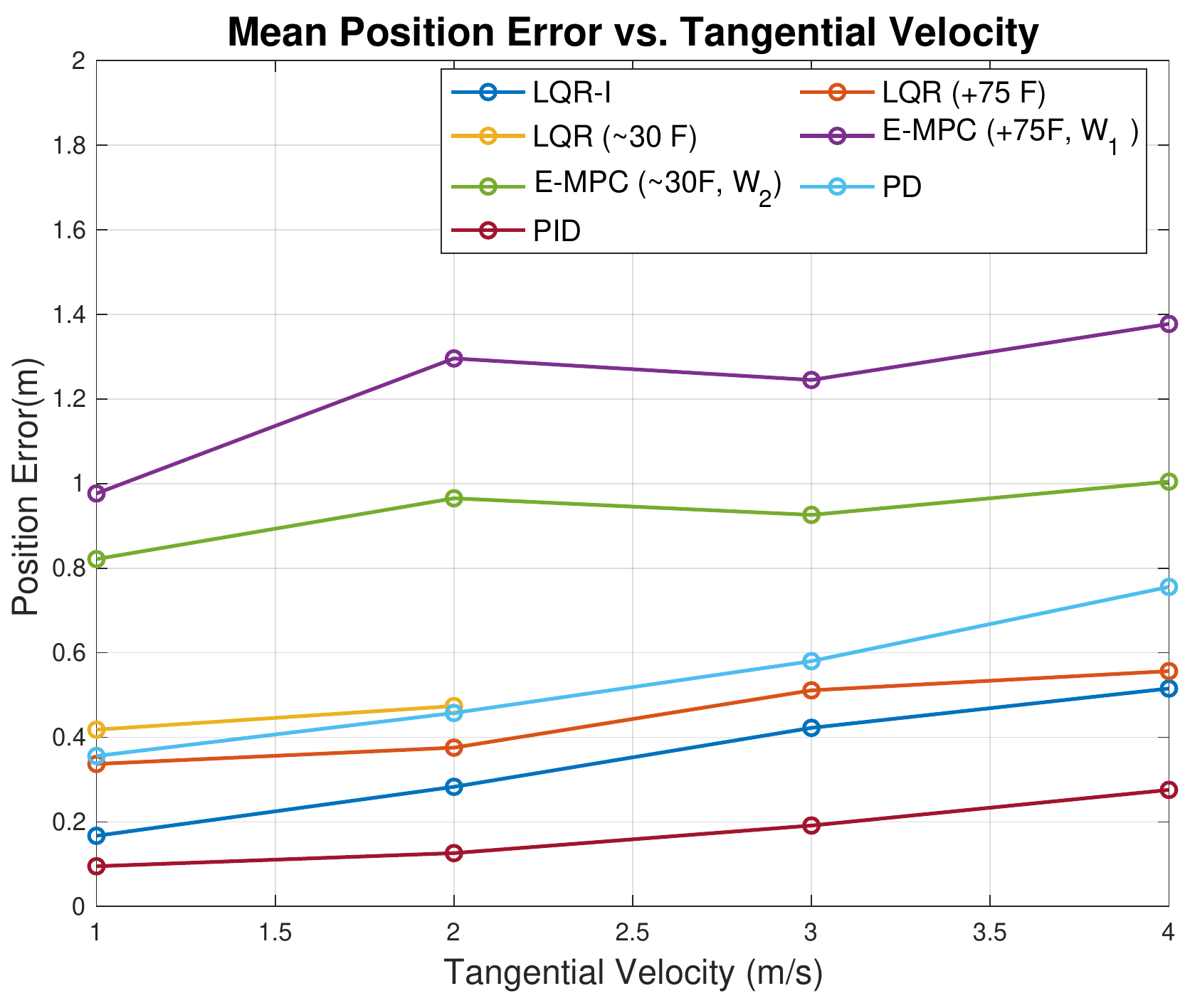}
\caption{Mean position error plot }
\label{Fig:position_error}
\end{subfigure}
\begin{subfigure}{\columnwidth}
\includegraphics[width=\linewidth, height=7cm]{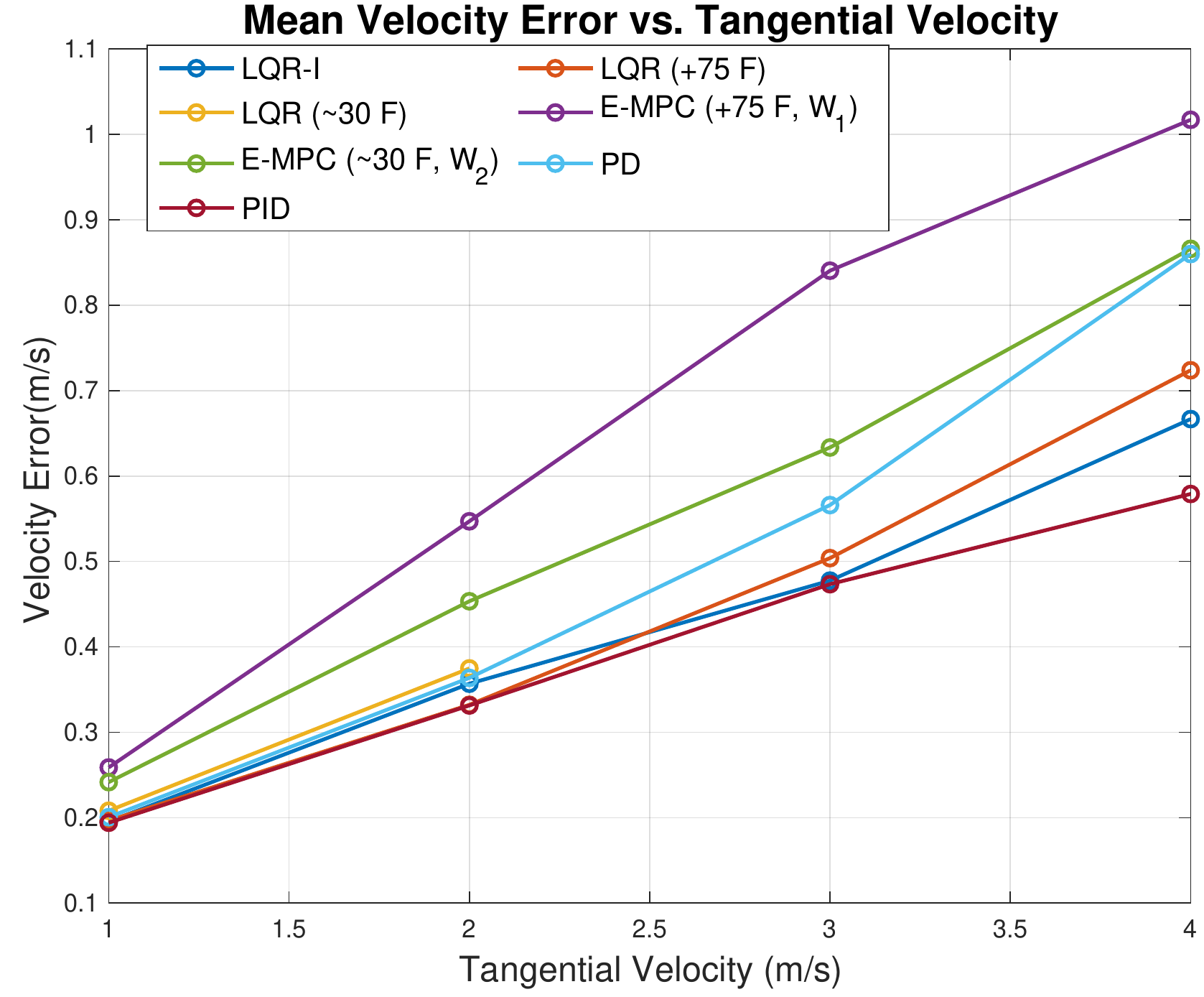}
  \caption{Mean velocity error plot}\label{Fig:velocity_error}
\end{subfigure}  
\caption{Results from outdoor quadcopter experiments, with different ambient temperatures for LQR and E-MPC.}
\label{Fig:exp_results}
\end{figure}

Fig. \ref{Fig:exp_results} provides controller performance details at different tangential velocities. The position and velocity errors are calculated by taking the norm of the error vector. It can be seen that an increase in tangential velocity along the circle led to an increase in position and velocity error. 

The control effort of each controller was calculated using the following formula:
\begin{equation}
    {\rm Control} \: {\rm Effort} = \frac{||u||_2}{T_{\rm total}},
\end{equation}
where $u$ is the control input to the motors and $T_{\rm total}$ is the total time to execute the circular trajectory.

\begin{figure}[h!]
    \centering
    \includegraphics[width=\columnwidth]{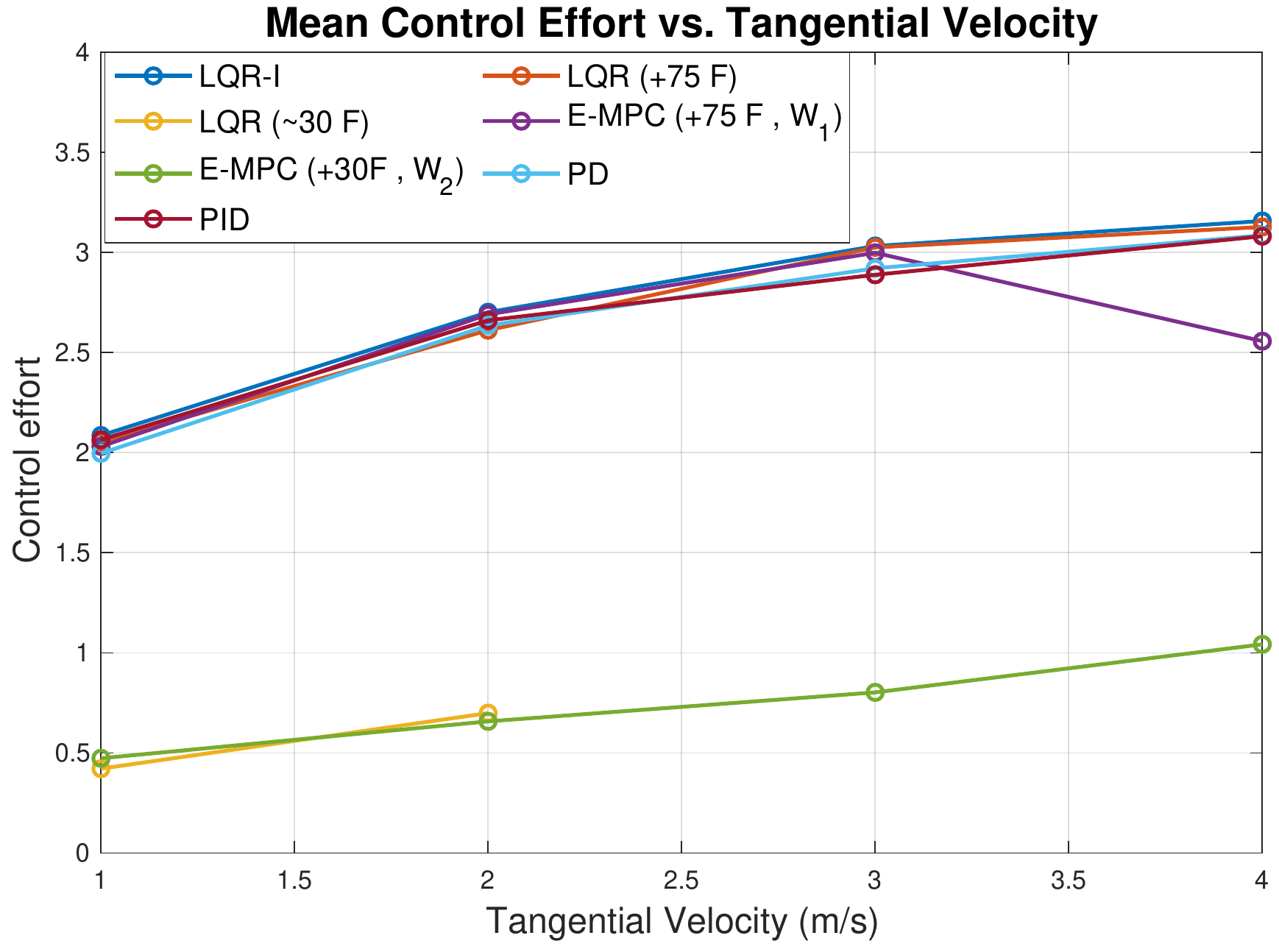}
    \caption{Control effort plot for quadcopter tracking an inclined circle, with different ambient temperatures for LQR and E-MPC.}
    \label{Fig:control_effort}
\end{figure}

Control effort is presented in Fig. \ref{Fig:control_effort}. Nearly all the controllers had similar control effort. However, a significant change in control effort was observed with a change in operating temperatures.

\begin{figure}[h!]
    \centering
    \includegraphics[width=\columnwidth]{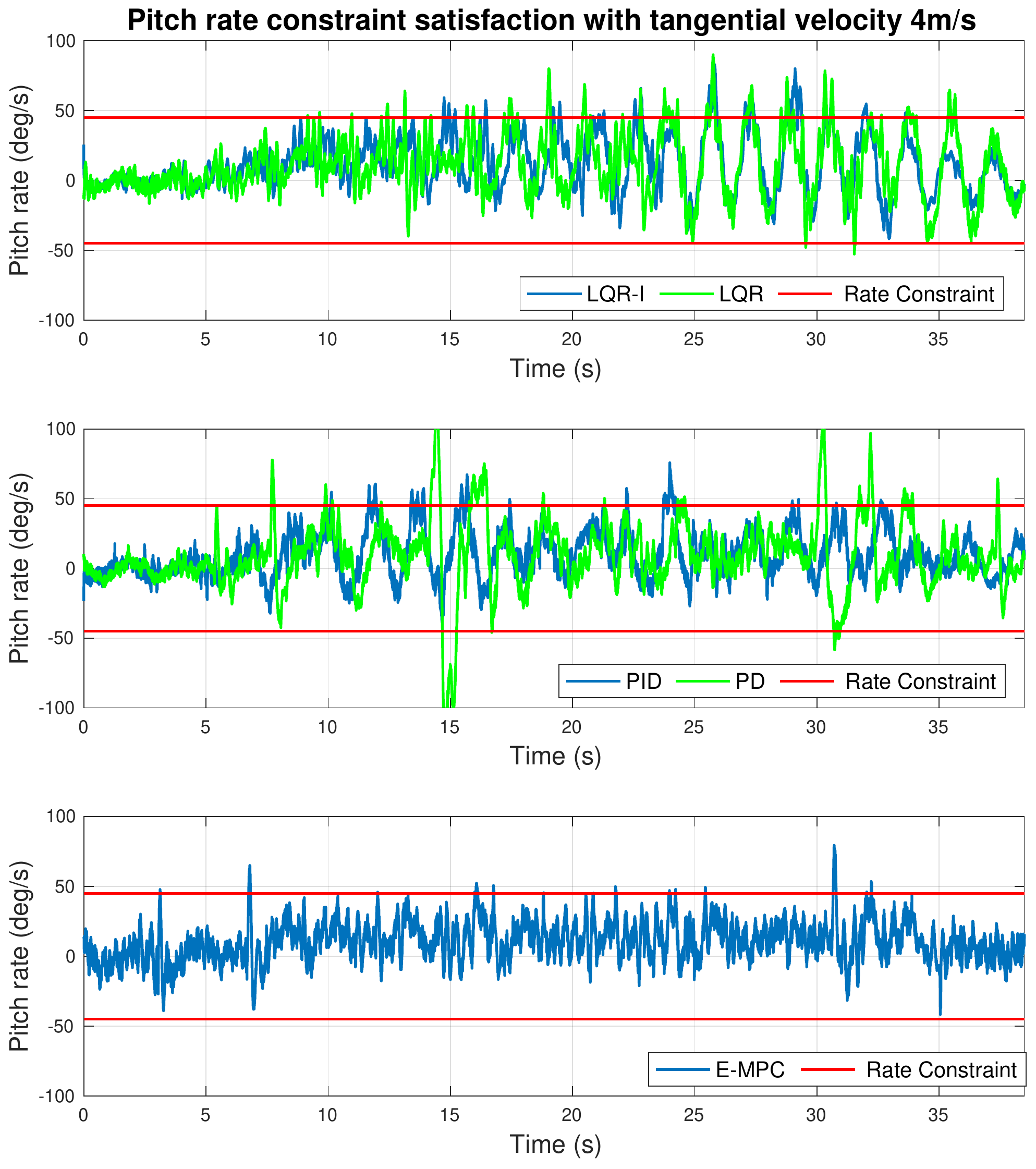}
    \caption{Pitch rate constraint satisfaction supported by E-MPC compared to other control schemes, while executing an inclined circle with tangential velocity of 4 m/s}
    \label{Fig:constraint_pitchRate}
\end{figure}

\begin{figure}[h!]
    \centering
    \includegraphics[width=0.9\columnwidth]{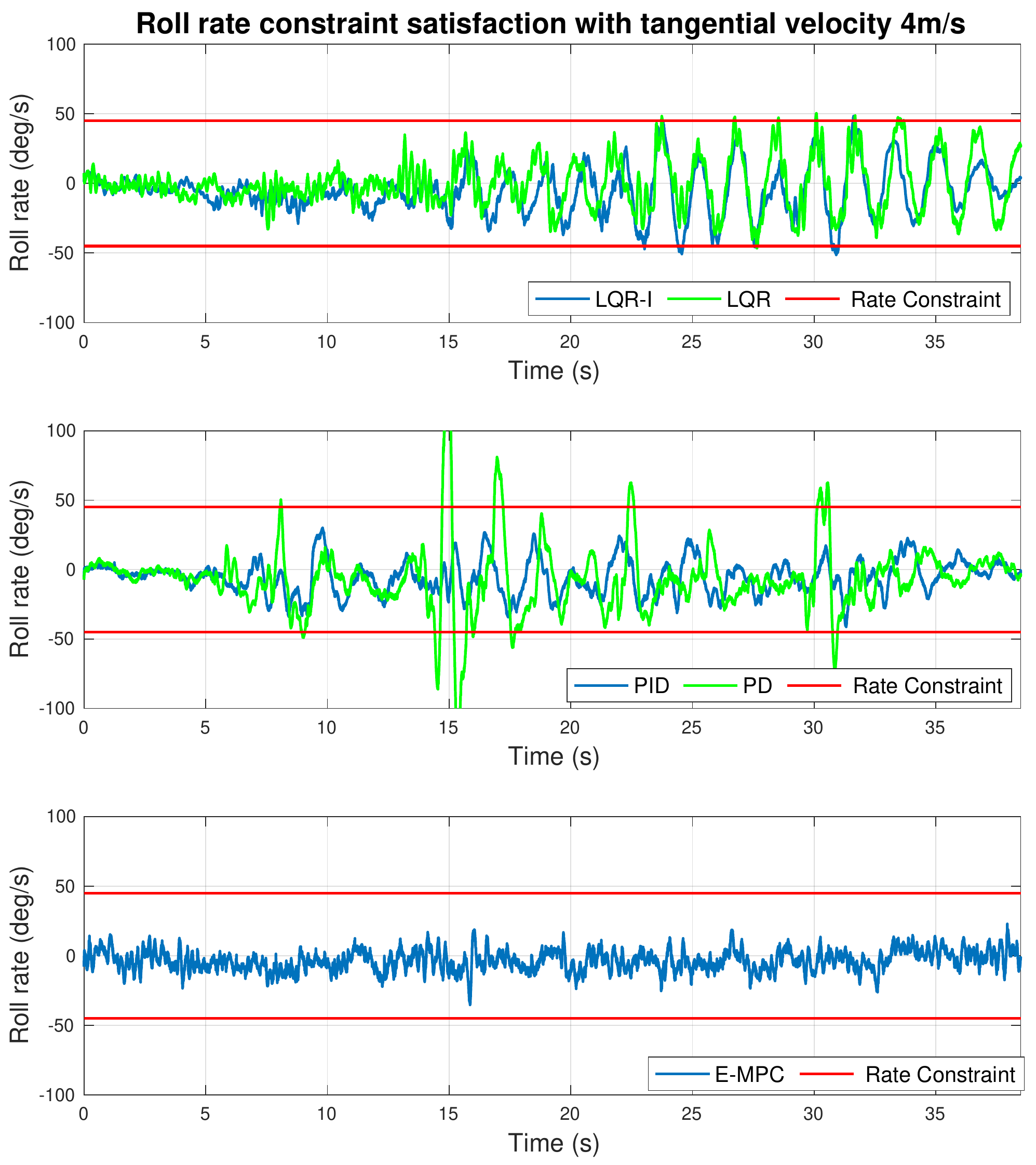}
    \caption{This plot illustrates the roll rate constraint satisfaction capability of E-MPC compared to other control schemes, while executing an inclined circle with tangential velocity of 4m/s}
    \label{Fig:constraint_rollRate}
\end{figure}

 The E-MPC control scheme offers the capability to handle state constraints.
 In the conducted experiments, roll rate and pitch rate constraints were imposed to restrict aggressive attitude changes that might destabilize the system.
 It can be seen in Figs. \ref{Fig:constraint_pitchRate} and \ref{Fig:constraint_rollRate} that E-MPC satisfies a maximum 45 deg/s magnitude rate constraint
 %
 %
 for a tangential velocity of $4$ m/s.
 In the the case of the other control schemes, constraints were violated as there was no mechanism to enforce them.
 %
 %
Example results presented here are for one flight; however, similar outcomes were observed in the other experiments.
 %
 %
 The benefits of constraint satisfaction could only be observed at higher velocities since, at lower velocities, the rates were within limits for all controllers. Reference trajectories with a tangential velocity of 5 m/s could only be completed when using the E-MPC controller. When the other control schemes (PD, PID, LQR and LQR-I) were used, the higher speeds required for trajectory tracking resulted in crashes when moving downwards due to the increased momentum, which E-MPC managed to constrain. This pattern suggests that  detuning the gains of the other control schemes to slow down the translational response could have prevented the crashes at the expense of tracking performance; however, that study is left for future work.

\subsection{Data Playback for Controller Timing Analysis}
In order to benchmark onboard controller computation times, experimental data was played back using the Beaglebone Blue on a benchtop.
In these simulations, only the controllers were executed to determine the amount of time required to generate a control command in each loop.
The results from the data playback simulations are presented in Table \ref{Table:cntrltime}.
It was observed that, on average, the LQR controller took the least amount of time to compute the control command while E-MPC took the most.
Note that the mean computation times of all controllers were less than the time between system interrupts ($T_{\rm s} = 5000$ $\mu$s) and that only the worst case computations times of PID and LQR-I exceeded this time.
%
%

\begin{table}[hbt!]
\caption{Controller computation time from data playback}
\centering
\resizebox{\columnwidth}{!}{%
\begin{tabular}{lccc}
\toprule
 \multirow{ 2}{*}{\textbf{Controller}} &  \multicolumn{2}{c}{\textbf{Computation Time}} \textbf{(}$\bm{\mu}$\textbf{s)} & \multirow{2}{*}{\begin{tabular}{c}\textbf{Worst Case} \\($\bm{\mu}$\textbf{s})\end{tabular}} \\ \cmidrule{2-3}
  & \textbf{Mean} & $\bm{\sigma}$  &  \\\midrule
 LQR   & 5.2468  &  3.0878   &  314\\
 LQR-I & 6.7263  &  35.0383  &  8163 \\
 PD    & 8.8461  &  7.6955   &  1424\\
 PID   & 9.3786  &  29.3469  &  5071\\
 E-MPC & 13.6239 &  6.6481   &  1301\\
\bottomrule
\label{Table:cntrltime}
\end{tabular}
}
\end{table}

\section{Discussion}
\label{section:discussion}

Fig. \ref{Fig:exp_results} shows that, out of all controllers, the implemented PID controller had the best tracking performance.
However, while the same controller parameters were used in both the simulations and experiments for the LQR, LQR-I, and MPC cases, different parameter sets were required for the PD and PID simulations and experiments (as was explained in Section \ref{subsec:PID_PD}).
Hence, it can be argued that LQR-I offered a better compromise between tracking performance and maintaining consistent simulation and experimental performance.

With an increase in tangential velocity, an increase in position and velocity error was observed.
This can be attributed to the inertia of the vehicle which resists changes in attitude and position, especially at higher velocities.
Furthermore, due to their low KV, the motors lacked the control authority to rapidly change quadcopter attitude.
However, an increase in motor KV may require different battery characteristics, which may increase the weight and require a different set of propellers.
Hence, an appropriate combination of motors and propellers is required for satisfactory flight performance.
%

%
Fig. \ref{Fig:control_effort} shows insignificant control effort variation across controllers under similar flight conditions.
%
%
However, lower ambient temperatures reduced the required control effort.
This can be attributed to
%
%
the higher density of freezing air (-1 °C during the experiments), which required lower control effort to propel the quadcopter.
Similar results were presented in \citet{juan2017}, where more energy was consumed at higher altitudes with lower air density while less energy was consumed at lower altitudes with higher air density (for similar flight times).

Fig. \ref{Fig:constraint_pitchRate} shows E-MPC controller constraint violations at approximately $7$ s and $31$ s.
These spikes were attributed to random delays in MOCAP information, thus acting as noise for the system. Overall, the E-MPC scheme proved helpful in situations where aggressive trajectory tracking was required while satisfying constraints.

The data playback simulations showed that, on average, LQR took the least amount of time to compute control commands while E-MPC took the most.
This was due to the fact that E-MPC determines gains based on a lookup table as opposed to other controllers which use predefined gains.
These simulations also showed that the mean computation times of all controllers were less than the time between system interrupts ($T_{\rm s} = 5000$ $\mu$s) and that only the worst case computation times of PID and LQR-I exceeded this time.
This implies that all controllers will usually meet the time constraint imposed by the embedded system.
Furthermore, since the PID and LQR-I controllers had the best tracking performance and completed all test flights, it can be inferred that the cases where these don't meet the time constraint are rare and don't have a significant impact on tracking performance.
Since the only controllers that exceeded the time constraint feature an integrator, improving the computational efficiency of the algorithm that implements the integrator might decrease the probability of time constraint violation.

\section{Conclusions}
\label{section:conclusion}
This paper reviews quadrotor system modeling, state estimation, and control.  Multiple digital controllers were successfully designed based on a linearized quadcopter model and were implemented on a computationally limited embedded platform.
Tested controllers included PD, PID, LQR, LQR-I, and E-MPC.
These controllers were evaluated in the M-Air outdoor motion capture facility during inclined, circular flight sequences.
Various challenges such as limited onboard computational power and loss of MOCAP tracking during outdoor flight were encountered with mitigation methods provided.
Reported results for E-MPC, LQR, LQR-I, PD, and PID controllers analyzed position tracking error, velocity tracking error, and control effort.
Overall, PID had the best tracking performance, LQR-I offered a better compromise between tracking performance and maintaining consistent simulation and experimental performance, and LQR had the fastest average computation time. 
All controllers satisfied the time constraint imposed by the embedded system.

If constraints must be met on a computationally limited platform, E-MPC provides an effective solution by computing gains offline and using a lookup table in-flight.  However, E-MPC must rely on a backup (e.g., LQR) when conditions stray from precomputed cases.
%
%
Controller selection should be based on factors such as available computational power, mission specifications (e.g. velocity constraints), and a designer's ability to determine optimal gains and performance.
 Appropriate motor and propeller selection is also a key element in quadcopter design.
%
%
For example, trajectories requiring aggressive velocities or quick changes in direction need higher KV motors.

Note that, other than PID, most of the control approaches discussed in this paper require modelling information. ``Model-free'' approaches to quadcopter control have been proposed and successfully implemented, as shown in \cite{al2016,bekcheva2018} for example, where an ultra-local model is estimated online and used to compensate for uncertainties and disturbances. The comparison with such model-free approaches is left as a subject for future research.
%
%
%

The authors hope this paper will serve as a comprehensive guide for engineers needing to select, design, implement, and tune a quadcopter digital control solution.

\section*{Appendix}

\begingroup
\begin{appendices}

\section{Propulsion Modeling} \label{propapp}
The thrust $T$ (in N) and torque M (in N $\cdot$ m) generated by a propeller spinning at $\omega$ RPM are given by,
\begin{equation}
\label{eqthrust}
\begin{aligned}
T=C_{\rm T}\omega^2,
\end{aligned}
\end{equation}
\begin{equation}
\label{eqmoment}
\begin{aligned}
M=C_{\rm M}\omega^2,
\end{aligned}
\end{equation}
where $C_{\rm T}$ and $C_{\rm M}$ are computed through a least squares fit on the RPM versus Thrust and RPM versus Moment curves shown in Fig. \ref{propulsion_plots}. The R-squared value of the fit for $C_{\rm T}$ was $0.9983$, whereas the R-squared value for $C_{\rm M}$ was $0.9988$. These high R-squared values show that the fits were very accurate.
%
The motors were driven by a normalized control signal $\sigma$ that goes from $0$ (motor not spinning) to $1$ (full throttle). The steady state angular velocity of the propeller $\omega$ (in RPM) is fit with a linear model with respect to $\sigma$,
\begin{equation}
\label{eqsigma}
\begin{gathered}
w_{\rm ss}=C_{\rm R}\sigma + \omega_{\rm b},
\end{gathered}
\end{equation}
where $C_{\rm R}$ and $\omega_{\rm b}$ were obtained with a least squares fit on the curve $\sigma$ versus RPM curve shown in Fig. \ref{propulsion_plots}. The R-squared value of this fit was $0.9911$, showing that the linear model fit the experimental data accurately. It is interesting to note that the fit was worse for $\sigma<0.2$ and $\sigma \approx 1$ due to the existence of a dead zone for low $\sigma$ and saturation for high $\sigma$.
%
%
This wasn't problematic since quadcopters do not usually operate in those zones while flying.

Finally, the propeller experienced some transient behavior until it reached the steady state $\omega_{\rm ss}$ given by Eq. \eqref{eqsigma}. This was modeled as a low pass filter with time constant $T_{\rm m}$:
\begin{equation}
\label{eqtm}
\omega=\dfrac{1}{T_{\rm m} {\rm s}+1}w_{\rm ss}.
\end{equation}
The time constant $T_{\rm m}$ was obtained by analyzing the step response of the system (see Fig. \ref{propulsion_plots}).
Note that $T_{\rm m}$ is the time that the system in Eq. \ref{eqtm} needs to reach $63.2\%$ of the final step value.

%
The following plots have been included in this section:
\begin{itemize}
\item RPM versus PWM, Thrust versus PWM, Efficiency versus RPM, Thrust versus RPM, Torque versus RPM, Step Response, $\sigma$ versus RPM for 3S batteries and 8 inch propellers.
\item RPM versus Efficiency compares different combinations of batteries and propeller size. The efficiency peak that appears at $2000$ RPM for the 12 V / 8 inch curve is because the motor's mechanical power output is very low at low RPM. Then, for a range of RPM around $2000$, it increases much faster than the electrical power provided to the motor, generating the peak.
\end{itemize}

From these plots, it was determined that a larger propeller (10 inches) was better for efficiency
because bigger propellers have a larger surface in contact with the air.
In general, larger propellers paired with lower KV motors are more efficient than shorter propellers matched with higher KV motors. The testing quadcopter had a constraint on propeller size due to the propeller guards. Thus, the 8 inch propeller was chosen. 

%
Regarding battery voltage, it is desirable that, should a designer choose a higher battery voltage, a smaller propeller size should be used.
The loss of efficiency at high RPM with the 16 V battery may have been due to the increased energy losses.

In summary, the pairing of a 12 V battery, motor, and 8-in propeller was adequate for the quadcopter as it provided a reasonable efficiency (around $70\%$) for the expected typical flight condition.


The mass and inertia were computed using two methods, namely a lumped mass model, and by measuring the physical properties using a scale and a bifilar pendulum. 

\begin{figure*}[t!]
    \centering
   \includegraphics[width=0.95\textwidth]{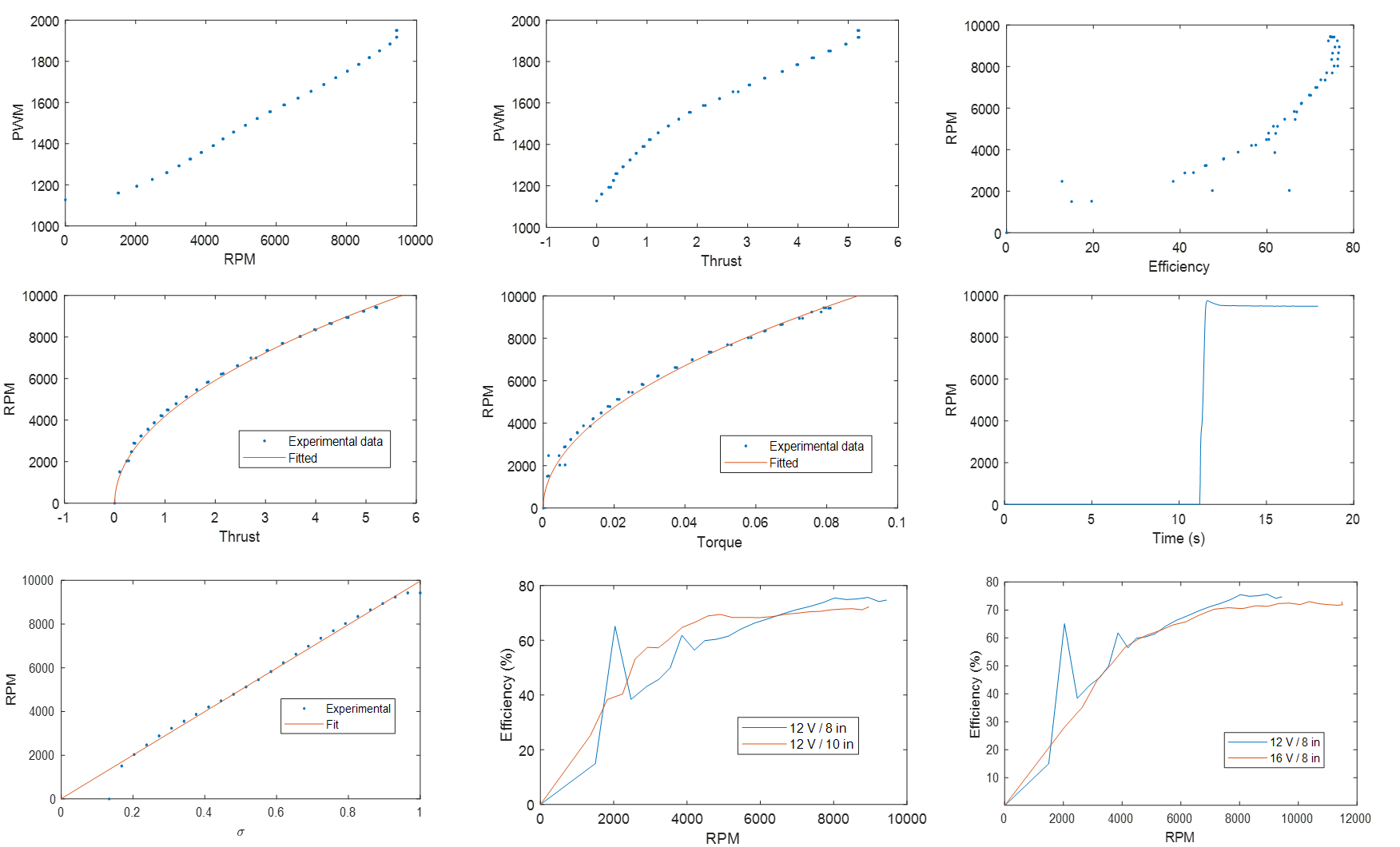}
    \caption{Dynamometer experimental results. Propulsion system parameters were obtained through a least squares curve fit.}
    \label{propulsion_plots}
\end{figure*}

The total mass of the quadcopter as determined by each method was the following:
\begin{equation*}
m_{\textrm{lumped model}}=0.997 \textrm{ kg}, \quad m_{\textrm{scale}}=1.062\textrm{ kg}
\end{equation*}

The inertia tensor (in $kg\cdot m^2$) computed with the lumped mass model was as follows

\vspace{-0.75em}
\footnotesize
\begin{equation*}
    J=\begin{bmatrix} 0.0109 & -4.3412\cdot 10^{-6} & 3.6716\cdot 10^{-5} \\
    -4.3412\cdot 10^{-6} & 0.0109 & -2.2186\cdot 10^{-5} \\
    3.6716\cdot 10^{-5} &  -2.2186\cdot 10^{-5} & 0.0200
    \end{bmatrix}.
\end{equation*}
\normalsize

The products of inertia were assumed to be zero due to the symmetry in the mass configuration of the quadcopter. The moments of inertia were computed using the bifilar pendulum method. If the pendulum period is $T_{\rm pend}$, then the moment of inertia will be given by
\begin{equation}
    J=\dfrac{mgd^2}{16\pi^2L}T_{\rm pend}^2.
\end{equation}
To determine the pendulum period, 
%
%
two methods were used
(see Fig. \ref{period_pendulum}):
\begin{enumerate}
    \item Inspection of the Time versus Orientation plot.
    \item Computation of the periodogram of the Time versus Orientation plot and identification of the pendulum period from the peak.
\end{enumerate}
\begin{figure}[t!]
    \centering
    \includegraphics[width=0.82\columnwidth]{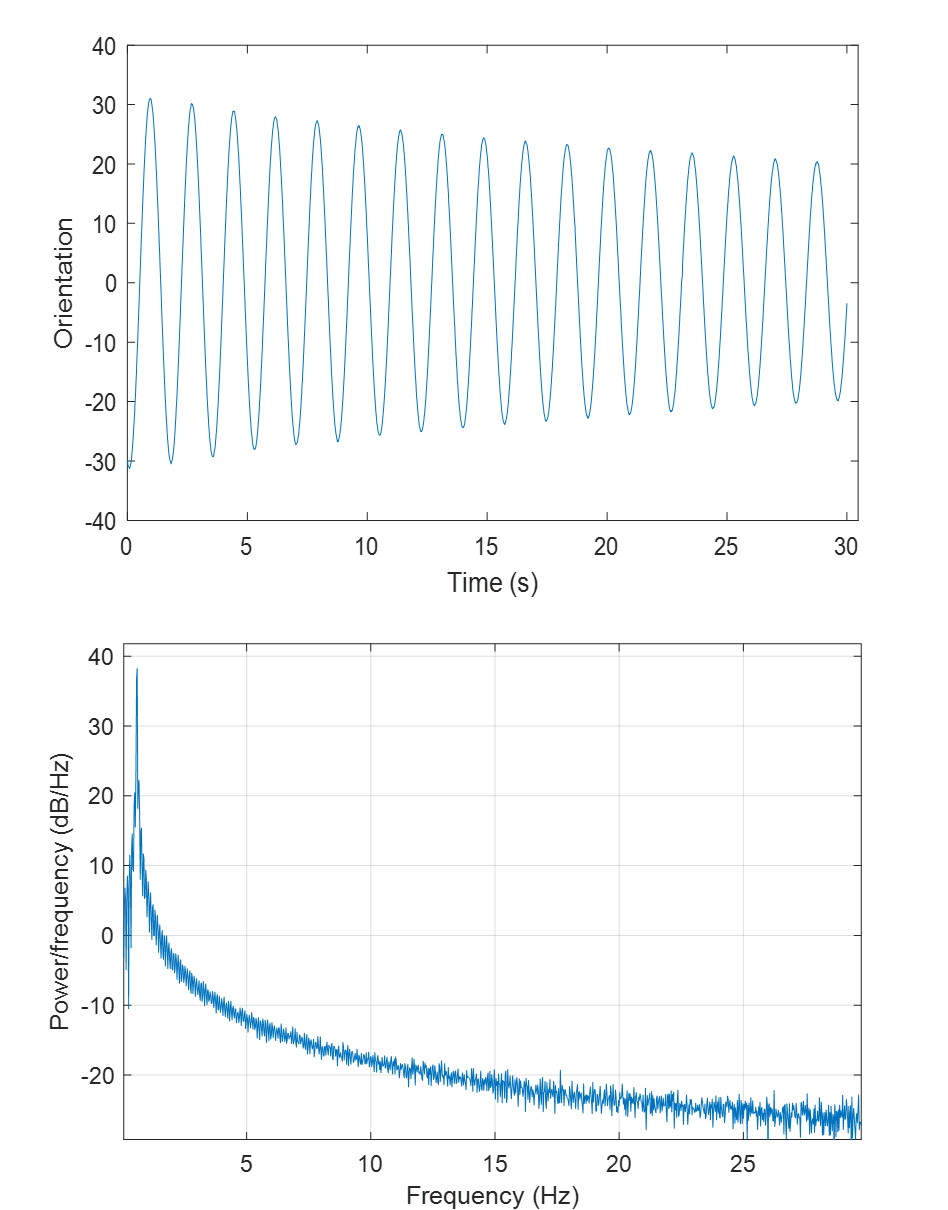}
    \caption{Results of the bifilar pendulum experiment for the characterization of $J_{zz}$. The top plot is the orientation of the drone against time, the bottom plot is the periodogram of the time series given by the top plot.}
    \label{period_pendulum}
\end{figure}
Both methods yielded very similar results,
%
%
so, in the end, 1) was arbitrarily chosen.
The resulting moments of inertia (in kg $\cdot$ $\text{m}^2$) were:
\begin{equation*}
    J_{xx}=0.0107, \quad J_{yy}=0.0111, \quad J_{zz}=0.0229.
\end{equation*}
Note how similar the results were between the lumped mass model and the bifilar pendulum experiments. The differences among them were attributed to erroneous and noisy measurements of distance, weight, and pendulum period.

\section{Vertical Position \& Velocity Estimator}
\label{append_vert_estimator}
\subsection{Discrete Gauss-Markov Prediction Model} 

\vspace{-0.75em}
\footnotesize
\begin{align}
    \renewcommand*{\arraystretch}{1.5}
    \begin{bmatrix} z_{k+1} \\ \dot{z}_{k+1} \\ {\ddot{z}}_{\text{bias},k+1} \end{bmatrix} &= 
    \begin{bmatrix} 1 & T_s & 0 \\ 0 & 1 & -T_s \\ 0 & 0 & 1 \end{bmatrix}
    \begin{bmatrix} z_{k} \\ \dot{z}_{k} \\ {\ddot{z}}_{\text{bias},k} \end{bmatrix} \nn \\
    &+ \renewcommand*{\arraystretch}{1.5}\begin{bmatrix} \frac{T_s^2}{2m} \\ \frac{T_s}{m} \\ 0 \end{bmatrix}
    (-\ddot{z}_{k} - g) +
    \renewcommand*{\arraystretch}{1.5}\begin{bmatrix} \frac{T_s^2}{2m} \\ \frac{T_s}{m} \\ 0 \end{bmatrix}
    w_{\ddot{z}_{k}},
\end{align}
\normalsize
where ${\ddot{z}}_{\text{bias},k}$ is the estimated accelerometer bias in the $z$-axis at step $k$ and $w_{\ddot{z}_{k}}$ represents the process noise that acts upon the acceleration $\ddot{z}$ at step $k.$

\subsection{Discrete Gauss-Markov Measurement Model}

\vspace{-0.75em}
\footnotesize
\begin{align}
    \renewcommand*{\arraystretch}{1.5}
    \begin{bmatrix} z_{\text{baro},k} \\ z_{\text{rf},k} \\ z_{\text{MOCAP},k} \end{bmatrix} &= 
    \begin{bmatrix} -1 & 0 & 0 \\ -1 & 0 & 0 \\ -1 & 0 & 0 \end{bmatrix}
    \begin{bmatrix} z_k \\ \dot{z}_k \\ \ddot{z}_{\text{bias},k} \end{bmatrix} \nn \\
    &+ \begin{bmatrix} \text{b}_{\text{baro}} \\ \text{b}_{\text{rf}} \\ 0  \end{bmatrix} +
    \begin{bmatrix} n_{\text{baro}} \\ n_{\text{rf}} \\ n_{z, \text{MOCAP}}  \end{bmatrix}
\end{align}
\normalsize
where $z_{\text{baro},k},$ $z_{\text{rf},k},$ and $z_{\text{MOCAP},k}$ are, respectively, the barometer, range finder, and MOCAP altitude measurements at step $k,$ $\text{b}_{\text{baro}}$ and $\text{b}_{\text{rf}}$ are, respectively, the barometer and range finder biases, and $n_{\text{baro}},$ $n_{\text{rf}},$ and $n_{z, \text{MOCAP}}$ represent, respectively, the barometer, range finder, and MOCAP sensor noise that acts upon the altitude measurements.

\subsection{Covariance Matrices}
The process covariance matrix $W$ was given by
\begin{equation*}
W = \text{diag}([ 0, \quad 0.1428, \quad 0.1 ]).
\end{equation*}
The sensor noise covariance matrix $V$ was given by
\begin{equation*}
V =  \text{diag}([ 1.00 \cdot 10^6, \quad 1.20 \cdot 10^{-6}, \quad 10^{-4} ]).
\end{equation*}

\section{Horizontal Position \& Velocity Estimator} \label{append_horiz_estimator}

\subsection{Discrete Gauss-Markov Prediction Model}

\vspace{-0.75em}
\footnotesize
\begin{align}
    \renewcommand*{\arraystretch}{1.5}
    &\begin{bmatrix} x_{k+1} \\ y_{k+1} \\ \dot{x}_{k+1} \\ \dot{y}_{k+1} \\ \ddot{x}_{\text{bias},k+1} \\ \ddot{y}_{\text{bias},k+1} \end{bmatrix}= 
    \begin{bmatrix} 1 & 0 & T_s & 0 & 0 & 0 \\ 0 & 1 & 0 & T_s & 0 & 0 \\ 0 & 0 & 1 & 0 & -T_s & 0
    \\ 0 & 0 & 0 & 1 & 0 & -T_s \\ 0 & 0 & 0 & 0 & 1 & 0 \\ 0 & 0 & 0 & 0 & 0 & 1\end{bmatrix}
    \begin{bmatrix} x_{k} \\ y_{k} \\ \dot{x}_{k} \\ \dot{y}_{k} \\ \ddot{x}_{\text{bias},k} \\ \ddot{y}_{\text{bias},k} \end{bmatrix} \nn \\
    &+ 
    \begin{bmatrix} \frac{T_s^2}{2m} & 0 \\ 0 & \frac{T_s^2}{2m} \\ \frac{T_s}{m} & 0 \\ 0 & \frac{T_s}{m}\\ 0 & 0 \\ 0 & 0 \end{bmatrix}
    \begin{bmatrix} \ddot{x}_{k} \\ \ddot{y}_{k} \end{bmatrix} +
    \begin{bmatrix} \frac{T_s^2}{2m} & 0 \\ 0 & \frac{T_s^2}{2m} \\ \frac{T_s}{m} & 0 \\ 0 & \frac{T_s}{m}\\ 0 & 0 \\ 0 & 0 \end{bmatrix}
    \begin{bmatrix} w_{\ddot{x}_k} \\ w_{\ddot{y}_k} \end{bmatrix},
\end{align}
\normalsize
where ${\ddot{x}}_{\text{bias},k}$ and ${\ddot{y}}_{\text{bias},k}$ are the estimated accelerometer biases at step $k$ in the $x$-axis and $y$-axis respectively, and $w_{\ddot{x}_{k}}$ and $w_{\ddot{y}_{k}}$ represent the process noise at step $k$ that acts upon the accelerations $\ddot{x}$ and $\ddot{y}$ respectively.

\subsection{Discrete Gauss-Markov Measurement Model}

\vspace{-0.75em}
\footnotesize
\begin{align}
    \renewcommand*{\arraystretch}{1.5}
    \begin{bmatrix} x_{\text{MOCAP},k} \\ y_{\text{MOCAP},k} \end{bmatrix} &= 
    \begin{bmatrix} 1 & 0 & 0 & 0 & 0 & 0 \\ 0 & 1 & 0 & 0 & 0 & 0 \end{bmatrix}
    \begin{bmatrix} x_k \\ y_k \\ \dot{x}_k \\ \dot{y}_k \\ \ddot{x}_{\text{bias},k} \\ \ddot{y}_{\text{bias},k} \end{bmatrix} \nn \\ 
    &+ 
    \begin{bmatrix} n_{x, \text{MOCAP}} \\ n_{y, \text{MOCAP}}  \end{bmatrix}
\end{align}
\normalsize
where $x_{\text{MOCAP},k}$ and $y_{\text{MOCAP},k}$ are, respectively, the MOCAP $x$-axis and $y$-axis position measurements at step $k,$ and $n_{x, \text{MOCAP}}$ and $n_{y, \text{MOCAP}}$ represent the MOCAP sensor noise that acts upon the $x$-axis and $y$-axis position measurements respectively.

\subsection{Covariance Matrices}
%
%
The process covariance matrix $W$ was given by
\begin{equation*}
W = \text{diag}([ 0, \quad 0, \quad 0.2749, \quad  0.0328, \quad 0.1, \quad 0.1 ]).
\end{equation*}
The sensor noise covariance matrix $V$ was given by
\begin{equation*}
V =  \text{diag}([10^{-4}, \quad 10^{-4} ]).
\end{equation*}

\section{Yaw \& Yaw Rate Estimator} \label{append_yaw_estimator}

\subsection{Discrete Gauss-Markov Prediction Model}

\vspace{-0.75em}
\footnotesize
\begin{equation}
    \renewcommand*{\arraystretch}{1.5}
    \begin{bmatrix} \psi_{k+1} \\ \omega_{z,k+1}^{\rm b} \\ \omega_{z, \text{bias}, k+1}^{\rm b} \end{bmatrix} = 
    \begin{bmatrix} 1 & T_s & -T_s \\ 0 & 1 & 0 \\ 0 & 0 & 1 \end{bmatrix}
    \begin{bmatrix} \psi_k \\ \omega_{z,k}^{\rm b} \\ \omega_{z, \text{bias}, k}^{\rm b} \end{bmatrix},
\end{equation}
\normalsize
where $\omega_{z, \text{bias}, k}^{\rm b}$ is the rate gyro bias around the $z$-axis at step $k.$

\subsection{Discrete Gauss-Markov Measurement Model}

\vspace{-0.75em}
\footnotesize
\begin{equation}
    \renewcommand*{\arraystretch}{1.5}
    \begin{bmatrix} \psi_{\text{MOCAP},k} \\ \omega_{z, \text{gyro},k}^{\rm b} \end{bmatrix} = 
    \begin{bmatrix} 1 & 0 & 0 \\ 0 & 1 & 0 \end{bmatrix}
    \begin{bmatrix} \psi_k \\ \omega_{z,k}^{\rm b} \\ \omega_{z, \text{bias}, k}^{\rm b} \end{bmatrix} + 
    \begin{bmatrix} n_{\psi, \text{MOCAP}} \\ n_{\text{gyro}}  \end{bmatrix},
\end{equation}
\normalsize
where $\psi_{\text{MOCAP},k},$ and $\omega_{z, \text{gyro},k}^{\rm b}$ are, respectively, the MOCAP yaw and the rate gyro yaw rate measurements at step $k,$ and $n_{\psi, \text{MOCAP}},$ and $n_{\text{gyro}}$ represent, respectively, the MOCAP and rate gyro sensor noise that acts upon the yaw and yaw rate measurements.

\subsection{Covariance Matrices}
%
%
The process covariance matrix $W$ was given by
\begin{equation*}
W = \text{diag}([ 0, \quad 0, \quad 0.1 ]).
\end{equation*}
The sensor noise covariance matrix $V$ was given by
\begin{equation*}
V =  \text{diag}([10^{-4}, \quad -0.7822 ]).
\end{equation*}


\section{Sensor Calibration Results}
\label{append_calib_results}

Covariances used in-flight for motion capture measurements were set to higher than measured values to assure the Kalman Filter correction step would not operate on near-singular matrices.
The motion capture biases were assumed to be zero because the camera system was calibrated before use.
Accelerometer and gyro biases and slopes were originally determined from linear least squares fits on rate table experimental data as shown in Table \ref{parameters_sensor_calibration}. 
In practice, the biases were estimated on-the-fly with Kalman Filters.

\begin{table*}[t!]
\caption{Sensor Calibration Results}
\label{parameters_sensor_calibration}
\centering
\resizebox{\textwidth}{!}{%
\begin{tabular}{lccccc}
\toprule
\textbf{Sensor} & \textbf{Covariance Measured} & \textbf{Covariance Used} & \textbf{Bias} & \textbf{Slope}\\ \midrule
Accelerometer $\ddot{x}^{\rm b}$  & 0.2749 $\text{m}^2$ & Same & 7 MPU Counts & 417.30 ($\text{m}/\text{s}^2$)/MPU Count \\
Accelerometer $\ddot{y}^{\rm b}$  & 0.0328 $\text{m}^2$ & Same & 73 MPU Counts & 417.83 ($\text{m}/\text{s}^2$)/MPU Count \\
Accelerometer $\ddot{z}^{\rm b}$  & 0.1428 $\text{m}^2$ & Same & -36 MPU Counts & 419.84 ($\text{m}/\text{s}^2$)/MPU Count \\
Rate Gyro $\omega_{x}^{\rm b}$  & 0.8747 $(\text{deg}/\text{s})^2$ & Not Used & -1 MPU Count & 16.325 ($\text{deg}/\text{s}$)/MPU Count \\
Rate Gyro $\omega_{y}^{\rm b}$  & 0.7486 $(\text{deg}/\text{s})^2$ & Not Used & 0 MPU Counts & 16.394 ($\text{deg}/\text{s}$)/MPU Count \\
Rate Gyro $\omega_{z}^{\rm b}$  & 0.7822 $(\text{deg}/\text{s})^2$ & Same & -11 MPU Counts & 16.378 ($\text{deg}/\text{s}$)/MPU Count \\
Barometer $z$  & $0.0552$ $\text{m}^2$ & $10^{6}$ $\text{m}^2$ & Avg. of 50 samples & N/A \\
Range Finder $z^{\rm b}$  & $1.20 \cdot 10^{-6}$ $\text{m}^2$ & Same & Avg. of 10 samples & N/A \\
Motion Capture $x^{\rm i}$  & $2.25 \cdot 10^{-10}$ $\text{m}^2$ & $10^{-4}$ $\text{m}^2$ & 0 & N/A \\
Motion Capture $y^{\rm i}$  & $2.89 \cdot 10^{-10}$ $\text{m}^2$ & $10^{-4}$ $\text{m}^2$ & 0 & N/A \\
Motion Capture $z^{\rm i}$  & $7.84 \cdot 10^{-10}$ $\text{m}^2$ & $10^{-4}$ $\text{m}^2$ & 0 & N/A \\
Motion Capture $\psi$  & Not measured & $10^{-4}$ $\text{m}^2$ & 0 & N/A \\
\bottomrule
\end{tabular}
}
\end{table*}

\end{appendices}
\endgroup



\bibliographystyle{elsarticle-harv} \biboptions{authoryear}

\par\noindent 
\parbox[h!]{\linewidth}{
\noindent\parpic{\includegraphics[height=1.5in,width=1in,clip,keepaspectratio]{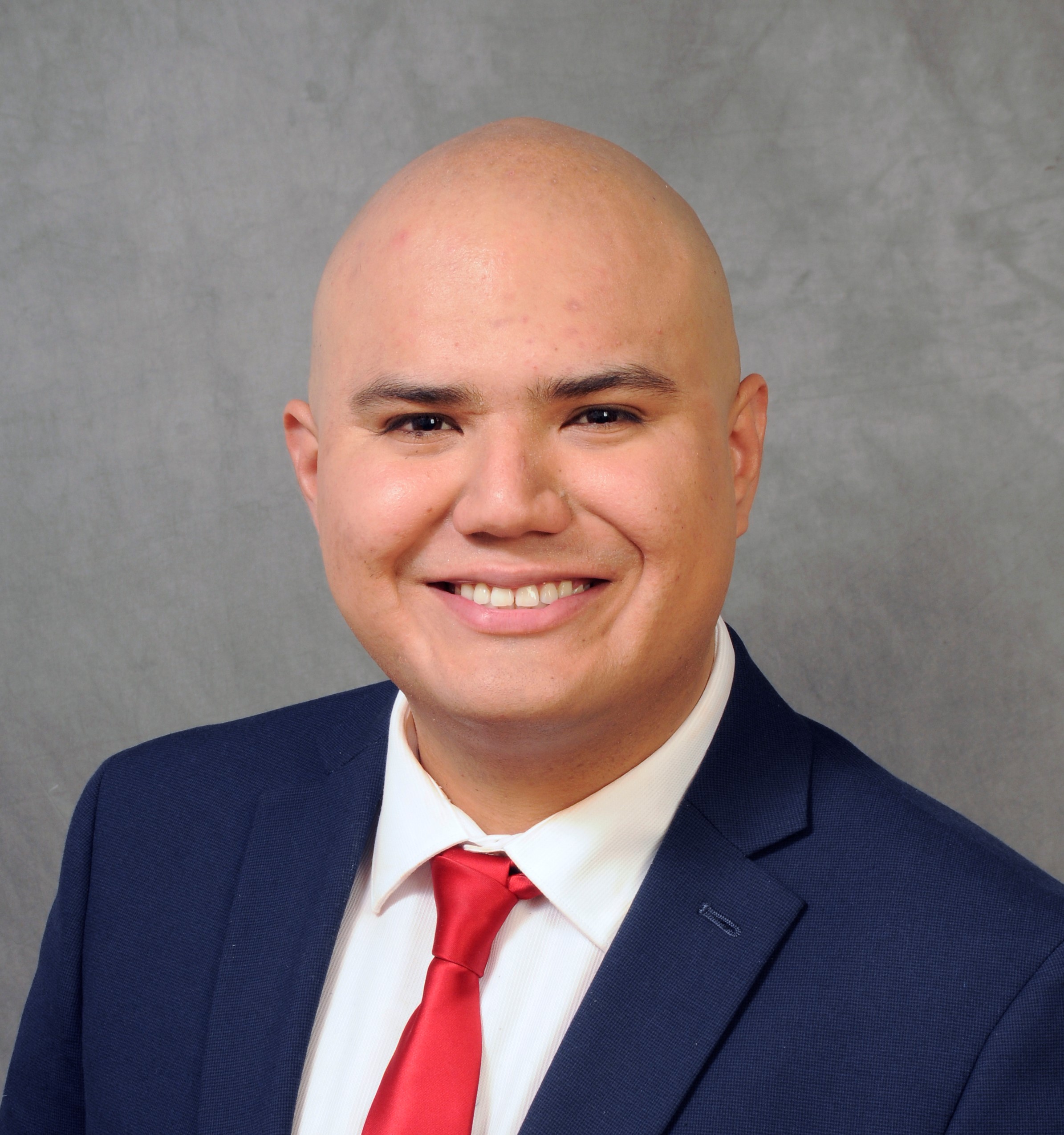}}
\noindent {\bf Juan Paredes}\
received the B.Sc. degree in mechatronics engineering from the Pontifical Catholic University of Peru and a M.Sc. degree in aerospace engineering from the University of Michigan in Ann Arbor, MI. He is currently a PhD candidate in the Aerospace Engineering Department at the University of Michigan.  His interests are in autonomous flight control and control of combustion.}

\par\noindent 
\parbox[t]{\linewidth}{
\noindent\parpic{\includegraphics[height=1.5in,width=1in,clip,keepaspectratio]{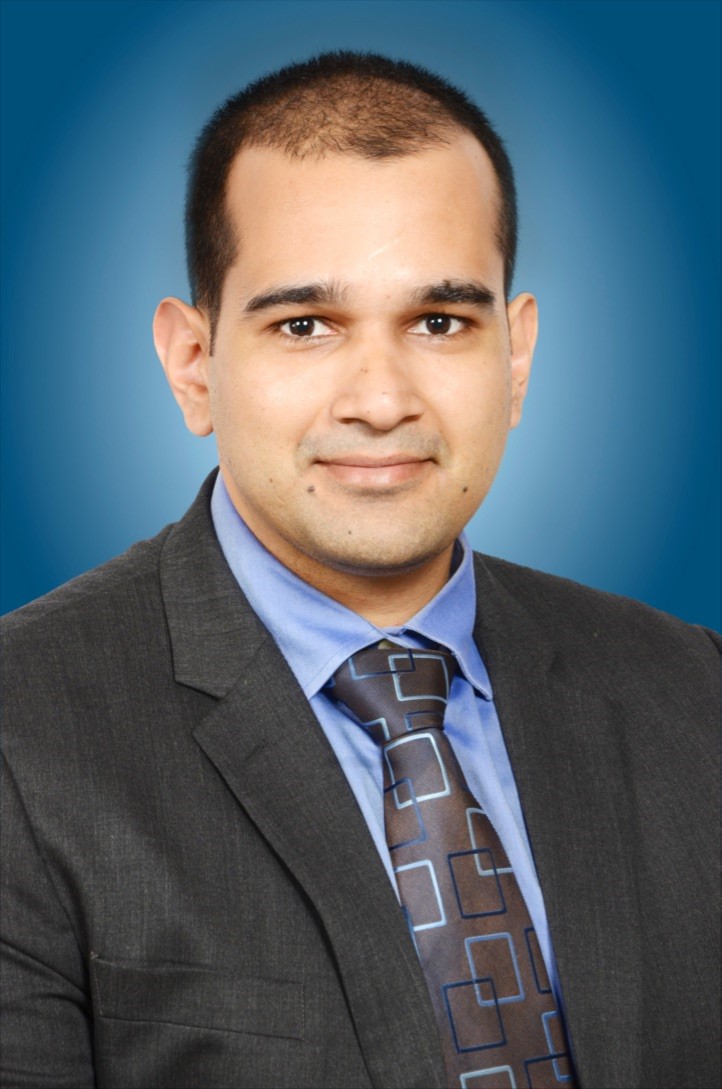}}
\noindent {\bf Prashin Sharma}\
is currently pursuing his PhD in Robotics from University of Michigan, Ann Arbor. His research is focused on prognostics informed decision making for safe autonomous drone flights. He has a combined work experience of six years in industries spanning from flat panel OLED displays to security. He holds a BE in Mechanical Engineering from Mumbai University and MS in Mechanical Engineering from Carnegie Mellon University.}

\par\noindent 
\parbox[t]{\linewidth}{
\noindent\parpic{\includegraphics[height=1.5in,width=1in,clip,keepaspectratio]{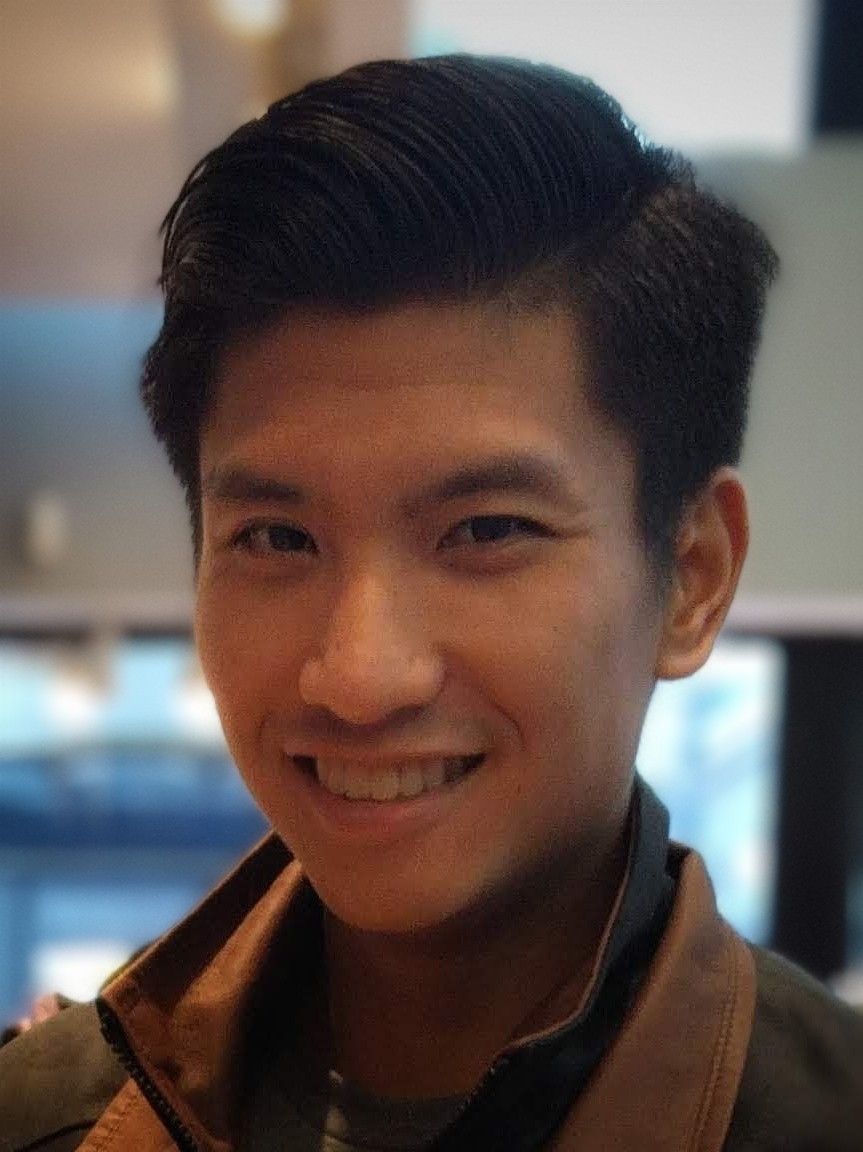}}
\noindent {\bf Brian Ha}\
is a professional aerospace engineer. His work experience includes satellite flight operations with Google Inc. and launch vehicle design with Northrop Grumman. He received his B.S. in Aerospace Engineering from San Jos\'e State University in 2015 and his M.S.E. in Aerospace Engineering from the University of Michigan in 2019. He specializes in the practical implementation of state estimation, constrained feedback control, and real-time optimization on embedded systems. }

\par\noindent 
\parbox[t]{\linewidth}{
\noindent\parpic{\includegraphics[height=1.5in,width=1in,clip,keepaspectratio]{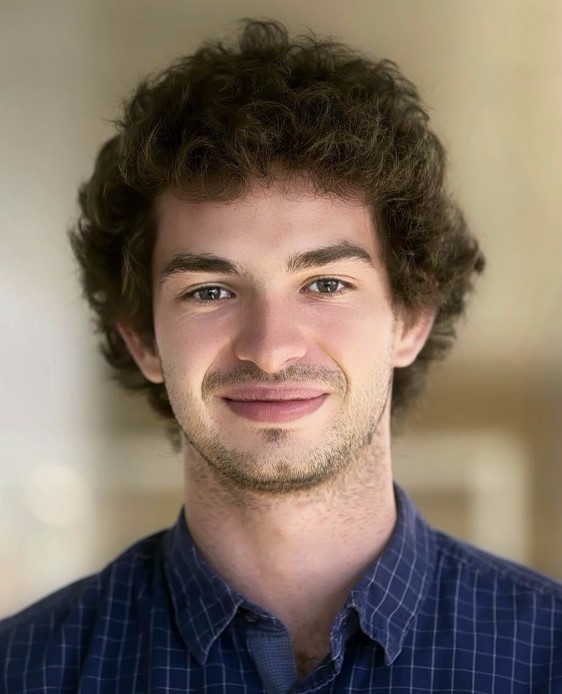}}
\noindent {\bf Manuel Lanchares}\
 is a PhD student in Aerospace Engineering at the Georgia Institute of Technology, where his research is focused on stochastic nonlinear control. His work experience includes computation of aircraft performance specifications and implementation of nonlinear methods for orbital covariance propagation. He received the BS in Aerospace Engineering from the Technical University of Madrid in 2016, the MS in Mathematical Engineering from Complutense University of Madrid in 2017, and the MSE in Aerospace Engineering from the University of Michigan in 2019.}

\par\noindent 
\parbox[t]{\linewidth}{
\noindent\parpic{\includegraphics[height=1.5in,width=1in,clip,keepaspectratio]{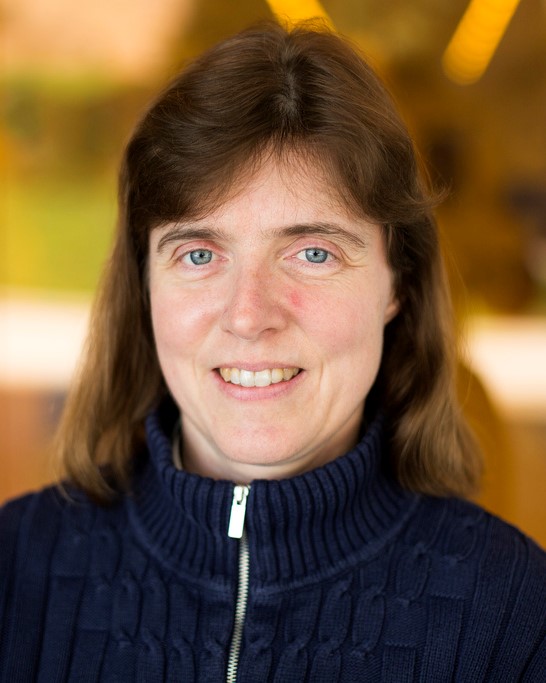}}
\noindent {\bf Ella Atkins}\
is a Professor of Aerospace Engineering at the University of Michigan, where she directs the Autonomous Aerospace Systems Lab.  Dr. Atkins holds B.S. and M.S. degrees in Aeronautics and Astronautics from MIT and M.S. and Ph.D. degrees in Computer Science and Engineering from the University of Michigan.  She is Editor-in-Chief of the AIAA Journal of Aerospace Information Systems (JAIS) and has served on the National Academy's Aeronautics and Space Engineering Board, the Institute for Defense Analysis Defense Science Studies Group, and in multiple AIAA technical committees. She pursues research in Aerospace system autonomy and safety and is currently engaged as a Technical Fellow at Collins Aerospace.}

\par\noindent 
\parbox[t]{\linewidth}{
\noindent\parpic{\includegraphics[height=1.5in,width=1in,clip,keepaspectratio]{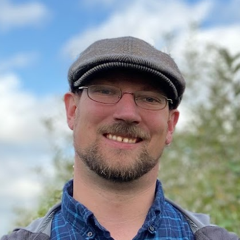}}
\noindent {\bf Peter Gaskell}\
 is a Lecturer and Adjunct Researcher at the University of Michigan Robotics Institute, where he develops robotics platforms for both education and research.  He researches efficient algorithms for localization and navigation of autonomous systems. Dr. Gaskell holds a B.S. degree in Physics from the University of Oregon, and M.Eng. and Ph.D. degrees in Electrical Engineering from McGill University.}

\par\noindent 
\parbox[t]{\linewidth}{
\noindent\parpic{\includegraphics[height=1.5in,width=1in,clip,keepaspectratio]{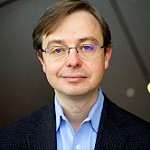}}
\noindent {\bf Ilya Kolmanovsky}\
is a professor in the department of aerospace engineering at the University of Michigan, with research interests in control theory for systems with state and control constraints, and in control applications to aerospace and automotive systems.  He received his Ph.D. degree from the University of Michigan in 1995.  He is a Fellow of IEEE and is named as an inventor on over 100 United States patents.}

\end{document}